\documentclass[aps,floatfix,superscriptaddress,twocolumn,nofootinbib,pra,10pt]{revtex4-2}

\usepackage{graphicx}
\usepackage{amsmath,amssymb,amsfonts}
\usepackage{url,hyperref}
\usepackage[version=4]{mhchem}
\usepackage{ulem}
\usepackage[caption=false]{subfig}
\usepackage{verbatim}
\usepackage{titlesec}
\usepackage{enumitem}
\usepackage{setspace}
\usepackage{bm}
\usepackage{makecell}
\usepackage{tabularx}
\usepackage{thmtools}
\usepackage{tikz}
\usetikzlibrary{calc,matrix,positioning}

\usepackage{algpseudocode}
\usepackage{algorithm}

\usepackage{array}      
\usepackage{multirow}   
\usepackage{pifont}

\newcommand{\cmark}{\ding{51}}
\newcommand{\xmark}{\ding{55}}

\newcommand{\AlgRule}{\Statex \hrulefill}
\newcommand{\AlgSection}[1]{\Statex \textbf{#1}}
\newcommand{\AlgSubsection}[1]{\Statex \textit{#1}}

\usepackage{color}
\usepackage{xcolor}
\usepackage{booktabs}

\begin{document}

\title{Design Space of Self-Consistent Electrostatic Machine Learning Interatomic Potentials}

\author{William J. Baldwin}
\email{wjb48@cam.ac.uk}
\affiliation{Engineering Laboratory, University of Cambridge, Trumpington St, Cambridge, UK}

\author{Ilyes Batatia}
\affiliation{Engineering Laboratory, University of Cambridge, Trumpington St, Cambridge, UK}

\author{Martin Vondr\'{a}k}
\affiliation{University of Bayreuth, Bavarian Center for Battery Technology (BayBatt), Bayreuth, Germany}
\affiliation{Fritz-Haber-Institute of the Max-Planck-Society, Berlin, Germany}

\author{Johannes T. Margraf}
\affiliation{University of Bayreuth, Bavarian Center for Battery Technology (BayBatt), Bayreuth, Germany}

\author{G\'{a}bor Cs\'{a}nyi}
\email{gc121@cam.ac.uk}
\affiliation{Engineering Laboratory, University of Cambridge, Trumpington St, Cambridge, UK}

\date{\today}

\begin{abstract}
Machine learning interatomic potentials (MLIPs) have become widely used tools in atomistic simulations. For much of the history of this field, the most commonly employed architectures were based on short-ranged atomic energy contributions, and the assumption of locality persists in many modern foundation models. While this approach has enabled efficient and accurate modelling for many use cases, it poses intrinsic limitations for systems where long-range electrostatics, charge transfer, or induced polarization play a central role. A growing body of work has proposed extensions that incorporate electrostatic effects, ranging from locally predicted atomic charges to electrostatically self-consistent models. While these models have demonstrated success for specific examples, their relationships, underlying assumptions, and fundamental limitations are not yet well understood. In this work, we present a systematic framework for treating electrostatics in MLIPs by viewing existing models as coarse-grained approximations to density functional theory (DFT). This perspective makes explicit the approximations involved, clarifies the physical meaning of the learned quantities, and reveals connections and equivalences between several previously proposed models. Using this formalism, we identify key design choices that define a broader design space of self-consistent electrostatic MLIPs. We implement salient points in this space using the MACE architecture and a shared representation of the charge density, enabling controlled comparisons between different approaches. Finally, we evaluate these models on two instructive test cases: metal–water interfaces, which probe the contrasting electrostatic response of conducting and insulating systems, and charged vacancies in silicon dioxide, where the correct distribution of excess charge between spatially separated defects represents a difficult test case. Our results highlight both the capabilities and limitations of existing approaches and demonstrate how more expressive self-consistent models are needed to resolve failures of simpler models.
\end{abstract}

\maketitle

\begingroup
\makeatletter
\let\l@subsubsection\@gobbletwo
\makeatother
\tableofcontents
\endgroup

\section{Introduction}

Machine learning interatomic potentials (MLIPs) are becoming a mainstream tool in atomistic simulations. Early MLIP architectures were formulated as short-range models, expressing the total energy as a sum of atom-centred contributions that depend only on the local geometric environment of each atom \cite{Behler2007GeneralizedSurfaces, Bartok2010GaussianElectrons, Drautz2019AtomicPotentials, MACE, grace, nequip, ANI12017, Schnet2017, MTP}. This locality assumption became a foundational design principle and has shaped much of the subsequent development of the field. There has been considerable work to address this limitation, but currently many widely used architectures, including most materials science foundation models \cite{MP0, M3GNET, aimnet2019, mattersim} and even some organic foundation models \cite{ESEN, rhodes2025orbv3atomisticsimulationscale}, are still short ranged. 

Examples of systems which cannot be handled with short-ranged models are easy to find. Any system which is interesting to study in more than one charge state (such as defects in some semiconductors \cite{Janotti2007NativeZnO, Milton2023DifferenceSiO2}) cannot be modelled with  MLIP architectures that depend on geometry alone, because geometry does not determine the state of the system. The structure of liquids with charged or dipolar components is also known to depend subtly on the long-range interactions between the particles \cite{Rodgers2008InterplayWater}. One might hope that short-range models can infer the correct behaviour only from the local geometry, but this is not always true: It has been shown that while short-range MLIPs can get bulk properties of polar liquids correct, they are systematically wrong for liquid-vapour interfaces compared to models with explicit electrostatics \cite{Niblett2021LearningInterfaces, parker2026falsemetallizationshortrangedmachine}. Finally, in biochemistry, classical force fields all have electrostatic energy terms. While this fact alone is not proof of necessity, electrostatic interactions are known to play an important role in the large-scale dynamics of biomolecular systems \cite{Zhou2018ElectrostaticCondensation}. Large biomolecules are often highly charged, and exist in a solution containing counter-ions where the balance between different electrostatic energy terms affects processes like protein folding and conformational change \cite{Zhou2005InteractionsEffect, Green1932STUDIESCONCENTRATION}. There is evidence that these effects are delicate and must be modelled in a thorough, physically correct way. An example of this is the stability of lipid bilayer membranes, where even truncation of Coulomb interactions in classical force fields leads to simulation artifacts \cite{PATRA20033636}.

Much progress has been made in addressing the electrostatics problem in the context of MLIPs, which will be reviewed below. The earliest milestone was the introduction of electrostatic energy terms from atomic partial charges, which are predicted from the local geometry. This was done in the ``3\textsuperscript{rd} generation'' HDNN models as well as other more recent works \cite{Morawietz2012ACharges, Artrith2011, Veit2020PredictingDipoles, popellier_multipoles_2008, physnet2019, tensormol, sherrill_electronpassing_2021, Zhang2024, Cheng2025LatentInteractions, Zhang2022AInteractions}. These models are sufficient for many applications, but there remain important phenomena that they cannot describe. This includes mutual induction between separated molecules, long-range charge transfer, or consistent treatment of external and internally generated fields. An alternative route is to construct more sophisticated models, often based on charge equilibration or incorporating elements of electronic structure theory \cite{4gnn_ko_2020, scfnn, eMLP, bpopnn, aimnet_nse, pqeq_hu_2025}. Such models have been shown to capture phenomena such as induced polarization and long-range charge transfer, but they also introduce new challenges. Classical charge equilibration has fundamental failure modes such as over-polarization and incorrect fragmentation of clusters \cite{Jensen2023UnifyingModels, conducting_molecules, LeeWarren2008OriginMethods}, and (as will be discussed) these problems can persist when the charge equilibration approach is applied to MLIPs. More complicated self-consistent models are significantly more cumbersome to train and use, and it is not clear whether the added flexibility is worthwhile.

Treating a wide variety of phenomena simultaneously is important because a major direction in the field of MLIPs is the development of foundation models. Foundation models have been transformative because one can perform qualitatively sound simulations of highly complex systems with off-the-shelf models \cite{MP0, sevennet, DPA2, CHGNET, mattersim}. The next generation of foundation models will be used to study everything from active interfaces to organic crystal structures, and hence any long-range components must be able to handle a wide range of phenomena. By their nature, foundation models need to extrapolate to new chemistries in order to fulfil their function, and we believe that this is best accomplished when the description of the long-range interactions has a physical basis.  

\subsubsection*{Existing Work on Electrostatic MLIPs}
\label{sec:existing_work}

We now provide a brief overview of existing MLIP models that include electrostatics; for more comprehensive discussions we refer to recent reviews \cite{andrea_perspective, Olexandr_lr_review, behler_4gnn_review_2021}. Many MLIP architectures incorporate electrostatic contributions via a coarse-grained charge density, typically represented by partial charges or atomic multipoles. While this is not the only approach to long-range electrostatics, our focus is on models that allow extraction of additional observables beyond dynamics, such as dipoles, polarizabilities, and responses to external electric fields. A central design choice in such models is how the coarse-grained charges are determined.

Numerous architectures predict partial charges or Wannier function centres directly from local geometry, often with a subsequent charge renormalization step \cite{Morawietz2012ACharges, Veit2020PredictingDipoles, popellier_multipoles_2008, Cheng2025LatentInteractions, Zhang2022AInteractions, physnet2019, fennix, SO3LR}. Examples include the ``3\textsuperscript{rd} Generation'' Neural Network \cite{Morawietz2012ACharges, Artrith2011}, Deep Potential Long-Range (DPLR) \cite{Zhang2022AInteractions, Zhang2024}, and the Latent Ewald Sum (LES) method \cite{Cheng2025LatentInteractions, Kim2025APotentials, Zhong2025MachineResponse}. This method, while simple, is already sufficient to make high quality foundation models for molecular science, such as the recent SO3LR \cite{SO3LR} model. These approaches also naturally yield properties such as dipole moments and first-order field responses.

However, predicting charge densities purely from local geometry has fundamental limitations: such models cannot capture induced polarization between well-separated subsystems or long-range charge transfer \cite{behler_4gnn_review_2021}. They also struggle in systems where a fixed amount of excess charge must be distributed over many sites \cite{kocer2024machinelearningpotentialsredox}. This has motivated alternative designs that either augment empirical charge-transfer models or incorporate elements of electronic structure theory, typically leading to self-consistent (SC) models. By ``self-consistent'' we mean that there is some surrogate charge distribution in the model which is obtained as a solution of coupled equations involving the entire system (either through minimization of an energy functional or via a self-consistent field loop) so that it is consistent with the potential it generates, rather than being predicted directly from geometry alone.

This paper focuses on self-consistent models, while acknowledging that simpler treatments may suffice in some regimes. We now summarize several key developments in this area.



\paragraph{CENT, 4GNN, kQEq.}
Several models determine partial charges using a machine-learned variant of classical charge equilibration (QEq) \cite{qeq1985, qeq1986, Rappe1991ChargeSimulations}, as first demonstrated in the CENT architecture \cite{cent2015, 4gnn_ko_2020, kqeq_og2022}. In these models, an energy functional is defined that is quadratic in the charges, with electronegativities and hardnesses predicted from local geometry. Charges are obtained by minimizing this energy, and the resulting Coulomb contribution is added to the total energy. Such models have been applied to challenging systems, including variable iron oxidation states in water \cite{kocer2024machinelearningpotentialsredox}. Also in this category, we highlight the recent foundation model from Hu \textit{et al.} \cite{pqeq_hu_2025} which uses a polarizable QEq charge model, utilising offset charges as well as variable atomic charges. However, it is known that QEq suffers from fundamental deficiencies, including incorrect fractional charge separation upon dissociation \cite{Jensen2023UnifyingModels, Perdew1982Density-FunctionalEnergy, Vondrak2025PushingLimits} and an unphysical cubic scaling of polarizability with system size for finite insulating systems \cite{LeeWarren2008OriginMethods, nonlinear_pol_fq}, leading to metal-like behaviour in large molecules \cite{conducting_molecules}. It is not yet clear how much machine learned models based on QEq, or ideas such as the polarizable QEq used in reference \cite{pqeq_hu_2025}, are affected by these features of classical QEq. 

\paragraph{ACKS2.}
Although not an MLIP, second-order atom-condensed Kohn–Sham DFT (ACKS2) \cite{acks2} is a relevant empirical development. ACKS2 and related schemes \cite{Jensen2023UnifyingModels} correct several failures of QEq, including restoration of integer charge fragmentation, by properly accounting for the long-range nature of the Kohn–Sham kinetic energy.

\paragraph{BpopNN.}
The BpopNN model constructs an energy functional of geometry and the coarse-grained charge (and spin) density, which is minimized during evaluation \cite{bpopnn}. One can think of BpopNN as an extension of the quadratic energy expression in CENT to more general machine-learned functionals. However, the model still contains a large quadratic component as a baseline, making the benefit of the added complexity unclear. Training the model relies on constrained DFT with non-ground-state charge distributions, which complicates practical use and requires a specific charge partitioning scheme.

\paragraph{SCFNN.}
The self-consistent field neural network (SCFNN) of Gao and Remsing represents the charge density using Gaussian charges at maximally localised Wannier function centres (MLWFCs) \cite{scfnn}. The MLWFC positions are initially predicted from local geometry and then iteratively updated in response to the local electric field, yielding a self-consistent loop similar to that of classical polarizable force fields. This enables long-range electrostatic induction, but the use of a fixed number of Wannier centres per molecule makes extension to charge-transfer processes, such as redox reactions, impossible.

\paragraph{eMLP.}
The eMLP model also represents charge density via MLWFCs, but formulates the potential energy by treating them as an additional elemental species \cite{eMLP}. The energy is minimized with respect to Wannier centre positions during evaluation. While conceptually appealing, the existence of stable minima is not guaranteed and caused a problem for this model: The authors of eMLP found that sometimes when minimizing the energy with respect to the Wannier centres, the Wannier centres left the molecule entirely, even for some structures within the training domain. It remains unclear whether this instability is intrinsic or related to the specific model architecture. The suitability of Wannier-based descriptions for metallic systems or fractional occupancies also remains unexplored.

\paragraph{AIMnet-NSE.}
AIMnet-NSE, and related architectures, \cite{aimnet_nse, aimnet2, aimnet2_nse} predict partial charges via message passing followed by redistributing excess charge globally at each step. This redistribution can be interpreted in terms of atom-condensed Fukui functions, and the model accurately captures where charge localizes upon electron addition or removal. However, the redistribution is not iterated to self-consistency, and only the total excess charge is treated globally, preventing spontaneous polarization between well-separated subsystems. While AIMnet-NSE is not a self-consistent model, a theoretical connection to self-consistent models can be established as discussed in Section \ref{sec:theory}. It is noteworthy that the Fukui function inspired charge distribution mechanism has already been reused in several other successful foundation models \cite{mace_polar, fennix_foundation}.

\paragraph{General Long-Range Operations and LODE.}
Finally, there are now several MLIP architectures which contain long-range interactions without a strict connection to charges or the Coulomb kernel. Notable examples are Spookynet and So3krates \cite{spookynet, so3krates}, among others \cite{neural_p3m, ewald_message_passing}. These make use of mechanisms like global attention to introduce long-range behaviour. An important point in this space is Long Distance Equivariant Descriptor (LODE). LODE (as well as the related LOREM model architecture \cite{lorem_2025}) constructs atomic features which contain global information by imagining each atom to have a fictitious charge of 1, computing the corresponding fictitious electric potential, and then projecting it onto atoms to create long-range equivariant features. The LODE model enforces physical long-range decay of interactions between atoms, and can be related to the new models and theory presented in the present paper. Recently, LODE features were used to predict the charge density and its response to applied fields in a step towards high fidelity simulations of electrochemical interfaces with induction effects \cite{grisafi_lode_salted, grisafi_metal_response, Rossi_2025_electron_response}. Part of the present paper could be viewed as a way to adapt LODE into a self-consistent model like those mentioned above. 

\subsubsection*{Focus of this Paper}

The purpose of this paper is to perform a systematic exploration of the treatment of electrostatics in MLIPs, albeit with a focus on applications in materials science, rather than biochemistry. We begin by describing two approaches to coarse-graining density functional theory and replacing the physical operations with machine-learned surrogates, leading to two self-consistent MLIP architectures. The two approaches are (i) learning an energy functional of a coarse-grained charge density, which is minimized to evaluate the model, and (ii) learning a function that maps a coarse-grained electric potential to a coarse-grained charge density, and using it in a fixed-point iteration to achieve self-consistency. We present a derivation of these methods from Hohenberg--Kohn DFT, where the required approximations are explicitly stated. This lets us state some conditions which should be met by degrees of freedom representing the charge density in the model, and allows us to write down precise \textit{ab initio} expressions for what the model is trying to learn. Following this, we explain how the two approaches are related, including when the approaches are directly equivalent.

We then explore the natural decisions researchers can take when implementing self-consistent models, and discuss which design choices have been made in existing models. This allows us to outline part of the ``design space'' of self-consistent electrostatic models, and categorize the existing models within our framework, making it more straightforward to understand their behaviour.

Given the formalism we present, many natural questions arise as to which components are necessary or optimal. The second half of the paper is dedicated to answering some of these questions. To do this, we implement the most general forms of the various self-consistent models in our MLIP software platform. By using the MACE architecture as the basis for our experiments and adopting a single common representation of the charge density, clear comparisons between different ways of treating electrostatics can be made.

Finally, we evaluate the models on two illuminating examples: Firstly, conducting and insulating behaviour in metallic and liquid water slabs. This test system allows one to explore contrasting behaviour of dielectric and conducting systems, and even make contact with the well known failures of some classical models. A metal--water interface is also the basis for a wide variety of interesting applications in the areas of heterogeneous catalysis and battery science.
In particular, we examine properties like the strict quantisation of charge on charged water clusters, as well as the different electronic screening one sees in a metallic as opposed a dielectric medium.
We show that while existing models are sufficient for describing some features of this system, there are deficiencies when it comes to the more challenging aspects. 

Secondly, we study charged vacancies in silicon dioxide, which is a system where the geometry alone does not determine the state of the system. Correctly modelling configurations of multiple defects with a specified overall charge is difficult because the total charge must be distributed unequally between different spatially distant defects. We will show that simpler versions of self-consistent models struggle at this task, while the most advanced models in our design space can succeed. Throughout the results we also compare to both local models and non-self-consistent long-range architectures.
A large part of our results also pertains to the methods used to train self-consistent models, which can be much more involved than for typical MLIPs. We show that the training dynamics is highly important, and go some way to explaining why this is the case for different architectures.

Overall, our theoretical framework and results shed light on many aspects of the design of self-consistent MLIP architectures, and on the training procedures which should be used for such models. We hope that this will open the door to a range of future architectures, and foundation models, with a much closer connection to electronic structure theory than current models.

\section{Theory}
\label{sec:theory}



We are interested in describing the design space of self-consistent models. In this section we present the theory behind two natural self-consistent model architectures, referred to as the energy functional approach and the fixed-point approach.


The section is organised as follows. Firstly, in section \ref{sec:definitions:cdft} we introduce some definitions and notation from conceptual density functional theory (CDFT), which will be used to understand what physical objects the models need to learn in order to be successful. Section \ref{sec:energy_min} then introduces our formulation of a machine learned energy functional for building electrostatic MLIPs. We present a derivation of the model from HK-DFT, where it is possible to show precisely what the model must learn including any short range terms which arise from the process of coarse-graining the charge density. This also reveals several conditions which must be satisfied by the coarse-grained representation of the charge density, and by any charge partitioning schemes which are used. Finally, since we have expressions for what must be learned in terms of the universal functional, one can make some comments on which known behaviours of the universal functional might impact the success of electrostatic MLIPs. Following this, section \ref{sec:theory:fixedpoint} defines the complementary method---the learned fixed-point approach---and repeats some of the analysis to understand the context and potential limitations of such a model. 

In section \ref{sec:scf:theory:equivalence}, we show that while both approaches can be viewed as approximating HK-DFT, when implemented using typical machine learning tools the two approaches are not equally expressive. Section \ref{sec:classification} classifies the existing schemes (including the models discussed in section \ref{sec:existing_work}), allowing us to see which models may have certain limitations.

In this study, we will not discuss the importance of having a well-defined electronic current or polarization density \cite{Resta2007TheoryApproach}. One motivation for using Wannier function centres to describe a charge distribution is that one can predict the polarization vector of a condensed phase system in a way which is consistent with the modern theory of polarization \cite{MarzariWF}. In order to allow us to directly compare different methods, we will use an atomic multipole description of the charge density throughout. 

\subsection{Definitions}

Consider a collection of nuclei and $N$ electrons. In this section, we will use $r$ to denote position, and throughout this paper we will use atomic units in equations. We will write the electronic energy of an electron density $n$ using the universal Hohenberg--Kohn functional $F$ and the external potential for the electrons, $v^e_{\mathrm{ext}}$, as:
\begin{equation*}
    E[n] = F[n] + \int v_{\mathrm{ext}}^e(r)n(r)\,dr
\end{equation*}
The reason for the notation $v^e_\text{ext}$ is that in this paper we will derive models which use positive \textit{charge} densities, not electron densities, and we therefore prefer potentials defined for positive test charges. With $v^e_\text{ext}$ defined as the potential of electrons, we therefore also introduce $v_\text{ext} = -v^e_\text{ext}$ where $v_\text{ext}$ is the energy of positive test charges. Furthermore, we consider the full external potential $v_\text{ext}$ as the sum of the nuclear potential and a component due to additional applied fields:
\begin{align}
    v_\text{ext}(r) = v_\text{nuc}(r) + v_\text{app}(r)
\end{align}
In which, again, $v_\text{nuc}(r)$ and $v_\text{app}(r)$ correspond to the potential energy of positive test charges. Along with the nuclear potential, we define the nuclear charge density $\nu$ as
\begin{align}
    \nu(r) = \sum_i \delta(r-r_i) Z_i,
\end{align}
from which $v_\text{nuc}(r)$ originates. We can then define the \text{total DFT energy} of the system as follows:
\begin{align}
    E_{\mathrm{tot}}^{\mathrm{DFT}}[n]
    =& \ F[n] + \int v_{\mathrm{ext}}^e(r)n(r)\,dr \nonumber \\
    &+ E_{\mathrm{nn}} + \int \nu(r) v_{\text{app}}(r) dr \nonumber \\
    =& \ G[n] + E_{\mathrm{Coulomb}}^{\mathrm{DFT}}[n]
    \label{eq:hk_dft}
\end{align}
Here, $E_\mathrm{nn}$ is the nuclear-nuclear energy and the second integral is the energy of the nuclei due to applied fields. We have grouped the energy into the kinetic, exchange-correlation functional $G$ (which is simply $F$ minus the Hartree energy of $n$) and $E_{\mathrm{Coulomb}}^{\mathrm{DFT}}[n]$ which contains all electrostatic pieces:
\begin{align}
    &E_{\mathrm{Coulomb}}^{\mathrm{DFT}}[n]
    = \frac{1}{2}\iint \frac{n(r)n(r')}{|r-r'|}\,dr\,dr' \nonumber \\
    &+ \int v_{\mathrm{ext}}^e(r)n(r)\,dr + E_{\mathrm{nn}} + \int \nu(r) v_{\text{app}}(r) dr
    \label{eq:E_coulomb_definition}
\end{align}
Equations \eqref{eq:hk_dft} and \eqref{eq:E_coulomb_definition} will be the central equations for working with the DFT energy in this section.

We denote the minimum of \eqref{eq:hk_dft} for a constant number of electrons $N$ as $\mathcal{E}_N$ and the minimum at constant (electron) chemical potential $\mu$ as $\mathcal{E}_\mu$:
\begin{align}
    \mathcal{E}_N &= \min_{n}\left[ E^{\mathrm{DFT}}_{\mathrm{tot}}[n]\right]  \ \ \text{s.t.} \ \ \int n(r) dr = N \label{eq:mathcal_E_N} \\
    \mathcal{E}_\mu &= \min_{n} \left[ E^{\mathrm{DFT}}_{\mathrm{tot}} [n] - \mu \int n(r)dr \right]  \label{eq:mathcal_E_mu}
\end{align}
Note that in \eqref{eq:mathcal_E_mu} the total charge will not necessarily be an integer. The canonical extension of density functional theory to non-integer number of electrons was proposed by Perdew \cite{Perdew1982Density-FunctionalEnergy}, and involves considering an ensemble of $N$ and $N+1$ electron density matrices, where $N$ is an integer. Note that with this extension, the energy of a fractional number of electrons in DFT is well defined even at zero temperature. The following notation will be used for the Hartree energy and Hartree potential of a density $n$:
\begin{align}
    E_{\text{H}}[n] &= \frac{1}{2}\iint \frac{n(r)n(r')}{|r-r'|}  drdr' \label{eq:E_Hartree} \\
    \qquad v_\text{H}[n](r) &= \int \frac{n(r')}{|r-r'|} dr' \label{eq:v_Hartree}
\end{align}

\subsubsection{Relevant Concepts from Conceptual DFT}
\label{sec:definitions:cdft}

Our aim is to carefully analyse the approximation steps needed to derive an MLIP architecture from HK-DFT. To do this, we will use many ideas and tools from \textit{conceptual density functional theory} (CDFT). Conceptual density functional theory provides quantitative, DFT-based definitions of reactivity concepts such as electronegativity and chemical hardness \cite{Geerlings_cdft_Review, Geerlings2022}. Beyond this definitional role, ideas from CDFT are also useful for performing perturbative analyses of ground-state properties with respect to variables such as the external potential. We now introduce several important quantities of CDFT for a chemical system in terms of derivatives of the universal functional $F$.

\paragraph{Hardness}
The molecular hardness is defined as $\eta = \partial^2 \mathcal{E}_N / \partial N^2$. Similarly, the \textit{hardness kernel} is defined as the second functional derivative of $F$:
\begin{align}
    \eta[n^0](r,r') = \frac{\delta^2 F[n]}{\delta n(r) \delta n(r')}\bigg|_{n^0} \label{eq:def_eta}
\end{align}
It is a functional of a reference density $n^0$, which is the density at which the derivative is evaluated. The hardness kernel and its higher-order generalizations are relevant because they provide a way to expand the universal functional in a functional derivative Taylor series. 
Assuming the functional derivatives exist, consider a small perturbation $n$ of a reference density $n^0$ and define $\Delta n(r) := n(r) - n^0(r)$. For a small enough $\Delta n$, $F$ admits a Taylor expansion about $n^0$:

\begin{align}
    F[n^0(r)+\Delta n(r)] &= F[n^0(r)] + \int \frac{\delta F}{\delta n(r)}\bigg|_{n^0} \Delta n(r) dr \nonumber \\
    + \frac{1}{2}\iint &\frac{\delta^2 F}{\delta n(r) \delta n(r')}\bigg|_{n^0} \Delta n(r) \Delta n(r') drdr' + ...
\end{align}
In which all the derivatives are taken at $n=n^0$ \cite{Ernzerhof1994Taylor-seriesFunctionals, nonlinear_senet_1996}. As will be discussed below, the universal functional is not always sufficiently smooth for these derivatives to exist. In particular, the energy as a function of $N$ for an isolated system is piecewise linear between integer numbers of electrons \cite{Perdew1982Density-FunctionalEnergy} (see also section \ref{sec:energy_min} and figure \ref{fig:lines_G_F}). This means that the derivative definition of the total hardness $\eta$ is not useful since it is undefined for integer electron numbers and zero otherwise.\footnote{A commonly used solution, introduced by Parr and Pearson \cite{ParrPearson83}, is to use a finite-difference approximation for $\eta$ which gives (IP-EA)$/2$ where IP and EA are the ionization potential and electron affinity, respectively. Readers are also referred to recent work on partition density functional theory, which resolves this problem in an atoms-in-molecules context \cite{Wasserman_pdft, Wasserman_foundations}.} The implications for this paper are that when using the hardness kernel in \eqref{eq:def_eta}, one must consider only changes in density which do not alter the total number of electrons, which we assume implicitly from now on except when stated otherwise. 

\paragraph{Softness}
If $n$ is the ground state density for the external potential $v_\text{ext}^e$, the softness kernel is defined as:
\begin{align}
    s(r,r') = \left(\frac{\partial n(r)}{\partial v_\text{ext}^e(r')}\right)_\mu.
\end{align}
Note that this sign convention will be used throughout, but the softness is more commonly defined as the negative of this derivative. The derivative is taken at constant chemical potential, rather than at constant number of electrons, as indicated by the $\mu$ subscript \cite{Geerlings2022}, meaning that $n$ is defined from \eqref{eq:mathcal_E_mu} and not \eqref{eq:mathcal_E_N}. The same derivative at constant number of electrons is called the linear response kernel $\chi(r,r')$. The softness kernel and the related higher-order derivatives \cite{Toon2016} appear when making a functional derivative Taylor expansion of the ground state density as a function of the external potential, at constant chemical potential:
\begin{align}
    \Delta n(r) &= \int \left(\frac{\delta n(r)}{\delta v^e_{\text{ext}}(r')}\right)_\mu \Delta v^e_{\text{ext}}(r') dr' + \nonumber \\
    + \frac{1}{2}\iint & \left(\frac{\delta^2 n(r)}{\delta v^e_{\text{ext}}(r')\delta v^e_{\text{ext}}(r'')}\right)_\mu \Delta v^e_{\text{ext}}(r') \Delta v^e_{\text{ext}}(r'') dr'dr'' + ...
\end{align}

\paragraph{Fukui Function}

Finally, the Fukui function is defined as the derivative of the ground state electron density with respect to the number of electrons $N$:
\begin{align}
    f(r) = \frac{\partial n(r)}{\partial N}
\end{align}
The Fukui function is a rationalization of Fukui’s frontier orbital reactivity theory and is a useful descriptor of chemical reactivity in molecules \cite{Fukui1952, Parr_frontier}. In reality, derivatives with respect to the number of electrons differ when adding or removing electrons, and there are actually two distinct Fukui functions, denoted $f_+$ and $f_-$, respectively, for these two processes. Throughout this paper we will work with smooth surrogate models, and thus discuss only a single Fukui function. 

\subsubsection{Coarse-Grained Charge Density}
\label{sec:definitions:cg_density}


We are interested in making machine learning models which use local, atom-centred descriptors. Local machine learning models are already able to describe a wide variety of chemical behaviour and typically have receptive fields which are large compared to bond lengths. Therefore, when building models, we will employ a coarse-grained representation of the charge density.

Firstly, in addition to the full DFT electron density $n(r)$ and the nuclear charge density $\nu(r) = \sum_i Z_i \delta(r-r_i)$, we define the DFT \textit{charge} density:
\begin{align}
    \tilde{n}(r) &= \nu(r) - n(r)
    \label{eq:define_n_tilde}
\end{align}
Then, we introduce a coarse-grained charge density $\rho(r)$ for use in our model, which will be constructed from a set of coefficients $\mathbf{p}=\{p_k\}$ and basis functions $\phi_k$ as:
\begin{align}
    \rho(r) = \sum_{k} p_{k}\phi_{k}(r)
    \label{eq:charge_density_decomp}
\end{align}
For instance, $\mathbf{p}$ may be simply a list of partial charges on atoms, in which case each $\phi_k$ is a delta function centred on an atom (or a smooth, localised distribution to provide damping at short distances). With the notation above, we include the possibility of having multiple coefficients per atom, such as atomic multipoles.  In this case the basis functions would be indexed by $k=\{i,lm\}$ with $i$ indexing the atom and $l$, $m$ indexing different spherical multipole components. The job of $\rho$ is to approximate the long-range effects of $\tilde n$ (in which one expects most of the nuclear charge to be screened by the surrounding electronic charge), not to approximate $n$ which describes only the electrons.

In this paper $n(r)$ will always refer to a fine-grained electron density, $\tilde n$ will refer to \eqref{eq:define_n_tilde}, and $\rho(r)$ will always be a coarse-grained density constructed via \eqref{eq:charge_density_decomp}. 

\subsubsection{Atoms-in-Molecules Quantities}
\label{sec:definitions:aim}

Since we will work with atom-centred functions and descriptors, we now discuss how one can define atom-centred (`atoms-in-molecules') \cite{Bader_qtaim, Hirshfeld1977} versions of concepts such as the Fukui function and hardness kernel. To start, suppose that we have chosen a concrete projection $\mathbf{p} = \mathcal P_{\mathbf R}[\tilde n]$ (depending on the nuclear positions $\mathbf{R}$) which maps the full DFT charge density $\tilde n$ to a set of model expansion coefficients $\mathbf p$. For example, one could use a Hirshfeld partitioning to assign partial atomic multipoles $\mathbf p$, or determine $\mathbf p$ by minimizing a distance metric between the real density and the coarse-grained version, such as $\int |\widetilde n(r)-\sum_k p_k\phi_k(r)|^2\,dr$. In other words, we are choosing some kind of charge partitioning scheme, but generalised to allow for more than just partial charges if required. There is no unique way to do this, but we will later state some conditions which define acceptable methods. Since the nuclei are fixed, this process will be written compactly as $n\mapsto\mathbf p$. One can now define the DFT energy of a set of coefficients $\mathbf p$ by a constrained search approach:
\begin{align}
    E_{\mathrm{pop}}^{\mathrm{DFT}}(\mathbf p)
    = \min_{n\mapsto\mathbf p} E_{\mathrm{tot}}^{\mathrm{DFT}}[n].
    \label{eq:constrained_min}
\end{align}
Here and throughout, $\min_{n\mapsto\mathbf p}$ means minimization over densities satisfying $\mathcal P_{\mathbf R}[\tilde n]=\mathbf p$. This has previously been called the population-constrained DFT energy when $\mathbf p$ is a vector of atomic partial charges \cite{acks2}. The \textit{minimizer} of this expression also defines a map\footnote{Over a domain of `$n$-representable $\mathbf p$ vectors'} from a set of coefficients to a full density $n$, which will be denoted $n[\mathbf p]$:
\begin{align*}
    n[\mathbf{p}] 
    = \operatorname*{arg \ min}_{n\mapsto\mathbf p} E_{\mathrm{tot}}^{\mathrm{DFT}}[n].
\end{align*}
Note that the minimum of \eqref{eq:constrained_min} over $\mathbf p$ gives the true minimum energy and the minimizing density.

Now that we have a connection between a full density and a set of density coefficients $\mathbf p$, we can define Atoms-in-Molecules (AIM) versions of some of the above concepts. Specifically, in this paper we define the AIM Fukui function as the derivative of the population constrained electron density $n[\mathbf{p}]$ (at a reference population $\mathbf{p}^0$) when the coefficient of basis function $k$ is changed, holding all other coefficients constant:
\begin{align}
    f_k(r) = \left.\frac{\partial n[\mathbf p](r)}{\partial p_k}\right|_{\mathbf p^0}
    \label{eq:def_fk}
\end{align}
This quantifies how much the density everywhere changes if we constrain it to change in one place, and is key to much of the logic below. Reference \cite{acks2} gives an extended discussion of the AIM Fukui function. A useful intuition is to imagine forcing the charge coefficient on atom $k$ to change from $p_k^0$ to $p_k^0+\delta p_k$ and viewing \eqref{eq:constrained_min} as a constrained DFT calculation. Then, the full density will change (at first order) from $n^0=n[\mathbf p^0]$ to:
\begin{align}
    n[p_k^0+\delta p_k](r) = n^0(r)+\delta p_k f_k(r) \label{eq:intuition_for_fk}
\end{align}
It should be noted that there is more than one way to define an AIM Fukui function, and the above concept is distinct from the `atom-condensed' Fukui function used in some other works \cite{aimnet_nse}. 

One can also consider second derivatives of the density with respect to the constraint values, and we will refer to these as `higher-order' AIM Fukui functions, denoted $f_{kj} = \partial^2 n / \partial p_k \partial p_j$, with an obvious generalization to more than two indices. 
By the same logic, one could define an AIM hardness kernel as $\partial^2 F[n[\mathbf{p}]] / \partial p_k \partial p_j$, and these concepts will be discussed further below. 

\subsection{The Energy Functional Approach}
\label{sec:energy_min}

We now define the first class of self-consistent machine learning models to be studied in this paper. Take the geometry of a molecule or crystal structure where each atom $i$ has several basis functions $\phi_k$, and therefore several coefficients $p_k$ describing the electronic state of the atom. Denote the charge density degrees of freedom for atom $i$ by $\mathbf{p}_i = \{p_{k}\}_{k\in A(i)}$, where $A(i)$ simply selects the basis functions $k$ which are associated with atom $i$. The state of each atom is characterized by a vector $\mathbf{u}_i = (z_i, r_i, \mathbf{p}_i)$ where $z_i$ and $r_i$ denote the chemical element and position, respectively, of atom $i$. Then, the model's energy prediction is defined as:
\begin{align}
    E_{\mathrm{model}}(\mathbf p)
    &=E_{\mathrm{local}}+G_{\mathrm{ML}}(\{\mathbf u_i\}_i)
    +E_{\mathrm{Coulomb}}^{\mathrm{CG}}(\mathbf p), \nonumber\\
    \mathcal E_{\mathrm{model}}
    &=\min_{\mathbf p}E_{\mathrm{model}}(\mathbf p)
    \label{eq:energy_model_energy}
\end{align}
Where $E_{\text{Coulomb}}^{\text{CG}}$ is the model's Coulomb energy, and `CG' stands for coarse-grained:
\begin{align*}
    E_{\mathrm{Coulomb}}^{\mathrm{CG}}(\mathbf p)
    &=\frac{1}{2} \iint \frac{\rho(r) \rho(r')}{|r-r'|} drdr'
    +\int v_{\mathrm{app}}(r)\rho(r)\,dr 
\end{align*}

Where $\rho$ is the coarse-grained charge density from \eqref{eq:charge_density_decomp}, and $G_\text{ML}$ is a machine learned function of the local geometry and charge variables. $E_\text{local}$ is the output of a local MLIP, which sees only the geometry of the system and does not have access to information about the charge density. Note that the final term couples the coarse-grained density to only the applied fields, and not the nuclear potential. Also note the sign convention: $\rho$ is defined positive for positive charge density, as is $v_\text{app}$. Equation \eqref{eq:energy_model_energy} is currently just a definition of a model architecture, which will be derived from HK-DFT in the following section. 

The reason for separating $E_\text{local}$ and $G_\text{ML}$ is that given the success of local MLIPs, one might hope that most of the energy can be captured by the local piece, and a simpler (and therefore cheaper) functional form can be used for $G_\text{ML}$. To evaluate the model, one must minimize the energy with respect to $\mathbf{p}$, the free parameters of the charge density, which may be possible analytically depending on the form of $G_{\text{ML}}$. If the total charge is specified, the minimization is done with a constraint on the total charge.

If the total energy is predicted exactly by \eqref{eq:energy_model_energy}, the model is variational. We note however that some models have instead used the minimization only to define the charge density, with the actual energy prediction including additional functions of the final charge density not used during the minimization \cite{4gnn_ko_2020}. In this case the predicted $\mathbf{p}$ will not be minimizers of the predicted total energy and we say that the model is not variational in $\mathbf{p}$.

\subsubsection*{Relationship to HK-DFT}

It is possible to derive the machine learned energy functional \eqref{eq:energy_model_energy} from HK-DFT \eqref{eq:hk_dft} given some conditions on the quality of the coarse-grained density, and on the locality of the underlying electronic structure. The full derivation and additional discussion of the conditions are presented in appendix \ref{appendix:cg_hkdft}; in this section we give an overview of the key argument and results. The derivation here is not needed to understand the implementations and results in section \ref{sec:implementation} and beyond, which only depend on the model architecture as defined in \eqref{eq:energy_model_energy}.

We start by choosing, as discussed in section \ref{sec:definitions:aim}, a concrete way of mapping a density to a set of density coefficients, which gives us a constrained DFT energy and density as functions of $\mathbf p$, denoted $E_{\mathrm{pop}}^{\mathrm{DFT}}(\mathbf p)$ and $n[\mathbf p]$. On the constrained minimum, we write
\begin{align}
    G(\mathbf p)&:= G[n[\mathbf p]], \\
    E_{\mathrm{Coulomb}}^{\mathrm{DFT}}(\mathbf p)
    &:= E_{\mathrm{Coulomb}}^{\mathrm{DFT}}[n[\mathbf p]],
    \label{eq:E_coulomb}
\end{align}
so that $E_{\mathrm{pop}}^{\mathrm{DFT}}(\mathbf p)=G(\mathbf p)+E_{\mathrm{Coulomb}}^{\mathrm{DFT}}(\mathbf p)$. Now consider a perturbation about a set of reference coefficients $\mathbf p^0$, with $\delta\mathbf p=\mathbf p-\mathbf p^0$, corresponding to the reference density $n^0=n[\mathbf p^0]$. The reference density is not necessarily a ground-state density; it is simply some convenient reference density.
\begin{align}
    E_{\mathrm{pop}}^{\mathrm{DFT}}(\mathbf p)
    &=E_{\mathrm{pop}}^{\mathrm{DFT}}(\mathbf p^0)
    +\Delta G+\Delta E_{\mathrm{Coulomb}}^{\mathrm{DFT}},
    \label{eq:perturbed_energy}
\end{align}
where $\Delta G=G(\mathbf p)-G(\mathbf p^0)$ and similarly for the Coulomb term. For the Coulomb energy term, we can consider a Taylor series expansion in $\delta\mathbf p$:
\begin{align}
    \Delta E_{\mathrm{Coulomb}}^{\mathrm{DFT}}
    &=\sum_k\delta p_k
    \left.\frac{\partial E_{\mathrm{Coulomb}}^{\mathrm{DFT}}}{\partial p_k}\right|_{\mathbf p^0}\nonumber\\
    &+\frac{1}{2}\sum_{kl}\delta p_k\delta p_l
    \left.\frac{\partial^2E_{\mathrm{Coulomb}}^{\mathrm{DFT}}}{\partial p_k\partial p_l}\right|_{\mathbf p^0}
    +\ldots
    \label{eq:taylor_series}
\end{align}
One can compute the partial derivatives in this series by applying the chain rule to $E_{\mathrm{Coulomb}}^{\mathrm{DFT}}$ (equations \eqref{eq:E_coulomb_definition} and \eqref{eq:E_Hartree}). For instance, the first-order term is
\begin{align}
    v_k^0
    &=\left.\frac{\partial E_{\mathrm{Coulomb}}^{\mathrm{DFT}}}{\partial p_k}\right|_{\mathbf p^0} \\
    &=\int\left(v_{\mathrm H}[n^0](r)+v_{\mathrm{ext}}^e(r)\right)f_k(r)\,dr.
    \label{eq:def_vk0}
\end{align}
Here we have used the definition of the AIM Fukui function $f_k$. One can continue to compute all the partial derivatives, after which the Taylor series \eqref{eq:taylor_series} can be written in the following form:
\begin{align}
    \Delta E_{\mathrm{Coulomb}}^{\mathrm{DFT}}
    &=\sum_k \delta p_k v_k^0 + \frac{1}{2}\sum_{kj} \delta p_k \delta p_j (C_{kj}+v^0_{kj}) \nonumber \\
    &\qquad + \frac{1}{3!}\sum_{kjl} \delta p_k \delta p_j \delta p_l (C_{kjl} + v^0_{kjl}) + ... ,
    \label{eq:coulomb_taylor}
\end{align}
where the second-order Coulomb interaction between density responses is
\begin{align}
    C_{kj} &= \iint \frac{f_k(r)f_j(r')}{|r-r'|}\,dr\,dr'. \label{eq:def_Ckl}
\end{align}
At each order above first order, there are two terms which we denote $v_{jkl\ldots}^0$ and $C_{jkl\ldots}$. The general expressions for these terms can be found in the appendix, equations \eqref{eq:app_higher_coulomb_response}--\eqref{eq:app_higher_C}. From examining equations \eqref{eq:def_vk0}--\eqref{eq:def_Ckl}, one can see that $\delta p_kv_k^0$ looks like the electrostatic interaction of $\delta p_k$ with the mean field, while $\delta p_k\delta p_jC_{kj}$ looks like a Coulomb energy term between variables $p_k$ and $p_j$.

Our objective now is to turn the constrained DFT energy (\ref{eq:perturbed_energy} and \ref{eq:coulomb_taylor}) into something which looks like the machine learning model \eqref{eq:energy_model_energy}, starting with the Coulomb energy part. In other words, we want to replace the DFT Coulomb-energy change \eqref{eq:coulomb_taylor} with the coarse-grained model Coulomb-energy change, plus controlled local corrections. This can be done given three conditions on the coarse-grained density $\rho$ and the Fukui functions $f_k$.

\subsubsection*{Conditions for Coarse-Graining the Coulomb Energy}

\begin{figure}
    \centering
    \includegraphics[width=1.0\linewidth]{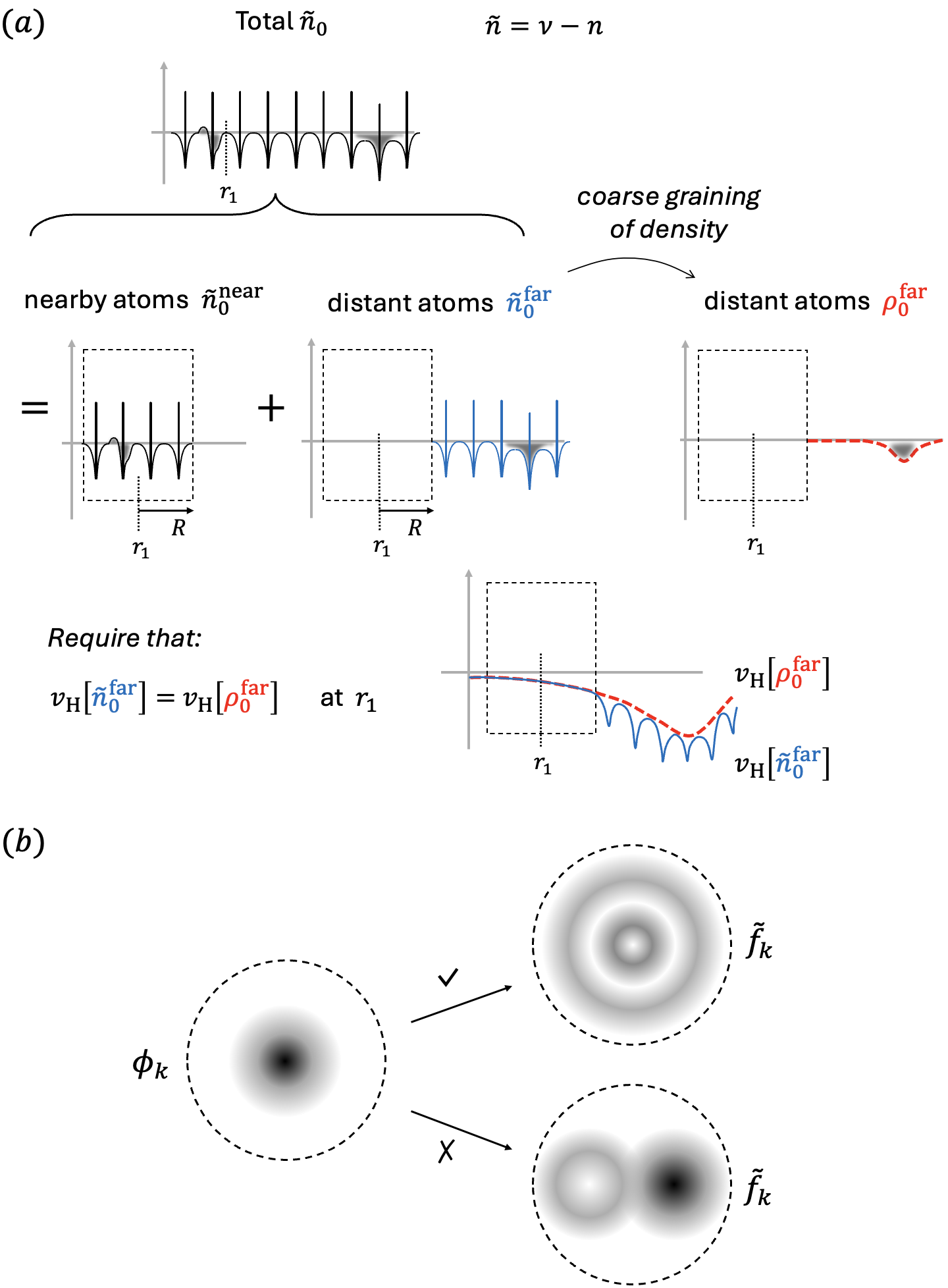}
        \caption{Required properties of the density expansion and AIM Fukui functions. (a): The potential at $r_1$ can be calculated by replacing all the nuclei and electrons outside of some radius $R$ by the coarse-grained density. The plots of $\tilde{n}$ and $\rho$ show charge density, with the repeating pattern coming from negative electrons and positive nuclei. The shaded regions indicate some excess charge due to, for example, defects. The coarse-grained density smooths out all the detail of the atoms, leaving only the net charge or dipole around the defects. (b): The AIM Fukui function $\tilde{f}_k$ must give rise to the same far-field potential as the basis function $\phi_k$. A spherically symmetric basis function associated with $p_k$ is shown on the left, and two possible AIM Fukui functions for this variable are shown on the right. The one at the top has more structure, but will still give rise to the same electric potential outside the dashed line as the $\phi_k$ so long as their multipole moments agree at all orders. The one at the bottom has a significant dipole moment, meaning the potential outside the dashed line will not match that of $\phi_k$.}
    \label{fig:conditions}
\end{figure}

\begin{enumerate}
    \item \textit{The coarse-grained density must be able to approximate the far-field electric potential and reproduce the coupling to applied fields.}
\end{enumerate}

Firstly, we require that the process of computing the charge coefficients $\mathbf p$ from $n$ and using them to construct $\rho$ via \eqref{eq:charge_density_decomp} is an accurate way to represent the long-range part of the Hartree potential. Specifically, consider a length scale $R$, which is less than or equal to the receptive field of our machine learning model. The electric potential at a point $r_1$ due to nuclei and the reference electron density is (for a positive test charge):
\begin{align*}
    v_{\mathrm H}[-n^0](r_1)+v_{\mathrm{nuc}}(r_1)=v_{\mathrm H}[\widetilde n^0](r_1).
\end{align*}
Here we simply used the definition of the total DFT charge density $\widetilde n$ in equation \eqref{eq:define_n_tilde}. The first condition is that one can approximate $v_{\mathrm H}[\widetilde n^0](r_1)$ by replacing the nuclei and electrons of all atoms far away from $r_1$ (meaning $|r-r_1|>R$) by the coarse-grained density $\rho$. Mathematically, if $\widetilde n^0=\widetilde n^{0,(\mathrm{near})}+\widetilde n^{0,(\mathrm{far})}$, where the near part represents only the electrons and nuclei within a distance $R$ of $r_1$, and oppositely for the far part, we have
\begin{align}
    v_{\mathrm H}[\widetilde n^{0,(\mathrm{far})}](r_1)
    =v_{\mathrm H}[\rho^{0,(\mathrm{far})}](r_1).
    \label{eq:main_density_condition}
\end{align}
Intuitively, we are saying that one can compute the electric potential due to $\widetilde n$ at $r_1$ using the coarse-grained density $\rho$ to account for all distant atoms. The principle is shown visually in Figure~\ref{fig:conditions}a.

Condition 1 also requires the coarse-grained density to reproduce the coupling to the class of applied fields considered. For every admissible fine-grained charge density satisfying $\mathcal P_{\mathbf R}[\widetilde n]=\mathbf p$, the corresponding coarse-grained density must satisfy
\begin{align}
    \int \widetilde n(r)v_{\mathrm{app}}(r)\,dr
    =\int \rho(r)v_{\mathrm{app}}(r)\,dr.
    \label{eq:main_field_coupling_condition}
\end{align}
For a constant potential and a uniform electric field, this is satisfied when the total charge and dipole of the coarse-grained density match those of the full charge density. As shown in Appendix~\ref{appendix:cg_hkdft}, this condition ensures that the constrained density and $G(\mathbf p)$ have no separate dependence on $v_{\mathrm{app}}$.

What does this `condition' imply for developers of electrostatic MLIPs, and can we, by a proper choice of coarse-grained density expansion in equation \eqref{eq:charge_density_decomp}, ensure that \eqref{eq:main_density_condition} is true? We will not answer this question in general, but we note that the requirement above is satisfied by some natural charge partitioning schemes and choices for the model coarse-grained density. For instance, consider using a Hirshfeld partitioning to split the electron density into atom-centred contributions, then compute atomic multipole moments from the density of each atom. This is the map $n \mapsto \mathbf p$. If one uses these partial multipole moments as point multipoles in the coarse-grained density (using a sum of point multipoles to construct $\rho$), then the above condition is met in the limit of using all orders of multipole moments (but large, even diverging, errors may be incurred when only using partial charges) as explained in Appendix \ref{appendix:hirshfeld}.

\begin{enumerate}
  \setcounter{enumi}{1}
    \item \textit{The AIM Fukui Function $f_k$ must give the same far-field potential as the corresponding basis function $\phi_k$, and both functions must be localised.}
\end{enumerate}


The first condition related the long-range part of the electric potential in the model to that of DFT, in the reference state $n^0$. Our second condition is a similar relationship, but about perturbations around the reference state.

Take a basis function $\phi_k$ which is associated with an atom at $r_k$. Then, consider the electric potential arising from the coarse-grained density $\rho$ when we put one unit of charge into the basis function $\phi_k$. This is:
\begin{align}
    v_\text{H}[\phi_k] = \int \frac{\phi_k(r')}{|r-r'|}dr'
    \label{eq:potential_from_phi_k}
\end{align}
The Fukui function $f_k(r)$ describes how the actual, fine-grained, electron density $n$ changes if the charge constraint $p_k$ is changed (see equation \eqref{eq:intuition_for_fk} and related discussion). Therefore, at first order the change in the actual DFT electron density when $p_k$ increases by one unit of charge is $f_k(r)$, and the electric potential arising due to this additional charge density is:
\begin{align}
    v_\text{H}[-f_k] = \int \frac{-f_k(r')}{|r-r'|}dr'
\end{align}
The minus sign occurs because electrons are negative and we want to write the potential of a positive test charge for consistency with equation \eqref{eq:potential_from_phi_k}.
The second condition for deriving the model from DFT is that these two electric potentials must match at all points which are far away from $r_k$:
\begin{align*}
    v_\text{H}[\phi_k](r) &= v_\text{H}[- f_k](r) \qquad \text{when } |r-r_k| > R
\end{align*}
Since $f_k$ is the derivative of the DFT electron density with respect to $p_k$, we can introduce $\tilde f_k = -f_k$, as the derivative of the DFT charge density, $\tilde{n}$, also with respect to $p_k$. The condition is then that:
\begin{align}
    v_\text{H}[\phi_k](r) &= v_\text{H}[\tilde f_k](r) \qquad \text{when } |r-r_k| > R
\end{align}
This should be read as ``when the value of a charge constraint $p_k$ changes slightly, the change in the full electric potential which happens in DFT, $v_\text{H}[\tilde f_k]$, should match the change in potential in the model, $v_\text{H}[\phi_k]$, at all points far from atom $k$".

In principle, for a given charge partitioning scheme one could actually compute $f_k(r)$ in a real system using constrained DFT calculations. The shape of $f_k(r)$ could then be compared to the basis $\{ \phi_k \}_k$ to check the requirement. Further discussion of the AIM Fukui function defined in this paper can be found in \cite{acks2}. 

Examples of AIM Fukui functions which would and would not satisfy this requirement are shown in figure \ref{fig:conditions}b. The simple implication of this condition is that the Fukui functions must be `localised': For instance, if $p_k$ represents a partial charge on an atom, and we use a spherically symmetric basis function to represent this charge in the model, then the corresponding Fukui function should not be a highly non-local function. A more subtle aspect of this condition is demonstrated in figure \ref{fig:conditions}b, where the lower Fukui function is localised, but will give rise to a different far-field electrostatic potential compared to $\phi_k$.

More precisely, we require both $\widetilde f_k$ and $\phi_k$ to be localised around $r_k$.

\begin{enumerate}
  \setcounter{enumi}{2}
    \item \textit{All higher-order AIM Fukui functions vanish}
\end{enumerate}

Finally, we make the simplifying assumption that
\begin{align}
    f_{k_1\ldots k_m}(r)=0, \qquad m\geq2.
\end{align}
This is an overly restrictive but transparent sufficient condition; appendix \ref{appendix:cg_hkdft} explains the role of the higher-order Fukui functions in the derivation from HK-DFT.

\subsubsection*{What the Model Must Learn}

By using these three conditions, one can evaluate coefficients such as $v_k^0$ and $C_{kj}$ in \eqref{eq:coulomb_taylor} using the coarse-grained basis functions $\phi_k$, plus \textit{short-range correction terms} that depend on the reference density $n^0$ and the Fukui functions $f_k$. Writing $\rho^0$ for the coarse-grained density associated with $\mathbf p^0$, one finds that the DFT Coulomb-energy change can be written as
\begin{align}
    \Delta E_{\mathrm{Coulomb}}^{\mathrm{DFT}}
    &=\Delta E_{\mathrm{Coulomb}}^{\mathrm{CG}}
    +L_{\mathrm{sr}}(\mathbf p^0,\delta\mathbf p),
    \label{eq:convert_coulomb}\\
    \Delta E_{\mathrm{Coulomb}}^{\mathrm{CG}}
    &=E_{\mathrm{Coulomb}}^{\mathrm{CG}}(\mathbf p)
    -E_{\mathrm{Coulomb}}^{\mathrm{CG}}(\mathbf p^0).
\end{align}
Here $L_{\mathrm{sr}}$ collects all short-range electrostatic corrections. The coarse-grained density $\rho$ interacts only with $v_{\mathrm{app}}$, and not with $v_{\mathrm{nuc}}$. In other words, perturbatively one can replace changes in the full DFT Coulomb energy with changes in the coarse-grained Coulomb energy, plus $L_{\mathrm{sr}}$. These corrections are \textit{local} functions of the geometry, the reference state $\mathbf p^0$, and the perturbation coefficients $\delta p_k$, as shown in Appendix \ref{appendix:cg_hkdft} and the equations leading up to \eqref{eq:app_central_difference}. They are local in the sense that although there are terms involving $\delta p_k\delta p_j$, they are non-zero only when $k$ and $j$ refer to nearby atoms, and can therefore be folded into the machine-learned functional as mentioned below.


By combining \eqref{eq:convert_coulomb} with \eqref{eq:perturbed_energy}, one can now rewrite the population-constrained DFT energy using the coarse-grained Coulomb energy. The final result is
\begin{align}
    E_{\mathrm{pop}}^{\mathrm{DFT}}(\mathbf p)
    &=\left[E_{\mathrm{pop}}^{\mathrm{DFT}}(\mathbf p^0)
    -E_{\mathrm{Coulomb}}^{\mathrm{CG}}(\mathbf p^0)\right]
    +\Delta G \nonumber\\
    &\quad+L_{\mathrm{sr}}(\mathbf p^0,\delta\mathbf p)
    +E_{\mathrm{Coulomb}}^{\mathrm{CG}}(\mathbf p).
    \label{eq:amazing_energy_expansion}
\end{align}
This result is an order-by-order statement within the formal Taylor expansion about $\mathbf p^0$.

These terms can now be matched up with the definition of our machine learning model, $E_{\text{model}}(\mathbf{p})$ in equation \eqref{eq:energy_model_energy}.
The bracketed reference term is the quantity represented by $E_{\mathrm{local}}$:
\begin{align}
    E_{\mathrm{local}}
    &\simeq E_{\mathrm{pop}}^{\mathrm{DFT}}(\mathbf p^0)
    -E_{\mathrm{Coulomb}}^{\mathrm{CG}}(\mathbf p^0) \nonumber\\
    &=G(\mathbf p^0)
    +\left[E_{\mathrm{Coulomb}}^{\mathrm{DFT}}(\mathbf p^0)
    -E_{\mathrm{Coulomb}}^{\mathrm{CG}}(\mathbf p^0)\right].
    \label{eq:reference_local_energy}
\end{align}
The local reference term therefore contains the kinetic and exchange-correlation contribution of the \textit{reference} state, $G(\mathbf p^0)$, together with the difference between the fine-grained and coarse-grained reference Coulomb energies, also in the reference state only.

The next two terms, $\Delta G$ and $L_{\mathrm{sr}}$ are associated with $G_{\text{ML}}$ in the model. The correction term $L_{\text{sr}}$ is a local function of the geometry and charge variables if the conditions above are met, and hence can be learned together with $\Delta G$. Finally, the remaining electrostatic term in \eqref{eq:amazing_energy_expansion} is just the model, coarse-grained Coulomb energy.

We can now analyse how easy these terms are to learn, starting with the local model term \eqref{eq:reference_local_energy}. The reference state $\mathbf{p}^0$ at a given geometry could be a fixed value (such as neutral atoms, or formal oxidation states), or it could be a prediction of the local model, meaning that $p^0_k$ is a function of the geometry around atom $k$ (it does not need to have any connection to a ground state). For our model to work, we need $G(\mathbf{p}^0)$, plus the \textit{error} in the Coulomb energy, to be a function of the local geometry only. $G(\mathbf{p}^0)$ is the kinetic, exchange-correlation energy of a constrained DFT calculation in which the charge coefficients are held at $\mathbf{p}^0$. Since the Coulomb piece is subtracted, this energy does not have long-range components due to mean field electrostatic effects in the reference density. It may, however, not be local because of other quantum mechanical effects such as the kinetic or exchange energy in the reference state. The difference in model and DFT Coulomb energy, on the other hand, can be shown to be local given the conditions stated above, as discussed in the appendix.

The other machine-learned term in \eqref{eq:energy_model_energy} is $G_{\mathrm{ML}}$, which approximates $\Delta G+L_{\mathrm{sr}}$. We can gain some insight into what is being learned by Taylor expanding $G(\mathbf p)$ about $\mathbf p^0$. Define
\begin{align}
    \overline\eta_{k_1\ldots k_m}
    =\left.\frac{\partial^mG(\mathbf p)}
    {\partial p_{k_1}\cdots\partial p_{k_m}}\right|_{\mathbf p^0}.
\end{align}
Then
\begin{align}
    \Delta G
    &=\sum_k\delta p_k\overline\eta_k
    +\frac{1}{2}\sum_{kj}\delta p_k\delta p_j\overline\eta_{kj}
    +\ldots
    \label{eq:taylor_expand_G}
\end{align}
If we use a local, body-ordered model to learn $\Delta G$ in terms of the geometry and $\delta\mathbf p$, then the coefficients $\overline\eta_{k_1\ldots k_m}$ are what the model has to learn from the geometry. One can compute them in terms of $G$ and the AIM Fukui functions by applying the chain rule. For instance, the second-order partial derivative is
\begin{align}
    \overline\eta_{kj}
    :=&\left.\frac{\partial^2G(\mathbf p)}{\partial p_k\partial p_j}\right|_{\mathbf p^0} \nonumber\\
    =&\iint\left.\frac{\delta^2G}{\delta n(r)\delta n(r')}\right|_{n^0}
    f_k(r)f_j(r')\,dr\,dr' \nonumber\\
    &+\int\left.\frac{\delta G}{\delta n(r)}\right|_{n^0}f_{kj}(r)\,dr.
    \label{eq:def_etabar_2}
\end{align}

The general result for the order-$m$ derivative follows from the Fa\`a di Bruno formula, equation \eqref{eq:app_faa_di_bruno}. At all orders, the terms in the expression are combinations of functional derivatives of $G$ and AIM Fukui functions. Since $F[n]=G[n]+E_{\mathrm H}[n]$, the functional derivatives of $G$ are very closely related to the nonlinear hardness kernels $\eta(r,r',\ldots)$ \cite{Toon2016}, defined as (see section \ref{sec:definitions:cdft})
\begin{align*}
    \eta[n^0](r_1,\ldots,r_n)
    =\left.\frac{\delta^nF}{\delta n(r_1)\cdots\delta n(r_n)}\right|_{n^0}.
\end{align*}
In fact, the functional derivatives of $G$ are equal to the nonlinear hardness kernels minus the mean field contribution, which only appears at first and second order:
\begin{align*}
    \left.\frac{\delta G}{\delta n(r)}\right|_{n^0}
    &=\eta[n^0](r)-v_{\mathrm H}[n^0](r), \\
    \left.\frac{\delta^2G}{\delta n(r)\delta n(r')}\right|_{n^0}
    &=\eta[n^0](r,r')-\frac{1}{|r-r'|}, \\
    \left.\frac{\delta^nG}{\delta n(r_1)\cdots\delta n(r_n)}\right|_{n^0}
    &=\eta[n^0](r_1,\ldots,r_n),\qquad n>2.
\end{align*}

Combining this with \eqref{eq:def_etabar_2}, we can see that the coefficients $\overline\eta_{k_1\ldots k_m}$ are projections of the nonlinear hardness kernels $\eta$ onto AIM Fukui functions, with the exception of first and second order, where we first subtract the mean-field contribution to the hardness kernels. In practice, the model needs to learn both this expansion, as well as the short range correction terms in $L_{\mathrm{sr}}$.

Whether a local machine-learning model can be used for $G_{\mathrm{ML}}$ depends on how smooth the expansion in $\delta\mathbf{p}$ is in practice. Specifically, how small can the $\delta p_k$ be made, how local are the coefficients $\overline\eta_{k_1\ldots k_m}$, and, even when their indices refer to nearby atoms, are they functions only of the local geometry? All of these properties are related to the behaviour of the true functional $G[n]$, or equivalently the nonlinear hardness kernels.

\subsubsection*{Relevant Properties of $G$}

\begin{figure}
    \centering
    \includegraphics[width=0.9\linewidth]{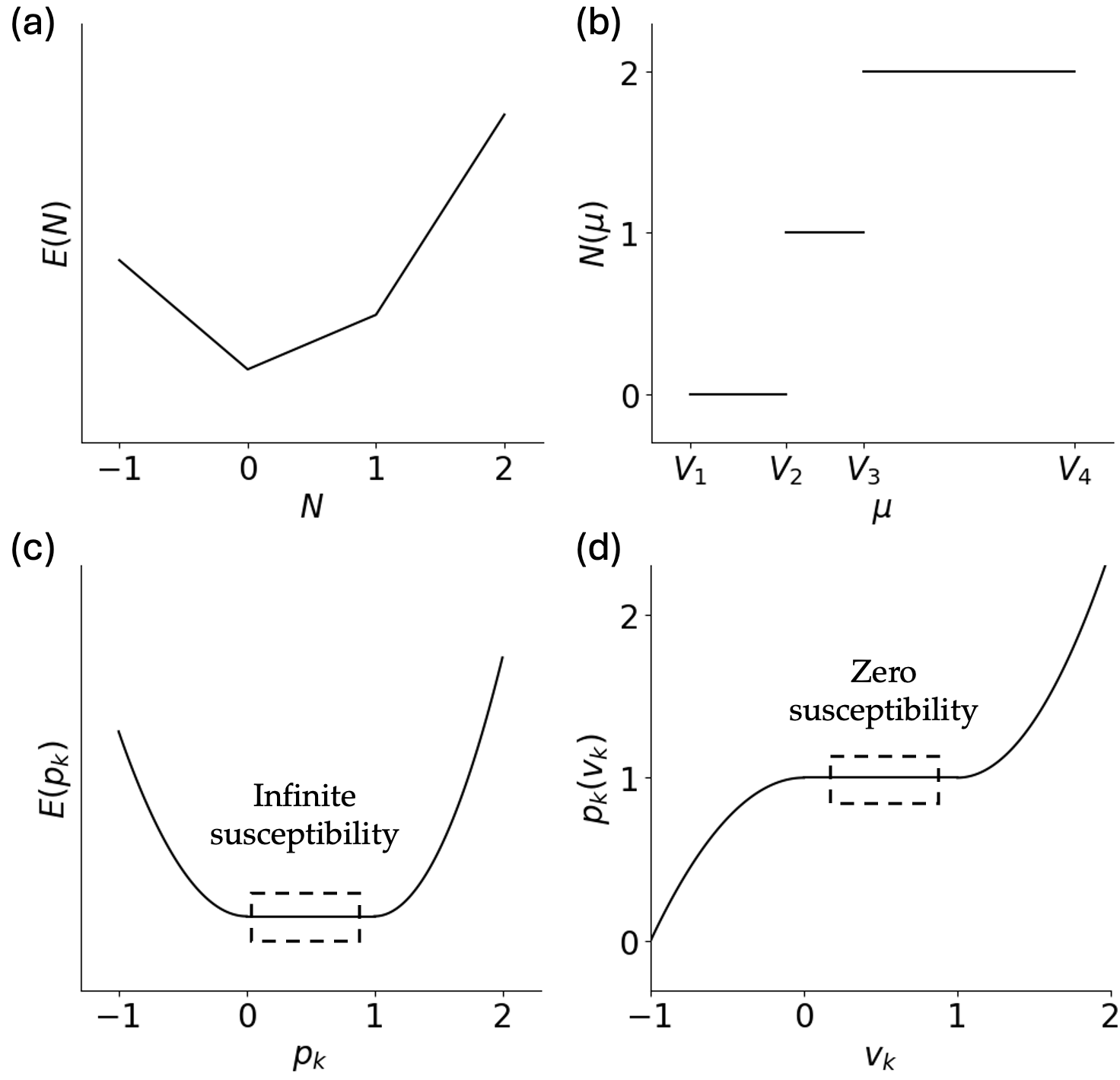}
    \caption{(a): Derivative discontinuity of the energy of an isolated system, (b) discontinuity of the total charge as a function of chemical potential for an isolated system. (c): Infinite susceptibility of one variable, $p_k$, as one would find in a classical perfect conductor, appears as zero curvature in the energy. (d): Similarly, zero susceptibility appears as zero response to a change in the potential corresponding to that variable.}
    \label{fig:lines_G_F}
\end{figure}

The above derivation, which uses many tools from conceptual density functional theory \cite{Geerlings_cdft_Review}, makes one strong assumption about the universal functional, which is false. Namely, by making a Taylor series of the constrained energy in \eqref{eq:constrained_min}, we assume that the energy is smooth with respect to our description of the charge density. One important case when this is not true is the energy of an isolated system as a function of the total (fractional) number of electrons. 

For the real universal functional, the energy, at zero temperature, of an isolated system considered at a fractional total charge is piecewise linear between integers \cite{Perdew1982Density-FunctionalEnergy} as sketched in figure \ref{fig:lines_G_F}a. For commonly used density functionals, the total energy is not piecewise linear, but does have discontinuous derivatives at integer numbers of electrons, arising from, for instance, the Kohn-Sham kinetic energy term. In exact DFT, the piecewise linearity is crucial for ensuring that any isolated fragment will always have an integer number of electrons when the energy is minimized with respect to the charge density at any chemical potential.

From the perspective of model fitting, the fact that both exact DFT and approximations in KS-DFT have discontinuous derivatives means that smooth polynomial approximations for $G_\text{ML}$ will not be able to reproduce some properties.

Setting aside problems of smoothness, another key property affecting these models is the locality of the coefficients $\bar{\eta}^{(n)}$ in \eqref{eq:taylor_expand_G}, which is in turn related to functional derivatives of the kinetic and exchange-correlation functional. While the exchange and correlation components of $G$ may decay rapidly when training from a local exchange-correlation functional, the kinetic energy, even in KS-DFT, is not typically a local functional of the density \cite{acks2}. Hence, we cannot argue that the $\bar{\eta}^{(n)}_{kj...}$ coefficients are local, in the sense that they do not necessarily become small when $k$ and $j$ refer to distant atoms. At first this appears to be a major limitation, but our experiments in section \ref{sec:results:metal_water} suggest that the model architectures do not necessarily suffer very much because of this fact. 

\subsection{The Fixed Point Approach}
\label{sec:theory:fixedpoint}

We now define a second way of constructing electrostatic MLIPs, taking inspiration from the self-consistency cycle of Kohn-Sham (KS) DFT. Generally in KS-DFT, one does not explicitly minimize the functional, but instead solves the Kohn-Sham equations self-consistently via an iterative process:
\begin{subequations}
    \begin{align}
    (-\nabla^2/2& + v_\text{eff}[n^{(t)}]-\mu)\phi_i^{(t+1)} = \epsilon_i \phi_i^{(t+1)} \label{eq:ks_equations_a} \\
    n^{(t+1)} &= \sum_{i=1}^{\infty} \theta(-\epsilon_i) |\phi_i^{(t+1)}|^2 \label{eq:ks_equations_b} \\
    v_\text{eff}[n](r) &:= \int \frac{n(r')}{|r-r'|}dr' + v_\text{xc}[n] + v^e_\text{ext}
\end{align}
\label{eq:KS_eq_equation}
\end{subequations}
Where $\{\phi_i\}_i$ is a set of auxiliary one-electron orbitals and $\mu$ is the Fermi level, or chemical potential of the electrons. $\theta$ is the Heaviside step function and $v_\text{xc}[n]$ is the exchange-correlation potential. Note how the Fermi level appears as an additive constant on the effective potential, which is equivalent to the normal way of writing KS-DFT\footnote{Because an additive constant $-\mu$ in the effective potential simply shifts all the eigenvalues down by $\mu$, filling more orbitals.}. The superscript $t$ indicates that one iteratively converges to a solution, even though in practice mixing schemes are used to accelerate convergence. 

In this section we define a model which uses the same kind of self-consistency loop to introduce electrostatic interactions. To illustrate how our proposed model works, consider starting from HK-DFT, equation \eqref{eq:hk_dft}, and make the energy stationary by computing the Euler-Lagrange equation with a Lagrange multiplier $\mu$ on the total charge as in \eqref{eq:mathcal_E_mu}. The Lagrange multiplier will end up taking the role of the Fermi level, so we will us the same symbol $\mu$. The resulting equation is:
\begin{align}
    0 &= \delta \left(E[n] -\mu \int n(r) dr \right) \nonumber \\
    \implies 0 &= \int \delta n(r) \left(\frac{\delta G}{\delta n(r)} + v_\text{H}[n](r) + v_\text{ext}^e(r) - \mu\right) dr
\end{align}
Since this holds for all $\delta n$, the term in brackets should be zero, and so we need:
\begin{align}
    \frac{\delta G}{\delta n} = -(v_\text{H}[n] + v_\text{ext}^e(r) - \mu)
\end{align}
Defining $h[n](r) = \delta G / \delta n(r)$, one can try to solve this equation using an iterative procedure:
\begin{align}
    h[n^{(t+1)}] = - ( v_\text{H}[n^{(t)}] + v_\text{ext}^e - \mu )
    \label{eq:blah}
\end{align}
Solving for $n^{(t+1)}$ requires inverting the functional $h$, which for simple orbital free density functionals such as Thomas-Fermi theory, is possible analytically. 

We can attempt to make a self-consistent machine learning model by replacing the terms in \eqref{eq:blah} with their coarse-grained equivalents, and replacing the inverse of $h$ with a flexible machine learned function, which we will denote $F_\text{ML}$. Specifically, we again construct a coarse-grained density as $\rho(r) = \sum_k p_k \phi_k(r)$ from a set of charge coefficients $\mathbf{p}$, and describe the geometry of the system by the nuclear positions $r_i$ and atomic numbers $z_i$. With this, the `fixed-point model' architecture is defined as follows:
\begin{subequations}
    \begin{align}
    \mathbf{p}^{(t+1)} &= F_\mathrm{ML}(\{(z_i, r_i)\}_i, \ v_\mathrm{eff}[\mathbf{p}^{(t)}]) \label{eq:37a} \\
    v_\mathrm{eff}[\mathbf{p}](r) &= \int \frac{\rho(r')}{|r-r'|}dr' + v_\text{app} + \mu \label{eq:37b}
\end{align}
\label{eq:field_model_no_mu}
\end{subequations}
These equations must be iteratively solved until one reaches a fixed point, hence the name of the model. In comparison to \eqref{eq:blah}, the model uses a coarse-grained charge density, not an electron density, which means that $\mu$ needs to change sign. The signs of the various terms are explained further in the derivation below. The electron external potential in \eqref{eq:blah} has also been exchanged for just the applied potential, exactly as in the energy functional approach \eqref{eq:energy_model_energy}. This is because, as in the energy functional model, the coarse-grained density $\rho$ captures the long-range effects of both the nuclei and the electrons.

The model \eqref{eq:field_model_no_mu} is similar, but not identical to the Kohn-Sham approach \eqref{eq:KS_eq_equation}. Comparing to \eqref{eq:KS_eq_equation}, we can see that $F_\text{ML}$ is approximating the diagonalization followed by the sum over orbitals to get the charge density. The effect of the exchange-correlation potential which appears in the effective potential in the Kohn-Sham equations is also approximated in the machine learned function $F_\text{ML}$. As for the function of the Fermi level in the two sets of equations, in \eqref{eq:KS_eq_equation}, when $\mu$ is raised, more electrons are included in the sum due to all the eigenvalues decreasing by $\mu$. In our model, the Fermi level should play a similar role. Once $F_\text{ML}$ is trained to reproduce the reference data, an increase in $\mu$ should cause the total predicted charge to decrease (increasing the number of electrons).

Iterating \eqref{eq:field_model_no_mu} gives a charge density, but not an energy. This formulation of the fixed-point model therefore uses a separate equation to compute the total energy:
\begin{align}
    \mathcal{E} = E_\text{local} &+ E_\text{nonlocal}(\{\mathbf{u}_i\}_i) \nonumber \\
    &+ \frac{1}{2}\iint \frac{\rho(r) \rho(r')}{|r-r'|} drdr' + \int v_\text{app}(r) \rho(r) dr
    \label{eq:field_model_energy}
\end{align}
Where $\mathbf{u}_i = (z_i, r_i, \{p_{ik}\}_k)$, as in section \ref{sec:energy_min}. The density $\rho$ is the final density after converging the loop in \eqref{eq:field_model_no_mu} and $E_\text{local}$ is an energy contribution from a local MLIP. $E_\text{nonlocal}$ is a separate energy contribution which depends on the final charge density. This term plays the same role as $G_\text{ML}$ in the energy functional model. This provides a separation between the complexity of the function used to model the charge density ($F_\text{ML}$) and the energy terms which depend on this density ($E_\text{nonlocal}$). 

Equations \eqref{eq:field_model_no_mu} and \eqref{eq:field_model_energy} define the `fixed-point approach' to self-consistent MLIPs in this paper. An approach like this has been demonstrated before in SCFNN, which uses maximally localised Wannier function centers to describe the charge density, and employs a self-consistency loop similar to equation \eqref{eq:field_model_no_mu}. Using a Wannier center description of the charge density means that the overall charge is fixed, given by the number of Wannier centres used. Using atomic partial charges, the natural way to control the overall charge is via the Fermi level, as is done in KS-DFT. Since the Fermi level is the Lagrange multiplier for the total charge constraint, one can simply raise or lower $\mu$ to obtain the desired total charge in equation \eqref{eq:field_model_no_mu}. This is the approach taken in our implementation and will be discussed more in sections \ref{sec:scf:evaluating_fixedpoint_model} and \ref{sec:scf:methods:scfloop_details}.

It is important to note that the fixed-point architecture defined by \eqref{eq:field_model_energy}, with $\rho$ the solution of \eqref{eq:field_model_no_mu}, is not variational. For the model to be variational in the charge density, $\rho$ would have to be a minimum of \eqref{eq:field_model_energy}, requiring an exact relationship between $E_\text{nonlocal}$ and $F_\text{ML}$, which may not be possible to fulfil in practice. This has potential downsides since information about energies and forces does not constrain the charge density as much, but the structure of $F$ and $E_\text{nonlocal}$ may be easier to learn than that of the single functional $G$.

\subsubsection*{Relationship between the model and HK-DFT}
\label{sec:theory:fixedpoint:theory}

We now explore how the fixed-point model \eqref{eq:field_model_no_mu} can be understood as a coarse-graining of DFT, similarly to the energy functional model.

The easiest way to connect the model to DFT is to start from equation \eqref{eq:amazing_energy_expansion}, which is the result of coarse-graining the HK energy functional and applying the assumptions about the coarse-grained density and Fukui functions. When working with the energy functional model, we have to minimize the energy over $\mathbf{p}$. To turn this into a fixed point problem, consider making the energy stationary with respect to the charge coefficients, with a Lagrange multiplier constraint on the total charge:
\begin{align}
    0 = \frac{\partial}{\partial p_k} \left( G_\text{ML}[\mathbf{p}] + E_\text{Coulomb}[\rho] + \mu \int \rho(r) dr \right)
\end{align}
The sign convention for $\mu$ matches that in \eqref{eq:mathcal_E_mu} and in KS-DFT above, where increasing $\mu$ increases the number of electrons (thus decreasing the total charge). After using the Taylor series expansion for $G$, this gives the following Euler-Lagrange equation for each $k$:
\begin{align}
    0 = &\,\mu \int \phi_k(r) dr + \int \phi_k(r) \left( v_\text{H}[\rho] + v_\text{app} \right) dr \nonumber \\
    &+ \bar{\eta}^{(1)}_k + \sum_l \bar{\eta}^{(2)}_{kl} \delta p_l + \frac{1}{2} \sum_{ln}\bar{\eta}^{(3)}_{kln}\delta p_l \delta p_n + ...
\end{align}
Note that the second integral term is the electric potential which $p_k$ is interacting with. Therefore, define $v_k$ as the effective potential for charge coefficient $k$, as follows:
\begin{align}
    v_k := \int \phi_k(r) \left( v_\text{H}[\rho] + v_\text{app} + \mu\right) dr
    \label{eq:fp_define_vk}
\end{align}
We can now rearrange to get:
\begin{align}
    \bar{\eta}^{(1)}_k + \sum_l \bar{\eta}^{(2)}_{kl} \delta p_l + \frac{1}{2} \sum_{ln}\bar{\eta}^{(3)}_{kln}\delta p_l \delta p_n + ...  = - v_k
    \label{eq:fp_series_in_q}
\end{align}
To make a set of equations which can be solved iteratively, we need to solve the above equation for $\delta \mathbf{p}$ in terms of the effective potential $v_k$. For now, suppose there exists a power series expansion which reverses this relation, with coefficients $\bar{s}^{(n)}$:
\begin{align}
    \delta p_k = \bar{s}^{(1)}_k + \sum_j \bar{s}^{(2)}_{kj} v_j + \frac12 \sum_{jl} \bar{s}^{(3)}_{kjl} v_j v_l + ...
    \label{eq:field_model_expansion_v}
\end{align}
This expansion, combined with \eqref{eq:fp_define_vk} is very close to the fixed-point model in equation \eqref{eq:field_model_no_mu}. The only difference is that in the definition of the model, we did not specify how, or whether, the effective potential should be coarse-grained into atom-centred quantities. In the above series expansion, however, the effective potential is projected into atom-centred quantities $v_k$ in \eqref{eq:fp_define_vk}. We note that the overall sign of the potential in \eqref{eq:37b} does not matter, since this quantity will be the input to a very flexible function. Therefore, we choose the sign so that $v_{\text{eff}}$ in the model is defined for positive test charges. The relative signs of the Fermi level and potential terms, however, must be as in \eqref{eq:fp_define_vk}.

We can understand the coefficients in the series, $\bar{s}$, using conceptual DFT. The model is learning the response of the charge density $\delta \mathbf{p}$ to the effective potential described by $\{v_k\}_k$. This is closely related to the nonlinear softness kernels, which can be used to compute the change in density due to a change in the external potential. As discussed in section \ref{sec:definitions:cdft}, for the electron density $n$ which is the ground state of the external potential $v_\text{ext}^e$, one can write:
\begin{align}
    \Delta n(r) &= \int s(r,r') \Delta v^e_{\text{ext}}(r')dr' \nonumber \\&+ \frac{1}{2}\iint s(r,r',r'') \Delta v^e_{\text{ext}}(r') \Delta v^e_{\text{ext}}(r'')dr'dr'' + ...
    \label{eq:perturb_dft_density}
\end{align}
Where the softness kernel $s(r_1,...,r_n)$ is defined as:
\begin{align}
    s(r_1, ..., r_n) = \frac{\delta n(r_1)}{\Delta v^e_{\text{ext}}(r_2) ... \Delta v^e_{\text{ext}}(r_n)}
\end{align}
This is evidently related to the expansion in \eqref{eq:field_model_expansion_v}, but not quite the same. The softness kernels are the derivative of the density with respect to an \textit{external} potential, when in the ground state. This is different from equation \eqref{eq:field_model_expansion_v} where we are expressing $\delta \mathbf{p}$ in terms of the \textit{effective} potential $v_k$, and in which the reference $\mathbf{p}^0$ is not a ground state, but instead just any convenient set of reference charges. This distinction is important, because the effective potential contains a contribution from the Hartree potential, which itself depends on the charge coefficients. Nevertheless, the terms in $\bar{s}$ are playing a similar role to the nonlinear softness kernels, and hence we will next discuss some relevant properties of the softness kernel. 

Finally, there is a question of when the inverse series expansion \eqref{eq:field_model_expansion_v} exists. We will not attempt to analyse this in the general case, but will note that relationships between nonlinear hardness and softness kernels have been investigated previously \cite{nonlinear_senet_1996}. The case of the inversion of the truncated series to the second order is discussed in Section~\ref{sec:scf:theory:equivalence}.

In this section we have presented a connection between the fixed-point model architecture and DFT. Specifically, we showed that one can understand the fixed-point model as the Euler-Lagrange equation of a coarse-grained energy functional. Therefore, if one can find a coarse-grained energy functional which approximates HK-DFT (as discussed in the energy functional section), then one can reformulate the minimisation problem as a self-consistent field iteration, which will also approximate HK-DFT. It is important to note, however, that the self-consistent field loop in our model is not the direct analogue of the KS-DFT loop, even though the iterative process looks similar. The SCF loop in KS-DFT comes from solving the auxiliary non-interacting electron problem, while the SCF loop in our model comes about from separating the Euler-Lagrange equation (of HK-DFT) into the Coulomb term and the remainder. This is the reason that the exchange correlation term in KS-DFT appears in the effective potential, but is included in the machine learned function $F_\text{ML}$ in our model. 

An alternative justification for machine learned fixed-point schemes, which directly analyses the diagonalization step one would find in KS-DFT, was recently proposed by Thomas \textit{et al} \cite{Thomas_2025}. This work considers the action of the Fermi-Dirac operator on a density functional tight binding Hamiltonian. The action of the Fermi-Dirac function at zero temperature is equivalent to the diagonalization followed by sum over orbitals in \eqref{eq:KS_eq_equation}. In that work it is shown that one can decompose the charge density into local pieces, and that there exists a function which predicts the local component of the density based on atom-centred expansion coefficients of the electrostatic potential. Crucially, it is shown that this function can be approximated by a local, body ordered polynomial function. While there are other differences in the analysis, this result can also be examined as a way to derive a self-consistent-field based MLIP from KS-DFT, where one can argue that the functions to be approximated are indeed local. While this result provides a more direct link between the fixed-point model and KS-DFT, in much of the analysis in this paper we will make use of the HK-DFT perspective since it lets one easily compare and contrast the limitations of the two different architectures.


\subsubsection*{Relevant Properties of the fixed-point function}

The terms in the series expansion which the fixed-point model is approximating are related to the nonlinear softness kernels. If the coefficients $\bar{s}$ are local, in that $\bar{s}_{kj}$ becomes small when $k$ and $j$ are far apart, then a local MLIP architecture is appropriate for $F_\text{ML}$ in \eqref{eq:field_model_no_mu}.

Thankfully, some properties of the softness kernel are known. Notably, it has been argued that locality of the softness kernel is directly related to the near-sightedness of electronic matter \cite{locality_kernels_Ayers_2017}. This was investigated numerically by computing various reactivity kernels, including the softness, for molecules and then integrating $s(r,r')$ onto atoms via a partitioning scheme, to create an `atom condensed' softness $s(i,j)$. It was shown, in DFT calculations of both saturated and conjugated molecules, that the atom condensed softness decays rapidly as a function of distance between the atoms \cite{Proft2024, locality_kernels_Ayers_2017, Geerlings2010}. This is in contrast to the hardness kernels that must be learned by the energy functional approach, which are non-local due to the kinetic energy term.

Unfortunately, like in the energy functional model, the case of a variable number of electrons on an isolated cluster or fragment presents a challenge because the underlying functions become non-smooth. In the energy functional approach this manifests as a discontinuous derivative in the kinetic-exchange-correlation energy of a cluster as a function of electron number. In the fixed-point model, we are instead learning a map from the effective potential, including the added Fermi level, to the coarse-grained charge density. The corresponding non-smooth behaviour is that this function should be a staircase function, with the total charge decreasing by exactly 1 each time the Fermi level passes the next electron affinity of the cluster---as sketched in figure \ref{fig:lines_G_F}b. This implies that the total charge in the model (as predicted by $F_\text{ML}$) needs to be a discontinuous function of $\mu$.

\subsubsection*{Connection To Other ML Approaches}

The way that the total charge is naturally controlled in the fixed-point model - by explicit use of the Lagrange multiplier - can be connected to other approaches in literature. The AIMnet-NSE model \cite{aimnet_nse} iteratively predicts partial charges via a message passing procedure and controls the total charge by redistributing the total excess charge in proportion to an atomic, learned function $f_i$, which is interpreted as the atom-centred Fukui function. In the fixed-point model above, the iterative updating of the Fermi level to reach the correct charge bares some similarities. The difference is that because the Fermi level is added to the effective potential (and it is the effective potential that controls charge redistribution), charge transfer is treated in the same way as induced polarization. The self-consistency loop therefore mediates electrostatic induction as well as controlling the total charge. 

Finally, there is a connection between the fixed-point model and the MACE-POLAR-1 foundation model \cite{mace_polar}. The MACE-POLAR-1 architecture inherits many of the ideas of the self-consistent fixed-point model, but with adaptations that make it more like a global message passing neural network and not a true self-consistent model. In MACE-POLAR-1, the same fixed-point style iteration is performed, wherein atomic density coefficients are updated based on electrostatic potential features, but this is not repeated until convergence. Instead, the model just makes two updates, and uses \textit{different} functions $F_\text{ML}$ in each update. Not iterating to self consistency means that one can no longer control the total charge with the Fermi level, and hence the model uses the Fukui function based charge distribution scheme, as described in AIMnet-NSE above, to achieve the correct total charge after just two steps. The MACE-POLAR-1 model is thus somewhere in between a self-consistent approach and a long-range message passing neural network, using some elements of the fixed-point model but in simpler, easier to train configuration.

\subsection{Relationship between the Energy Functional and fixed-point models}
\label{sec:scf:theory:equivalence}

\begin{figure*}[htb]
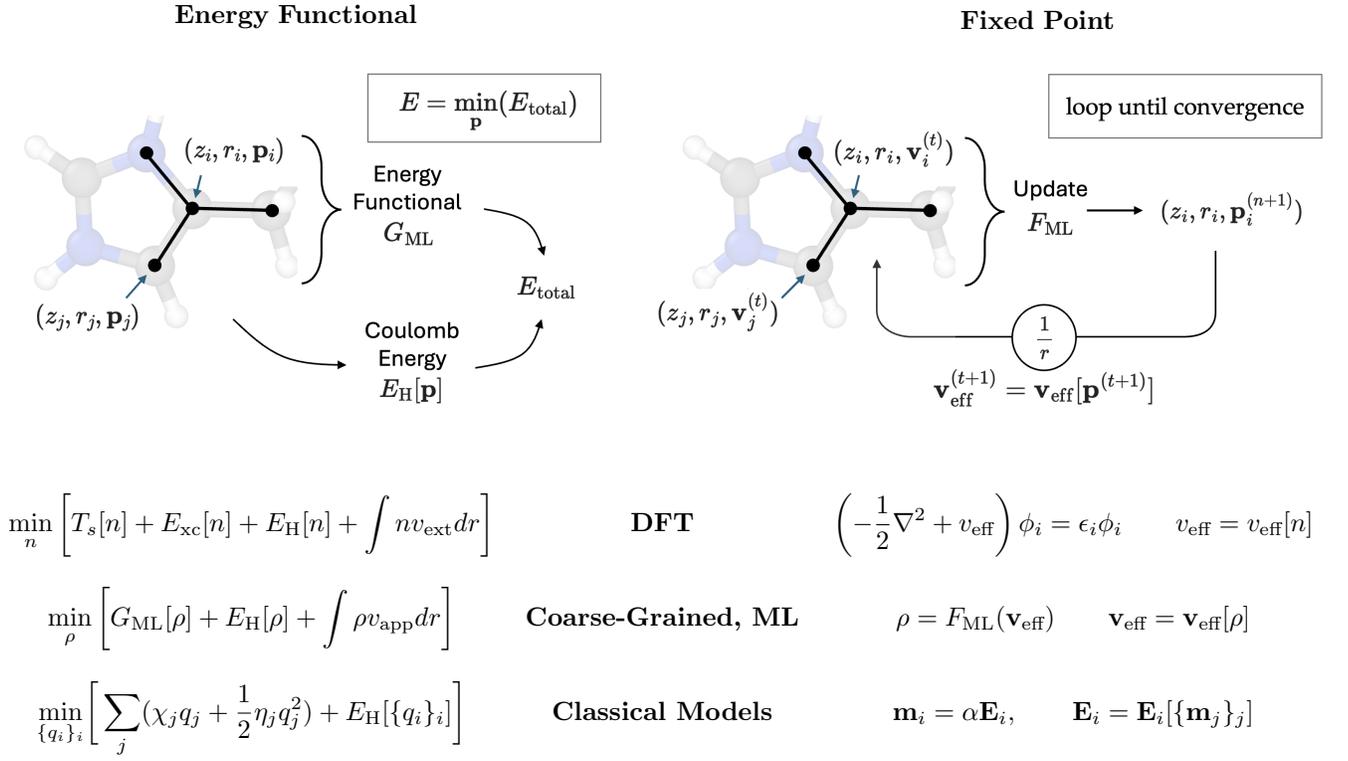

    \centering
    \include{tables/hand_waving_tikz_figure}
    \caption{This paper investigates the theory and practical aspects of two alternative methods for creating MLIPs with a rich description of electrostatics. The figure should be understood along with sections \ref{sec:energy_min}--\ref{sec:scf:theory:equivalence}. The two methods are referred to as the energy functional method and fixed-point method. In this cartoon, $\mathbf{p}_i$ represents a set of descriptors of the charge density associated with atom $i$, for instance a list of partial multipole moments. $\mathbf{v}_i$ represents a vector of descriptors of the effective potential around atom $i$. The lowest row, labelled `empirical models' compares the classical charge equilibration approach and a linear polarizable force field (in which $\mathbf{m}_i$ denotes an atomic dipole on atom $i$, and $\mathbf{E}_i$ denotes the electric field at atom $i$). In DFT, the left and right hand side are equivalent. However, when we approximate the functions involved with machine learned surrogates, and switch to a coarse-grained charge density, the two approaches are no longer equivalent in principle.}
    \label{fig:model_cartoons}
\end{figure*}

To summarize, the energy minimization approach attempts to learn an approximate functional of a coarse-grained density, analogous to the kinetic and exchange-correlation functional. The fixed-point method attempts to learn the corresponding part of the Euler-Lagrange equation. If one considers perturbations, the models are approximating generalized hardness and softness kernels. 

Figure \ref{fig:model_cartoons} presents a visual comparison of the two approaches, where the connection to both DFT and empirical force fields is also noted. For instance charge equilibration, as used in empirical and some machine learning models, naturally fits into the energy functional approach since one parametrizes an expansion of the energy functional. Likewise, polarizable force fields such as AMOEBA \cite{Ren2003PolarizableSimulation, Rathnayake2020EvaluationSolution} naturally fit into the other category. In AMOEBA, each atom is given a partial dipole moment $\mathbf{m}_i = \alpha \mathbf{E}_i$, which is related to the electric field at that point $\mathbf{E}_i$ by a polarizability $\alpha$. One must choose $\alpha$, explicitly parametrizing the response function, and then evaluate the model by iterating the computation of the partial dipoles and the electric fields to reach self-consistency.
 
In KS-DFT, the full kinetic, exchange-correlation energy is non-local (due to the KS kinetic energy) when considered as a function of the electron density. Similarly, the map from $v_\text{eff}$ to the electron density $n(r)$ used in the self-consistency cycle is also non-local as a function of $v_\text{eff}$. On the other hand, existing ML approximations in the literature have made extensive use of local and ``semilocal'' functions, the latter term being used here in the graph neural network context, meaning that a function's value on one atom depends on the state of the atom, and the states of neighbouring atoms. We now show that local approximations to either the energy functional or the fixed-point function are equally expressive, except in the case of perfectly conducting systems, but that semilocal approximations are not equivalent.

\subsubsection{Local Approximations are (almost) Equally Expressive}

Take for instance the functional used in CENT \cite{cent2015} and related approaches, where the charge density is defined as the minimizer of \eqref{eq:energy_model_energy}, with 
\begin{equation}
    G_{\text{ML}} = \sum_{i=1}^{N_\text{atoms}} (q_i \chi_i + \frac{1}{2}q_i^2 \eta_i)
    \label{eq:G_for_4gnn}
\end{equation}
The coefficients $\chi_i$ and $\eta_i$ are geometry dependent electronegativities and hardnesses for each atom. Assume the Hartree energy is evaluated by treating the charges as point charges.
In this case, one can derive a corresponding fixed-point function by solving analytically for the minimizer of \eqref{eq:energy_model_energy}. This is done by finding the stationary point with respect to $q_i$ while adding a Lagrange multiplier $\mu$ to control the total charge:
\begin{align}
0 = \frac{\partial}{\partial q_i} & \left( \sum_j \Big(q_j \chi_j + \frac{\eta_j q_j^2}{2}\Big) + \frac{1}{2} \sum_{k\neq j} \frac{q_k q_j}{r_{kj}} \right. \\
&+ \sum_k q_k v_k^{\text{app}}  + \left. \mu \left(\sum_j q_j - Q_\text{total}\right)\right) \nonumber \\
    \implies & q_i = -\frac{\chi_i + v_i + \mu}{\eta_i}
    \label{eq:4gnn_minimised}
\end{align}
where we defined $v_i$, which is the potential at site $i$ due to all the other charges and applied fields:
\begin{align}
    v_i(\mathbf{q}) = \sum_{j\neq i} \frac{q_j}{r_{ij}} + v_i^{\text{app}}
\end{align}
Equation \eqref{eq:4gnn_minimised} can now be interpreted as a fixed-point equation:
\begin{align}
   q_i^{(t+1)} = -\frac{\chi_i + v_i(\mathbf{q}^{(t)}) + \mu}{\eta_i}
    \label{eq:4gnn_as_field}
\end{align}
The key fact is that this fixed-point update function is local: the charge on atom $i$ only depends on the potential at atom $i$ (and the local geometry which determines $\chi_i$ and $\eta_i$), not on the potential at other atoms. The potential is a global function of the charge density, which is known and simple. 

The same holds if one replaces the quadratic dependence in \eqref{eq:G_for_4gnn} with an arbitrary function $g$ which depends on the local geometry $X_i$ around atom $i$, so that
\begin{equation}
    G_{\text{ML}} = \sum_{i=1}^{N_\text{atoms}} g_{X_i}(q_i).
    \label{eq:local_G}
\end{equation}
The Euler-Lagrange equation is
\begin{align}
    -\mu = g_{X_i}'(q_i) + v_i
    \label{eq:gradient_of_local_energy_method}
\end{align}
Rearranging this as a fixed-point iteration gives
\begin{equation}
    q_i^{(t+1)} = (g'_{X_i})^{-1}(- v_i^{(t)}-\mu)
    \label{eq:method_relation_for_local}
\end{equation}
if $(g'_{X_i})^{-1}$ exists. By the implicit function theorem, this is always true in some small neighbourhood around the solution to the fixed-point equation, provided $g''_{X_{i}}(q_i) \neq 0$. The condition $g''_{X_{i}}(q_i) = 0$ has physical meaning, which is when the degree of freedom $q_i$ has infinite susceptibility.

This reveals one key difference between the two approaches. An energy functional can model perfect classical conductors (with a zero screening length) in which the susceptibility is infinite, since the learned function $g_{X_i}(q_i)$ simply has to have zero curvature. The equivalent fixed-point model, however, is trying to learn the response of a charge to a change in potential (equation \eqref{eq:method_relation_for_local}). This goes to infinity as $g''$ goes to zero, and the model cannot learn a function with infinite gradient in practice. The opposite condition, zero susceptibility, where the gradient of $(g')^{-1}(v)$ goes to zero, is possible in the case of the fixed-point method, but impossible for a smooth energy functional, by similar logic. In real systems, the case of zero susceptibility does occur when considering the response of total charge to varying the potential for a non-periodic system with a band gap. If a constant is added to the electric potential everywhere, the total charge does not change and this variable therefore has zero susceptibility. Figure \ref{fig:lines_G_F}, panels c and d, show the above conditions visually. 

Intuitively, the main result is that \textit{local} energy functionals like \eqref{eq:local_G} can be re-expressed as \textit{local} fixed-point updates \eqref{eq:method_relation_for_local}. When using local functions, the two methods therefore have approximately the same expressivity. The difference is that the fixed-point Scheme cannot formally describe perfectly conducting systems, where electric susceptibility is infinite. By similar logic, the energy functional approach cannot describe perfectly stiff systems where some degrees of freedom have zero susceptibility. It is also worth noting that a quadratic energy functional \eqref{eq:G_for_4gnn} became a linear fixed-point equation \eqref{eq:4gnn_as_field}.

Here we have shown the result for point charges, but appendix \ref{appendix:exp_basis} shows that the same holds for smeared charges and for higher atomic multipoles. 



\subsubsection{Semi-Local Approximations have Different 
Expressivity}

Suppose instead that $G_{\text{ML}}$ is semilocal, in the sense that
\begin{align}
    G_{\text{ML}} = \sum_{i=1}^{N_\text{atoms}} f(\{\mathbf{u}_j\}_{j\in \mathcal{N}(i)})
    \label{eq:semilocal_G}
\end{align}

Where $\mathcal{N}(i)$ is the set of indices of the atoms within some fixed cutoff distance of atom $i$. This is a very natural choice given the success of semilocal features in message passing MLIPs. Now, however, the Euler-Lagrange equation cannot be re-expressed as a function which has semilocal dependence on the descriptors of the electric potential. As in equation \eqref{eq:gradient_of_local_energy_method}, the Euler-Lagrange equation is:
\begin{equation*}
    \nabla_{\mathbf{q}} G_\text{ML} = - \mu \mathbf{1} - \mathbf{v}
\end{equation*}
Where $\mathbf{1}$ is a vector of 1's and $\mathbf{v}$ is a vector of the electric potentials at each site, $\mathbf{v} = \{v_i\}_i$. Previously, because $G$ was a sum of local pieces, the equations became uncoupled in $i$. However if we instead have the semilocal form \eqref{eq:semilocal_G}, the gradient operation on the left hand side will give coupled equations. 

For instance, suppose one takes $G_{ML}$ to be quadratic, but including off diagonal terms like $\frac{1}{2}H_{ij} q_i q_j$, so that the total $G_{ML}$, \eqref{eq:semilocal_G}, is:
\begin{align}
    G_{ML} = \mathbf{q}^T \boldsymbol{\chi} + \frac{1}{2}\mathbf{q}^T H \mathbf{q}
\end{align}
Where $H$ is a matrix of hardness coefficients, with $H_{ii}$ being the familiar atomic hardness $\eta$, and $\bm{\chi}$ is a vector of electronegativities. The semilocal structure implies that the off diagonal terms are zero when they refer to spatially distant atoms.  This would give the following fixed-point equivalent:
\begin{align}
    H \mathbf{q}^{n+1} = - \mu \mathbf{1} -\mathbf{v}[\mathbf{q}^n] - \bm{\chi}
\end{align}
Even if the matrix $H$ is non-zero for only `nearby' pairs of atoms $i$ and $j$, in general $H^{-1}$ is not semilocal (but it may be still spatially decaying). The reverse also applies: semilocal fixed-point functions cannot be rewritten as strictly semilocal energy functionals, but may correspond to spatially decaying energy functionals. Therefore when using semilocal functions which are natural in MLIPs that pass atomic features to neighbouring atoms, the energy functional and fixed-point approaches have different expressivity.

Inverting the matrix $H$, to create a map from $\mathbf{v}$ to $\mathbf{q}$, is exactly the process which was discussed in section \ref{sec:theory:fixedpoint:theory}, in which we inverted a series in $\delta p_k$ to create a series in $v_k$. The above example shows that one can perform the inversion when the energy functional expansion is truncated at second order, but strict semi-locality in the coefficients $\eta$ is lost in the resulting inverted series. 

\subsection{Classification of Existing Schemes}
\label{sec:classification}

\begin{table*}[htb!]
\centering
\setlength{\belowcaptionskip}{3mm}
\caption{Comparison of methods for incorporating a model of the charge density into machine learning force fields. This table only presents models which are strictly self-consistent, and not message passing networks where the contributions with iterations get smaller but in which there is no `convergence criteria'. The column `Locality' describes whether the fixed-point function or the energy functional is local or semilocal, in the sense discussed in section \ref{sec:scf:theory:equivalence}. GAP refers to the Gaussian Approximation Potential MLIP \cite{Bartok2010GaussianElectrons}}
\label{table:model_cat}

\begin{tabular}{@{}l l c c l@{}}
\toprule
\thead{Category} 
  & \thead{Model} 
  & \thead{Description \\ of charge density} 
  & \thead{Locality} 
  & \thead{Functional form} \\
\midrule
\multirow{8}{*}{\makecell[l]{Energy \\ functional}}
  & \makecell[l]{QEq}         & Partial charges   & Local & Quadratic \\
  & \makecell[l]{4GNN \cite{4gnn_ko_2020}}      & Partial charges   & Local & Quadratic \\
  & \makecell[l]{CENT \cite{cent2015}}        & Partial charges   & Local & Quadratic \\
  & \makecell[l]{kQEq \cite{kqeq_og2022}}        & Partial charges   & Local & Quadratic \\
  & \makecell[l]{BpopNN \cite{bpopnn}}      & Spin resolved partial charges   & Semilocal & GAP+empirical baseline \\
  & \makecell[l]{ACE‐Q \cite{ace_q_published}}       & Partial charges   & Semilocal & Quadratic \\
  & \makecell[l]{eMLIP \cite{eMLP}}       & Wannier centers   & Semilocal & Schnet network \\
   & \makecell[l]{Hu et. al. \cite{pqeq_hu_2025}}       & Displaced charges  & Fixed Parameters & Quadratic \\
\addlinespace
\midrule
\addlinespace
 \multirow{2}{*}{\makecell[l]{Fixed \\ point}} & \makecell[l]{Classical \\ polarizable FF} & Partial charges/dipoles & Local & Linear \\
  & \makecell[l]{SCFNN \cite{scfnn}}      & Wannier centers   & Local & Linear \\
\bottomrule
\end{tabular}
\end{table*}

Given that the fixed-point and energy minimization schemes can, under some circumstances, be equivalent to one another, it is interesting to examine how the existing self-consistent models can be categorized, as shown in Table \ref{table:model_cat}. Although there are many ways in which researchers have introduced electrostatics, here we only compare those which are formally performing some kind of self-consistency procedure. Message passing models which iteratively construct a charge density, but with a fixed number of steps with no convergence criterion, can (and do) have different learned functions at each step or layer. This makes the comparison with true fixed-point models less useful, and thus we have not included such models in Table \ref{table:model_cat}.

It is clear to see that many energy functional based models use a quadratic functional, and both fixed-point models use a linear update. As can be seen by equations (\ref{eq:G_for_4gnn}-\ref{eq:method_relation_for_local}) a local, linear fixed-point function is equivalent to a quadratic energy functional and thus these models have, in principle, similar flexibility. The physical differences may therefore be apparent in how they models treat limiting cases of conducting systems and systems with near zero susceptibility.

These models have not been benchmarked on the same test systems and even if they were, a multitude of other differences in the model construction and training process mean that these models may actually perform very differently in practice.
 
\subsection{Arising Questions}

Given the above ideas, we would like to understand the following:
\begin{itemize}
    \item When the energy functional and fixed-point Schemes are implemented on an equal footing, do both approaches perform similarly, or are there key advantages to one or the other approach? 
    \item Existing schemes have a wide range of complexity of either the energy functional or fixed-point function, but since this operation is iterated during evaluation, cheaper functional forms are greatly preferred for this component. What level of flexibility is required for a given physical problem?
    \item Are there other practical aspects which make implementing one scheme inherently more difficult? This has certainly been suggested by previous work: Quadratic charge equilibration methods have been reproduced several times and, even without training on charges, have reliable behaviour. On the other hand, the nonlinear energy functionals of eMLP and BpopNN both required additional tricks during training to make well behaved energy functionals. These practical points are extremely important, but due to differences implementation and other aspects of model fitting and benchmarking, we do not have a clear picture of how the features discussed in this section influence these practical aspects. 
\end{itemize}

To answer these questions, we have implemented both the ML fixed-point and ML energy functional approaches in a directly comparable manner. In both cases, a MACE model is used for any local feature construction. Also, in both cases we vary the flexibility of the key part of the network: the energy functional or the fixed-point update function. Sections \ref{sec:implementation} and \ref{sec:scf:training} go over the details of our implementation, and the different ways one can train self-consistent models. Following that, sections \ref{sec:results:metal_water} and \ref{sec:results:sio2} present results on two interesting examples, answering the above questions.

\section{Implementations in the MACE Framework}
\label{sec:implementation}

In this section we will use $\mathbf{r}$ to denote position, since it will be useful to distinguish between the vector $\mathbf{r}$, its magnitude $r$ and its direction $\hat{\mathbf{r}}$.

\subsection{Representation of the Charge Density}
\label{sec:scf:rep_density}

In all the models in this paper, we represent the charge density via a multipolar expansion on atoms. Atomic multipoles are denoted in spherical tensor notation as $p_{i,lm}$ where $i$ is the atom index. $p_{i,lm}$ are equivalent to the $p_k$ used in the derivations of section \ref{sec:theory}. These atomic multipoles are associated with Gaussian type orbitals, to give a smooth charge density. Specifically, a set of atoms with some multipole coefficients $p_{i,lm}$ gives rise to the following smooth density:
\begin{align}
    \rho(\mathbf{r}) &= \sum_{i,lm} p_{i,lm} \phi_{\sigma lm}(\mathbf{r} - \mathbf{r}_i)
    \label{eq:atomic_multipole_expansion} \\
    \phi_{\sigma lm}(\mathbf{r}) &=  C_{l\sigma} r^{l} Y_{lm}(\hat{\mathbf{r}}) \exp \left(-\frac{r^2}{2 \sigma^2} \right)
    \label{eq:def_gto}
\end{align}
$C_{l\sigma}$ is a normalization constant and $\sigma$ is a smearing width. Unless otherwise stated, all models in this work use $\sigma=1.5\,\text{\AA}$ and $l\leq 1$, corresponding to atomic charges and dipoles. It was found that little improvement can be gained from higher multipole order, which comes at a significantly higher computational cost. We will use
\begin{align*}
    \mathbf{p} = \{p_{i,lm}\}_{i,lm}
\end{align*}
to denote the vector of all the model's atomic multipole moments, which may also simply be referred to as the model's charge density. 

 \begin{figure}[t!]
     \centering
     \includegraphics[width=\linewidth]{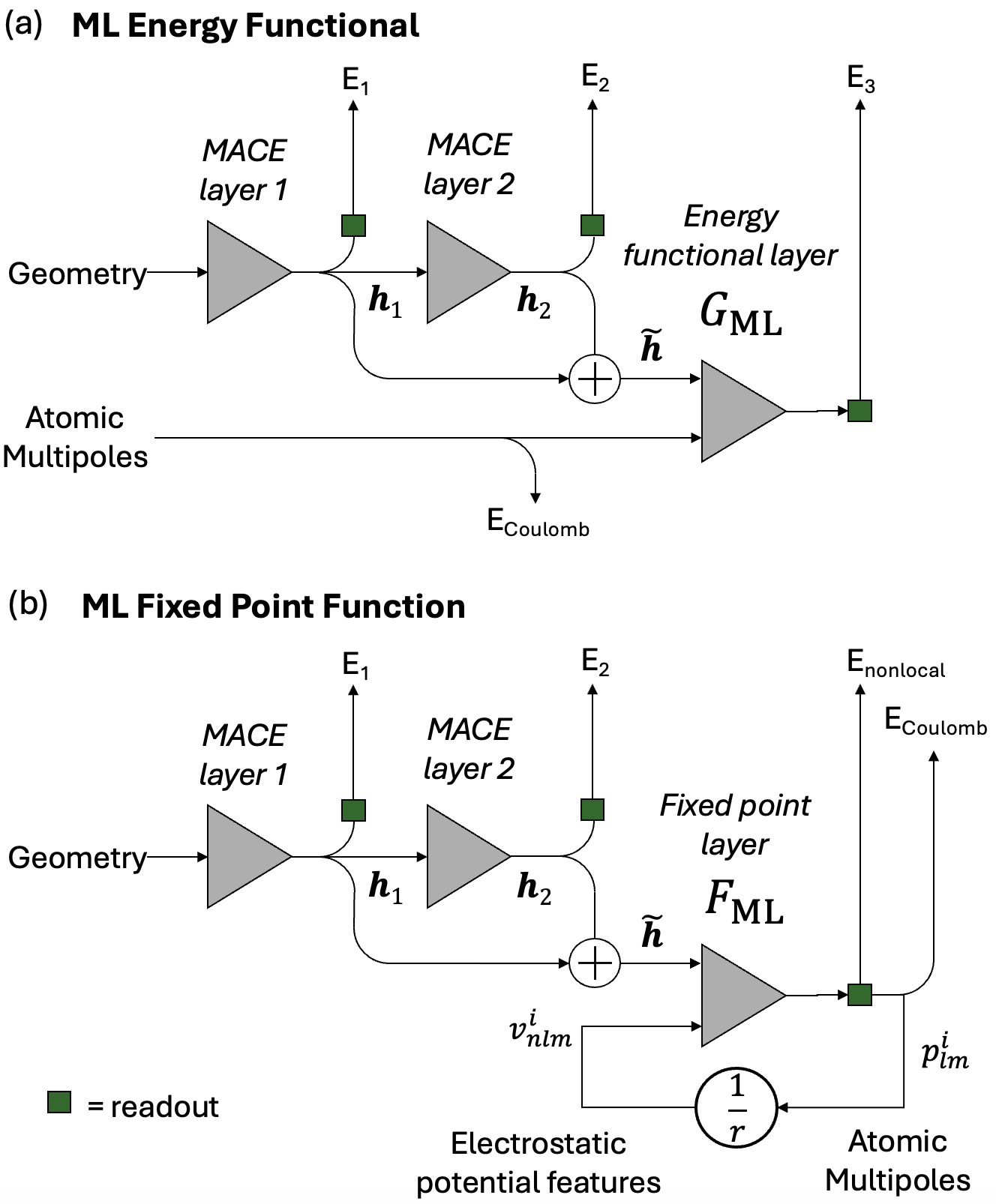}
     \caption{Overview of the implementation of each approach.}
     \label{fig:scf:model_diagram}
 \end{figure}

\subsection{fixed-point model}

In section \ref{sec:theory:fixedpoint} we defined the fixed-point approach to self-consistent electrostatics in MLIPs via equations \eqref{eq:field_model_no_mu} and \eqref{eq:field_model_energy}. To summarise, one obtains a set of charge density expansion coefficients via the following self-consistent loop:
\begin{align*}
    \mathbf{p}^{(t+1)} &= F_\text{ML}(\{(z_i, r_i)\}_i, \ v_\text{eff}[\mathbf{p}^{(t)}]) \tag{from \ref{eq:37a}} \\
    v_\text{eff}[\mathbf{p}](\mathbf{r}) &= \int \frac{\rho(\mathbf{r}')}{|\mathbf{r}-\mathbf{r}'|}d\mathbf{r}' + v_\text{app}(\mathbf{r}) + \mu \tag{from \ref{eq:37b}}
\end{align*}
And then computes and energy from the final set of coefficients $\mathbf{p}$ via:
\begin{align*}
    \mathcal{E} = E_\text{local} &+ E_\text{nonlocal}(\{\mathbf{u}_i\}_i) \nonumber \\
    &+ \frac{1}{2}\iint \frac{\rho(\mathbf{r}) \rho(\mathbf{r}')}{|\mathbf{r}-\mathbf{r}'|} d\mathbf{r}d\mathbf{r}' + \int v_\text{app}(\mathbf{r}) \rho(\mathbf{r}) d\mathbf{r}
    \tag{from \ref{eq:field_model_energy}}
\end{align*}

An outline of our implementation of a fixed-point model is shown in Figure \ref{fig:scf:model_diagram}b. Two normal MACE layers are used to make the local component of the energy ($E_\text{local} = E_1 + E_2$, in equation \eqref{eq:field_model_energy}), as well as descriptors of the geometry around each atom. The local geometry descriptors for atom $i$ after the first and second layers are denoted $h^{(1)}_i$ and $h^{(2)}_i$ respectively. These are the MACE node features. A learnable linear combination of these two vectors $\tilde{h}_i$ is used to represent the geometry in the following part of the model:
\begin{align}
    \tilde{h}_{i,klm} = \sum_{k'} W_{lkk'} h^{(1)}_{i,k'lm} + \sum_{k'} W'_{lkk'} h^{(2)}_{i,k'lm} \label{eq:h_tilde}
\end{align}
These local features on each atom are combined with a description of the electric potential around that atom, to predict a contribution to the charge density via $F_\text{ML}$. The following sections show how these descriptors are created, the form of the fixed-point function $F_\text{ML}$, and the practicalities of how the model is trained and evaluated. 

\subsubsection{Definition of Electrostatic Features}
\label{sec:scf:fixedpoint_imp_esp_features}

The charge density of the model is defined by equation \eqref{eq:atomic_multipole_expansion} and the potential is computed as
\begin{align}
    v_\text{eff}(\mathbf{r}) = \int \frac{\rho(\mathbf{r}')}{|\mathbf{r} - \mathbf{r}'|}d\mathbf{r}' + v_\text{app}(\mathbf{r}) + \mu,
    \label{eq:v_e_h}
\end{align}
in which $v_\text{app}$ is due to applied fields or embeddings and $\mu$ is the Fermi level. See section \ref{sec:theory:fixedpoint} for discussion of why the Fermi level appears here. From this potential, atom-centred features are derived by projecting onto atomic basis functions $\phi_{NLM}$.
\begin{align}
    v_{i,NLM} = \frac{1}{\mathcal{N}_{NL}} \int d\mathbf{r} \ \phi_{NLM}(\mathbf{r} - \mathbf{r}_i) v_\text{eff}(\mathbf{r})
    \label{eq:field_featurization}
\end{align}
This method for featurizing the effective potential is inspired by the long-distance equivariant descriptor (LODE) approach \cite{lode2019, lode2021}. The basis functions $\phi_{NLM}$ are Gaussian type orbitals, with the same form as those used for the charge density expansion \eqref{eq:atomic_multipole_expansion}.
In this case, $N$ indexes a set of different Gaussian widths $\sigma_N$. $\mathcal{N}_{NL}$ is a normalization constant for each $N$ and $L$ combination. We will write $\mathbf{v} = \{v_{i,NLM}\}_{i,NLM}$ for the vector of all electric potential features. The use of capital indices is to differentiate the functions used for creating the features ($\phi_{NLM}$) from those used to expand the charge density in \eqref{eq:atomic_multipole_expansion}. In both cases we use the same Gaussian type orbital form, but the smearing widths $\sigma$, and the range of $l$ or $L$ will in general be different.

The electric potential is not dimensionless and the value of the features may be far from 1, hence the normalization $\mathcal{N}_{NL}$ is required. Increasing either the number of Gaussians or the maximum order $L$ leads to a richer description. All the integrals are implemented for both open and periodic boundary conditions, as discussed in appendix \ref{appendix:electrostatics_implementation}.

In this paper, we also subtract a self-interaction term when computing electrostatic features. Appendix \ref{appendix:electrostatics_implementation} provides some more detail on self interaction terms.



\subsubsection{Fixed Point Update Function}
\label{sec:scf:fixedpoint_update_layer_details}

We need to design $F_\text{ML}$, which predicts atomic multipole moments based on the geometry and electric potential features. This function is broken down into a contribution depending only on the local geometry (analogous to $\mathbf{p}^0$ in section \ref{sec:theory}), and a term which also depends on the electric potential features (analogous to $\delta \mathbf{p})$:
\begin{subequations} \label{eq:prediction_of_multipoles} 
\begin{align}
    \mathbf{p}_{i} &= F_{\text{ML},i}(\tilde{\mathbf{h}}, \mathbf{v}) = \mathbf{p}_{i}^{(\text{local})} + \mathcal{F}_i(\tilde{\mathbf{h}}, \mathbf{v}) \label{eq:prediction_of_multipoles:sum} \\
    p_{i,lm}^{(\text{local})} &= \sum_{k} W_{lk} h^{(1)}_{i,klm} + \sum_{k} W'_{lk} h^{(2)}_{i,klm} \label{eq:prediction_of_multipoles:local}
\end{align}
\end{subequations}
In \eqref{eq:prediction_of_multipoles:sum}, $\tilde{\mathbf{h}}$ denotes all the elements of $\tilde{h}_{i,klm}$. The local part \eqref{eq:prediction_of_multipoles:local} is a learnable linear function of the MACE atomic features in which $W$ and $W'$ are weight matrices. 

The second part will be called the ``non-local contribution'', $\mathbf{p}^\text{(nonlocal)}:=\mathcal{F}_i(\tilde{\mathbf{h}}, \mathbf{v})$. We will use the term nonlocal because it depends on the potential $\mathbf{v}$ (which is a function of the entire system) even though the function $\mathcal{F}$ is generally a local function of $\mathbf{v}$ as discussed below. We compare three different choices for the functional form of $\mathcal{F}$. These are
\begin{itemize}
    \item \textbf{One-Body Linear.} $\mathcal{F}$ is an affine function of the electric potential features $\mathbf{v}$, dependent also on the geometric features $\tilde{\mathbf{h}}$. The non-local component of the density on atom $i$ only depends on the geometric and electrostatic features on atom $i$. This is an analogous to a polarizable force field with a linear polarizability:
    \begin{align}
    \mathbf{p}_i^{(\text{nonlocal})} = \mathcal{F}(\tilde{\mathbf{h}}_i, \mathbf{v}_i)
    \label{eq:one_body_F}
\end{align}
    \item \textbf{One-Body Nonlinear.} The function $\mathcal{F}$ is still one-body in both $\tilde{\mathbf{h}}$ and $\mathbf{v}$ (equation \eqref{eq:one_body_F}), but a nonlinearity is added to allow nonlinear dependence of the charge density on the electric potential features. As shown in the methods, section \ref{methods:implementation_fixed_point}, this takes the form of a 3 layer multi layer perceptron acting on invariant features of the electric potential.
    \item \textbf{Many-Body.} $\mathcal{F}$ predicts the non-local density on atom $i$ as a nonlinear function of the features of atom $i$ and those of neighbouring atoms.
    \begin{align}
    \mathbf{p}_i^{(\text{nonlocal})} = \mathcal{F}(\{(\tilde{\mathbf{h}}_j,\mathbf{v}_j)\}_{j\in\mathcal{N}(i)})
\end{align}
\end{itemize}

The implementation details for each of these choices are given in section \ref{methods:implementation_fixed_point}.

\subsubsection{Nonlocal Energy Contribution}

A fixed-point model in our formalism is not variational. Therefore, our architecture also includes a nonlocal energy contribution, labelled $E_\text{nonlocal}$ in Figure \ref{fig:scf:model_diagram}b and equation \ref{eq:field_model_energy}. One can think of this term as either restoring the missing variational energy contribution, or simply an additional energy term dependent on the final charge density. Unless otherwise stated, all the models presented in this paper use a simplified version of the a MACE layer for this nonlocal energy term. We found that for the metal--water system studied in section \ref{sec:results:metal_water}, the performance of all models is similar if this term is removed completely. For the second test case of defects in silicon dioxide, it was important to include this term. Section \ref{methods:implementation_fixed_point} gives more details on the implementation of the nonlocal energy contribution. 

\subsubsection{Evaluating the fixed-point model}
\label{sec:scf:evaluating_fixedpoint_model}

As outlined in Figure \ref{fig:scf:model_diagram}b, the model is evaluated by first performing two MACE layer iterations. This results in a local energy contribution, and a local contribution to the charge density \eqref{eq:prediction_of_multipoles}. In order to evaluate the non-local part, an initial guess for the electric potential $v_\text{eff}(\mathbf{r})$ is generated by computing the potential due to only the local contribution to the multipoles. The SC-cycle is then continued from this point and iterated until convergence. A linear mixing scheme is used when updating the charge density which is outlined in section \ref{sec:scf:methods:scfloop_details}.

\subsubsection{Enforcing a Specified Total Charge}
\label{sec:scf:enforcing_total_charge}

As discussed in section \ref{sec:theory:fixedpoint}, the natural way to enforce a total charge constraint in the fixed-point formalism is to introduce the Fermi level as a Lagrange multiplier which is added to the electric potential. There are two choices for how to evaluate the model. If $\mu$ is kept fixed, the SC loop will not necessarily converge to zero total charge, which could be useful for some applications where the grand canonical ensemble for the electrons is required. Alternatively, one can modify the Fermi level during the SC cycle to obtain a chosen total charge. Since the update function $\mathcal{F}$ is quite flexible, there is no a priori guarantee that the total charge will respond correctly to varying the Fermi level. 
In practice, we found in our experiments that when appropriate loss functions and training methods are used, the model does show the correct behaviour. The procedure for updating $\mu$ in principle does not affect the final answer, since the output does not depend on the path taken to converge the SC cycle. We have, however, attempted to implement an efficient scheme to quickly determine the correct Fermi level, which is discussed in section \ref{sec:scf:methods:scfloop_details}.

\subsection{Energy Functional Model}

The energy functional model is defined by equation \eqref{eq:energy_model_energy}:
\begin{align*}
    \mathcal{E} = \min_{\mathbf{p}}& \biggl[E_{\text{local}} + G_{\text{ML}}(\{\mathbf{u}_i\}_i) \nonumber\\
    &+ \frac{1}{2}\iint \frac{\rho(r) \rho(r')}{|r-r'|} drdr' + \int v_\text{app}(r) \rho(r) dr \biggr]
    \tag{from \ref{eq:energy_model_energy}}
\end{align*}
The implementation is outlined in Figure \ref{fig:scf:model_diagram}a. In the model, two normal MACE layers are used to produce local body-ordered contributions to the energy ($E_1 + E_2$) and descriptors of the geometry around each atom. These local features are then combined with the atomic multipoles in the energy functional layer to give the machine learned energy $G_\text{ML}$. 

As seen in the figure, the overall energy model implementation is comparable to the fixed-point model, but the individual elements of the model differ. We have put some effort into optimizing both architectures individually, since it is more useful to make comparisons between the best case scenarios of each approach. As in the fixed-point model, we compare three different architectures for the learned energy functional, which are equivalent to the one-body linear, one-body nonlinear, and many-body update functions in the fixed-point model. 

\subsubsection{Density Embedding}

As in the fixed-point model, a local component of the charge density $\mathbf{p}_{i}^{(\text{local})}$, is first predicted using equation \eqref{eq:prediction_of_multipoles:local}, but modified to use only the second layer local node features. The total charge density $p_{i,lm}$ and the locally predicted density $p_{i,lm}^{(\text{local})}$ are then linearly combined with element-dependent weights:
\begin{equation}
     \tilde{p}_{i,kl_2m_2} = W_{k, z_i}^{(1)} p_{i,l_2m_2} + W_{k, z_i}^{(2)} p_{i, l_2m_2}^{(\text{local})}
     \label{eq:energymodel_rhotilde}
\end{equation}
An equivariant product is then formed between these intermediate descriptors and the local geometric features $\tilde{\mathbf{h}}$:
\begin{equation}
    V_{i,kl_3m_3} = \sum_{\tilde{k}} W_{k\tilde{k}} \sum_{l_1m_1l_2m_2} C^{l_3m_3}_{l_1m_1l_2m_2} \tilde{h}_{i,\tilde{k}l_1m_1} \tilde{p}_{i,\tilde{k}l_2m_2}
    \label{eq:energymodel_density_embedding}
\end{equation}

The resulting descriptors $V_{i,klm}$ describe the geometry and charge density, and are the inputs to the following steps.

\subsubsection{Energy Functional}

The energy functional ($G_\text{ML}$) must predict an atomic energy contribution based on the geometry and embedded charge density around each atom. This term is decomposed into two pieces. We denote by $G_{\text{ML},i}$ the contribution from atom $i$ so that the total learned component of the energy is $\sum_{i=1}^{N_\text{atoms}} G_{ML,i}$. Then,
\begin{equation}
    G_{\text{ML},i} = Q_i + U_i
\end{equation}
The first term $Q_i$ is a learnable positive quadratic function of the charge density and $U_i$ is a flexible functional form. The free parameters in the positive quadratic can make this term zero during training. Nonetheless, it was found that in all cases this term helps to give stable models and better behaved training dynamics. We note that other approaches to making energy functional based models have also utilised quadratic terms like this \cite{scfnn}.

The term $U_i$ can be made more flexible to make accurate models. As in the fixed-point model, we test three different architecture choices for $U_i$, which are conceptually equivalent to those in the fixed-point model:
\begin{itemize}
    \item \textbf{One-Body Quadratic.} $U_i$ is a quadratic function (not restricted to have positive curvature) of the embedded charge density on atom $i$ only, dependent also on the geometric features. This model is formally equivalent to a machine-learned (quadratic) charge equilibration scheme generalized to higher-order atomic multipoles. 
    \item \textbf{One-Body Nonlinear.} $U_i$ is a nonlinear (and significantly more flexible than quadratic) function of the embedded charge density on atom $i$. 
    \item \textbf{Many-Body.} $U_i$ is a nonlinear function of the embedded charge density on atom $i$ and that of neighbouring atoms. 
\end{itemize}

Details of the implementation of the positive quadratic term and each of the above units are given in section \ref{methods:implementation_energy}.

\subsubsection{Evaluating the Energy Functional Model}

The energy functional model is evaluated by minimizing over the input charge density. The initial guess for the minimization is the locally predicted part of the density. The minimization is done using a trust-region newton conjugate gradient minimizer as described in section \ref{sec:scf:methods:scfloop_details}, and is always done with a constraint on the total charge. 

\section{Training}
\label{sec:scf:training}

So far we have introduced two ways of building self-consistent models, and described how each one has been implemented in the MACE architecture. 

Training self-consistent models is not straightforward, since during training one needs to compute the gradient of the model's output with respect to all the model's parameters. Unlike in a normal message passing neural network where the output is often an explicit, analytic function of the input, for a self-consistent model the output is defined implicitly as the solution of an equation (in the case of the fixed-point model), or the minimizer of a function (as in the energy model). 

Take, for example, the energy functional model. To minimize the energy, one needs to run a potentially long iterative optimization process, and stop once some convergence criterion is reached. During training, one needs gradients of the output (energies, forces, charge density coefficients) with respect to parameters. The naive way to train this model is to minimize the energy for a single training example, and compute the gradients for the weights by differentiating through the entire minimization process. This does not seem to be a sensible idea because of the computational expense, and because it requires the model to have a well defined and easily accessible minimum at all points during the training process. In practice, for the initial model (which is created with random weights at the start of training) it might be difficult or impossible to find the minimizing charge density for all training examples. 

In this study we consider three different training methods which can be applied to both model architectures, and compare how the different models behave when trained in different ways. 

\subsection{Direct Training}
\label{sec:scf:direct_training}

In the following, $\rho$ refers to the model's coarse-grained density \eqref{eq:atomic_multipole_expansion}, $\mathbf{p}$ is the vector of atomic multipole moments and $\mathbf{v}$ is the vector of electric potential features. We also introduce $\mathbf{v}[\mathbf{p}]$ to denote the set of electric potential features coming from atomic multipoles $\mathbf{p}$. Explicitly, $\mathbf{v}[\mathbf{p}, \mu]$ is:
\begin{align}
    \mathbf{p} &\rightarrow \rho(\mathbf{r}) \nonumber \\
    v_\text{eff}(\mathbf{r}) &= \int \frac{\rho(\mathbf{r}')}{|\mathbf{r} - \mathbf{r}'|}d\mathbf{r}' + v_\text{app}(\mathbf{r}) + \mu \nonumber \\
    v_\text{eff}(\mathbf{r}) &\rightarrow \mathbf{v} \label{eq:v[p]}
\end{align}
Where the first and last lines refer to equations  \eqref{eq:atomic_multipole_expansion} and \eqref{eq:field_featurization}. 

The objective is to train the function $F_\text{ML}$ in the fixed-point model, or $\mathcal{E}_\text{ML}$ in the energy model:
\begin{align}
    F_\text{ML}(\bm{\theta}, \mathbf{v}) &= \mathbf{p} \label{eq:fixed_point_fun_def} \\  \mathcal{E}_\text{ML}(\bm{\theta}, \mathbf{p}) &= E_\text{total} \label{eq:energy_model_energy_fun_def}
\end{align}
Where $\mathcal{E}_\text{ML}(\bm{\theta}, \mathbf{p}) := G_\text{ML}(\bm{\theta}, \mathbf{p}) + E_\text{local} + E_\text{Coulomb}(\mathbf{p})$ (see also equation \ref{eq:energy_model_energy}). $\bm{\theta}$ denotes the weights of the model. The simplest training method, which is the one taken in the BpopNN, SCFNN and eMLP approaches, is to extract both sides of the equation from DFT and fit the model to reproduce this. For the energy model, ignoring the total charge constraint, this means fitting:
\begin{align}
    L &= \|\mathcal{E}_\text{ML}(\bm{\theta}, \mathbf{p}_\text{DFT}) - E_\text{DFT}\|^2 + \|\nabla_{\mathbf{p}} \mathcal{E}_\text{ML}(\bm{\theta}, \mathbf{p})\|_{\mathbf{p}_\text{DFT}}^2 \label{eq:energy_direct_loss}
\end{align}
This says that \textit{at the DFT charge density} the model should predict the DFT energy, and that the DFT charge density should be a stationary point of the model's energy functional. To do this one has to extract atomic multipoles from DFT to make the vector $\mathbf{p}_{DFT}$. The second term is only valid if the charge density is indeed from a ground state DFT calculation. If non-ground state calculations are performed via constrained DFT, to better sample the landscape of $\mathcal{E}(\mathbf{p})$, then one cannot use this term. This was done in the BpopNN model. In practice, the model and DFT are valid only with a constraint on the total charge, so one must modify the second term in \eqref{eq:energy_direct_loss} to read $\|\nabla_{\mathbf{p}} \mathcal{E}_\text{ML}-\lambda \nabla_{\mathbf{p}} Q_{\mathrm{tot}}\|_{\mathbf{p}_\text{DFT}}^2$.

For the fixed-point model, the analogous loss is:
\begin{align}
    L &= \|F_\text{ML}(\bm{\theta}, \mathbf{v}_\text{DFT}) - \mathbf{p}_\text{DFT}\|^2
    \label{eq:fixedpoint_direct_loss_equation}
\end{align}
In other words, the charge density predicted from the DFT effective potential should equal the DFT charge density. In practice, while one could use the DFT effective potential, in this work we use the potential reconstructed from the (coarse-grained) DFT charge density:
\begin{align}
    \mathbf{v}_\text{DFT} := \mathbf{v}[\mathbf{p}_\text{DFT}] = \int \frac{\rho_\text{DFT} (\mathbf{r'})}{|\mathbf{r}-\mathbf{r'}|} \text{d} \mathbf{r}' + v_\text{app} + \mu_\text{DFT}
    \label{eq:vp_DFT}
\end{align}
where $\rho_\text{DFT}$ is computed from equation \eqref{eq:atomic_multipole_expansion} using the atomic multipoles extracted from DFT. Hence, the loss becomes:
\begin{align}
    L &= \|F_\text{ML}(\bm{\theta}, \mathbf{v}[\mathbf{p}_{DFT}]) - \mathbf{p}_\text{DFT}\|^2
    \label{eq:fixed_point_direct_loss}
\end{align}
One SC step of the fixed-point model can be written:
\begin{align}
    \mathbf{p}^{(n+1)} = F_\text{ML}(\bm{\theta}, \mathbf{v}[\mathbf{p}^{(n)}])
    \label{eq:fixed_point_scf_step}
\end{align}
Therefore, the loss \eqref{eq:fixed_point_direct_loss} says that the DFT density is a solution of the model's SC loop. 

Equations \eqref{eq:energy_direct_loss} and \eqref{eq:fixed_point_direct_loss} define `direct training' in this paper, and are equivalent to one another: If an energy model can be transformed into a fixed-point model, zero loss in equation \eqref{eq:energy_direct_loss} translates into zero loss in \eqref{eq:fixed_point_direct_loss}.

\subsection{Implicit Differentiation}
\label{sec:scf:implicit_diff}

The problem with direct training is that the model is not evaluated during training in the same way it will be used during inference. In training, the model only ever sees the DFT charge density (or the corresponding potential), but during inference the model must start from some guess and converge to its own solution. This causes several problems: 
\begin{enumerate}
\item \textbf{Stability}. Nothing in the above loss functions \eqref{eq:energy_direct_loss} and \eqref{eq:fixed_point_direct_loss} guarantees that the DFT density is a \textit{minimum} of the energy (or a stable fixed point). It can therefore be that during inference, the model converges to a nonsensical solution, or does not converge at all. 
\item \textbf{Compounding Errors}. When a (stable) model is evaluated, it will converge to a solution which is slightly different from the DFT solution. This means that other properties such as the energy and forces will have two sources of error. One is the actual fitting error from the DFT charge density to the energy (equation \eqref{eq:energy_model_energy_fun_def}). The second is because the input charge density is now the model's solution, not the DFT solution. Unless the model solution is very close to the DFT solution, this could lead to large errors during inference.
\item \textbf{Need for ab initio Charge Density}. In direct training, the reference (i.e. DFT) charge density must be partitioned to create atomic multipole moments or other descriptors in order to train the model. This is unavoidable, because the DFT charge density is an \textit{input} to the model during training (see equations \eqref{eq:energy_direct_loss} and \eqref{eq:fixed_point_direct_loss}). If one does not have a set of DFT descriptors of the charge density, direct training is simply not possible.
\item \textbf{No Cancellation of Errors}. Machine learning makes use of short cuts. MLIPs do not compute the full kinetic and Hartree energies of a molecule, since much of the variations in these terms cancel out. By forcing the model to be trained at the DFT solution, the model is unable to learn shortcuts or simplifications. For example, consider a bulk metal oxide with some point defects. Representing all the atoms in the bulk material as having zero charge, and the defects as haveing some other (non-zero) charge, might be a good way to handle the long-range electrostatic part of the energy since the bulk region is homogenous. By training with the real DFT partitioned charges, we force the model to use the actual charge density, rather than determining its own convenient description of the system's charge density.
\end{enumerate}

The solution to some of these issues is to train the model \textit{at the model's solution}, rather than at the DFT solution. This can be achieved by solving the model (converging the SC-cycle) during training, and then computing the gradients of the outputs (charge density 
$\mathbf{p}$) with respect to weights $\bm{\theta}$ but with the constraint that the model stays solved.

One can illustrate how this works with a fixed-point model. Define $\bm{f}$, a vector valued function, as follows:
\begin{align}
    \bm{f}(\bm{\theta}, \mathbf{p}) = F_\text{ML}(\bm{\theta}, \mathbf{v}[\mathbf{p}]) - \mathbf{p}
    \label{eq:define_f}
\end{align}
Once the SC loop is converged, we have
\begin{align}
    \bm{f}(\bm{\theta}, \mathbf{p}^\star(\bm{\theta})) = \mathbf{0}
    \label{eq:fixedpoint_constant_mu_eq}
\end{align}
The optimum density $\mathbf{p}^\star$ is a function of the weights since if the weights are changed, the solution to the SC-cycle will move. 

During training we need to compute the loss. For instance, if one is training on the charge density coefficients, one term in the loss is:
\begin{align}
    L = \lambda \| \mathbf{p}_\text{DFT} - \mathbf{p}^\star \|^2
\end{align}
After computing the loss, we need to compute the gradient of loss with respect to parameters, which in this case is:
\begin{align}
    \frac{\partial L}{\partial \bm{\theta}} = \frac{\partial L}{\partial \mathbf{p}^\star} \frac{\partial \mathbf{p}^\star}{\partial \bm{\theta}}
\end{align}
The first piece, $\partial L / \partial \mathbf{p}^\star$ is easy to compute, but the Jacobian matrix $\partial \mathbf{p}^\star /\partial \bm{\theta}$ is not, because $\mathbf{p}^\star$ is defined from $\bm{\theta}$ implicitly via equation \eqref{eq:fixedpoint_constant_mu_eq}. To compute this Jacobian, assume that during training the model has been converged for a given training example. Because for any $\bm{\theta}$ we define $\mathbf{p}^\star$ by \eqref{eq:fixedpoint_constant_mu_eq}, the total derivative $d \bm{f} / d \bm{\theta}$ is zero, so we can write:
\begin{align}
    \frac{d\bm{f}}{d \bm{\theta}} = \frac{\partial \bm{f}}{\partial \mathbf{p}^\star} \frac{\partial \mathbf{p}^\star}{\partial \bm{\theta}} + \frac{\partial \bm{f}}{\partial \bm{\theta}} = \mathbf{0} 
    \label{eq:implicit_derivative_1}
\end{align}
and rearrange,
\begin{align}
    \frac{\partial \mathbf{p}^\star}{\partial \bm{\theta}} = - \left(\frac{\partial \bm{f}}{\partial \mathbf{p}^\star}\right)^{-1} \frac{\partial \bm{f}}{\partial \bm{\theta}}.
    \label{eq:implicit}
\end{align}
Which gives an expression for the derivative we want in terms of things that can be calculated using auto-differentiation routines, because they are partial derivatives of $\bm{f}$ with respect to $\bm{\theta}$ and $\mathbf{p}$, which is an explicit, closed form operation. This is called implicit differentiation and has been used in other machine learning fields \cite{Bai2019DeepModels}. The equations above are uncoupled in $\bm{\theta}$ and $\partial \bm{f} / \partial \mathbf{p^\star}$ is $M\times M$ where $M$ is the number of density coefficients---the size of $\mathbf{p}$. The caveat is that calculating this derivative requires solving a system of linear equations, which scales cubically in the size of the vector $\mathbf{p}$. If the number of density coefficients is very small, solving the linear problem should not be expensive, although practical implementations differ in how the linear solve is actually performed.

In summary, training with implicit differentiation requires the following: For each training example, converge the model to its solution $\mathbf{p}^\star$, without keeping any information about the path taken during the self-consistent iterations. At the model solution, compute the Jacobian matrices on the right hand side of \eqref{eq:implicit}, and solve the linear problem to find the gradients of the charge density with respect to upstream parameters. This gradient matrix then enters into the computational graph of the auto-differentiation engine. The process can be done without any charge density information from the DFT, so long as one can converge the model. Throughout this paper we use the {\sc torchopt} package \cite{Ren2023TorchOpt:Optimization} to train using implicit differentiation, which performs all the above derivatives and matrix algebra behind the scenes. 

An exactly analogous procedure can be done for the energy functional architecture. Implementation details of implicit differentiation for both models, as well as the required extensions to handle a constraint on total charge, are presented in section \ref{sec:methods:implicit_diff}. 

For a variational energy functional model, the gradient of the energy and force predictions, with respect to model parameters, can be computed by the Hellmann-Feynman theorem without the need for implicit differentiation. However, in the results of section \ref{sec:results:metal_water} we explicitly fit to features of the charge density such as the total dipole moment, and therefore need to use implicit differentiation.

\subsection{Unrolling the SC loop}
\label{sec:scf:diff_the_scf}

Finally, we have also tested the naive method of running a fixed number of SC steps during training, and computing gradients by differentiating through the SC loop using back propagation. This can be thought of as `unrolling the SC loop'. The implementation of this scheme is very simple: one simply runs the self-consistency loop or minimization algorithm during training, and treats the entire process like a single, deep neural network. The gradients are then computed by backpropagation through the whole SC-process. 

\begin{figure}
    \centering
    \includegraphics[width=1.0\linewidth]{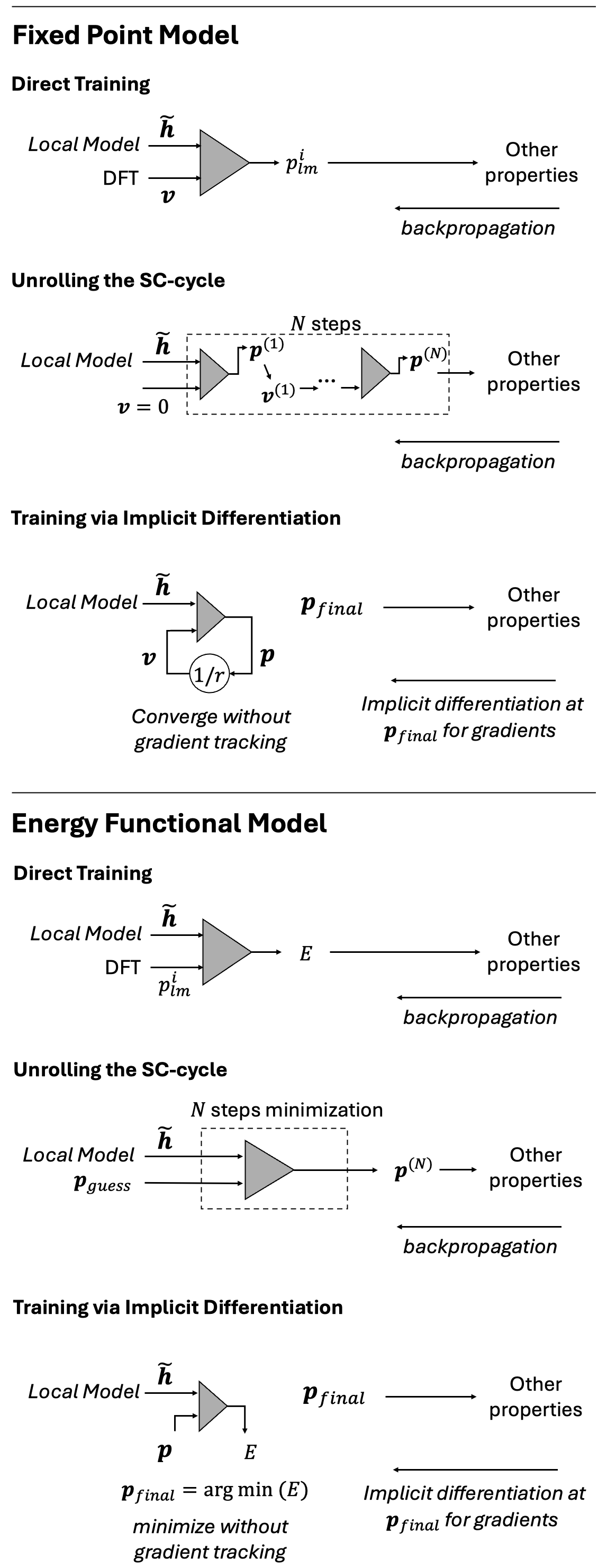}
    \caption{Training Methods for our self-consistent models.}
    \label{fig:training_methods_cartoon}
\end{figure}

With this training method, the way that SC loop is converged affects the training and this will be explored below. Figure \ref{fig:training_methods_cartoon} illustrates the three different training methods for both models. 

\section{Baselines and Comparisons}
\label{sec:scf:baseline_models}

We will compare our models to three other, simpler approaches for building MLIPs with charge information.

\subsection{Global Charge Embedding}

In the global embedding model, the total charge $Q$ is encoded as a one-hot vector. This one-hot vector is then passed into a one layer perceptron with a SiLU non-linear activation to form a learnable embedding feature vector for each total charge value of the same dimension as the initial node features of MACE,
\begin{align}
    e_{k} = \text{SiLU}(\sum_{\tilde{Q}} W_{k,\tilde{Q}} \delta_{\tilde{Q}, Q})
\end{align}
where $e_{k}$ is the embedding vector of the total charge and $\delta_{\tilde{Q}, Q}$ is the delta function selecting the row $Q$ of the weight matrix $W_{k,\tilde{Q}}$.
That total charge embedding is then summed to the initial node feature to form the starting node feature of the MACE model.
\begin{align}
    h^{(0)}_{i, k00} = h^{(0)}_{i, k00} + e_{k}
\end{align}
This approach has been used in many popular MLIPs including MACE-OMOL~\cite{levine2025openmolecules2025omol25}, UMA \cite{UMA}, OrbMol \cite{rhodes2025orbv3atomisticsimulationscale} and LES \cite{kim2025longrangeelectrostaticsmachinelearning}. However, it suffers from many intrinsic limitations, the most severe being that it can not extrapolate to charge states beyond those in the training set of the model.

\subsection{MACE-LocalCharges}
\label{sec:scf:local_charges}

A natural approach is to building an electrostatic MLIP, is to simply predict the charge density coefficients $\mathbf{p}$ based on the local environments of each atom. This was implemented in a model ``MACE-LocalCharges'', wherein multipole moments on each atom are computed from the node features after layers 1 and 2:
\begin{align}
    p_{i,lm} &= \sum_{k} W_{lk} h^{(1)}_{i,klm} + \sum_{k} W'_{lk} h^{(2)}_{i,klm}
    \label{eq:mace_localcharges}
\end{align}
This is exactly the same as the purely local part of the density prediction in both the energy functional and fixed-point models. These multipole moments are used in the same Gaussian type orbital basis as defined in section \ref{sec:scf:rep_density} and the Coulomb energy is computed in the same way as the fixed-point and energy functional models. The energy and forces are then derived from the sum of the MACE local energy and the Coulomb energy.

MACE-LocalCharges is almost identical to the MACE-LES architecture \cite{Cheng2025LatentInteractions}, except for the additional of higher-order atomic multipoles and that in MACE-LES the atomic charges are a nonlinear rather than linear functions of the node features.

\subsection{MACE-QEq}
\label{sec:scf:mace-qeq}

Finally, we compare to a MACE model augmented with the charge density from a classical charge equilibration scheme \cite{mace_qeq_arxiv}. Specifically, we consider a model where the total energy is 
\begin{align*}
    E(\{q_i\}_i) = E_\text{local} + \sum_i (\chi_i q_i + \frac{1}{2}\eta_iq_i^2) + E_\text{Coulomb}(\{q_i\}_i),
\end{align*}
and where the charges are the minimizers of this (the model is therefore variational). The local energy term is the output of a normal MACE model. Both the electronegativities and hardnesses are learnable functions of the local geometry. Similar to \eqref{eq:mace_localcharges}, these terms are predicted directly from the node features:
\begin{align*}
    \chi_i &= \sum_k W_k h_{i,k00}^{(1)} + \text{MLP}(h_{i,k00}^{(2)}) \\
    \eta_i &= \sum_k W'_k h_{i,k00}^{(1)} + \text{MLP}(h_{i,k00}^{(2)}) \\
\end{align*}
Where $W$ and $W'$ are weight matrices and `MLP' refers to a 1 layer multi-layer perceptron with 128 hidden neurons. Since both $\chi$ and $\eta$ are invariant, only the invariant ($l=m=0$) parts of the node features are needed.

Again, the partial charges are assigned to $p_{i,00}$ in section \ref{sec:scf:rep_density} and the Coulomb energy is computed in the same way as the other models. The electrostatics is thus consistent across the baselines and the new models.

MACE-QEq is designed to be representative of numerous models in literature which use a classical equilibration scheme to handle electrostatic effects.

\section{Results: Insulating and Conducting Behaviour in Slabs}
\label{sec:results:metal_water}

\subsection{Dataset and Model Hyperparameters}
\label{sec:scf:dataset}

\begin{figure}[t]
\centering
\includegraphics[width=0.95\columnwidth]{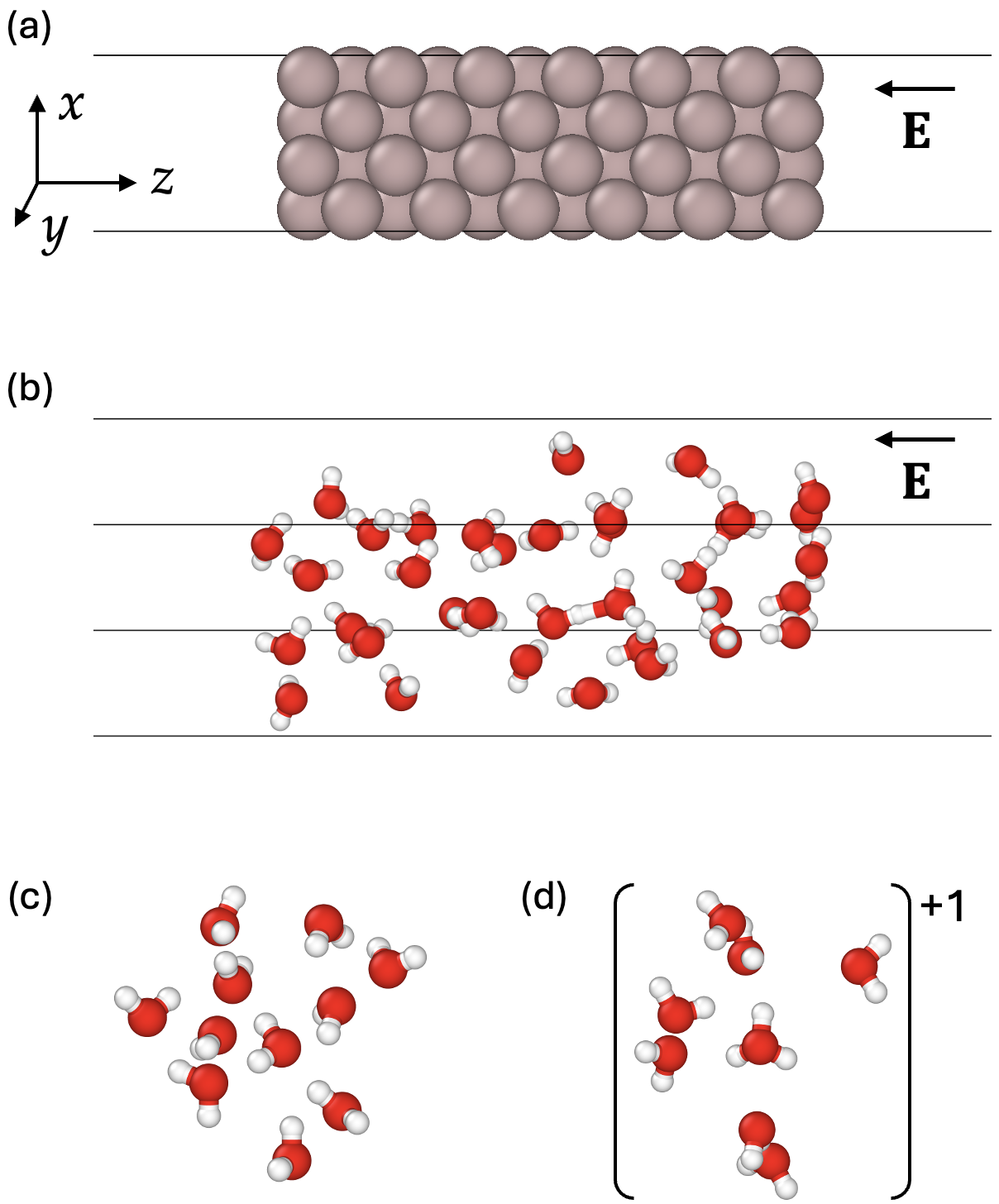}
\caption{
Representative configurations from the metal--water dataset.
(a) Aluminium slab where $z$ is the out-of-plane direction, with an applied field.
(b) Liquid water slab also with an applied field.
(c) Water cluster.
(d) Water cluster containing a hydronium ion, with a total charge of $+1$.
}
\label{fig:scf:datset_summary}
\end{figure}

A simple dataset has been developed to test some key behaviours of electrostatic and self-consistent models. It includes aluminium slabs with either (111) or (100) surfaces, slabs of liquid water and clusters of up to 12 water molecules. Half of the water clusters contain a hydronium ion and are positively charged. Some of the water and aluminium slabs have an applied electric field in the out of plane direction, with a magnitude randomly chosen between -0.1 and 0.1 V/\AA. A summary of the dataset is given in Table \ref{table:dataset_summary} and Figure \ref{fig:scf:datset_summary}.

All density functional theory calculations were done using the FHI-aims DFT code, with the PBE exchange-correlation functional \cite{PBE} and `tight' basis set and integration grids. Further details can be found in section \ref{sec:scf:methods:dft_settings}. For all calculations, atomic multipole moments were extracted from the DFT data. The atomic multipoles correspond to internal quantities used in FHI-aims to represent the Hartree potential, and are conceptually similar to Hirshfeld partitioned moments as discussed in section \ref{sec:scf:methods:aims_multipoles}.

\subsubsection{Model Hyperparameters}
\label{sec:scf:model_and_loss_hypers}

All of the models presented here are built on the MACE framework. The hyperparameters of the local component of all the MACE models are given in Table \ref{table:scf:model_hypers}. These hyperparameters correspond to `medium sized' MACE models, with roughly one million parameters. In the energy functional and fixed-point models, the latent features in the electrostatics layers are always the same size as those in the local part of the model. 

In all the fixed-point model results presented below, unless otherwise stated, the number of electric potential features (the range of $N$ and $L$ in \eqref{eq:field_featurization}) was kept constant. We used 2 radial channels and $L=0,1$. The supplementary information presents further details and experiments showing how this choice affects the performance.

\subsubsection{Loss Functions and Training Schedules}

Unless otherwise stated, all fits used the following loss function to train the models:
\begin{align}
    L = \ & w_E \left(\frac{E - \tilde{E}}{N_\text{atoms}}\right)^2 \nonumber \\
    & + w_{D} \sum_{\alpha=x,y,z} \left(\frac{P_{\alpha} - \tilde{P}_{\alpha}}{N_\text{atoms}}\right)^2  \nonumber \\
    & + \frac{w_F}{3N_\text{atoms}}\sum_i^{N_\text{atoms}}\sum_{\alpha=x,y,z} (F_{i,\alpha} - \tilde{F}_{i,\alpha})^2  \nonumber \\
    & + \frac{w_{\rho}}{N_\text{moments}} \sum_{ilm} (p_{i,lm} - \tilde{p}_{i,lm})^2
    \label{eq:scf:loss}
\end{align}
The components of this loss are total energy per atom, total dipole per atom, forces and finally atomic multipole moments. Symbols with a tilde represent DFT reference values, whereas those without represent the prediction of the model. When computing the dipole ($P$) loss, we only take into account the dipole in the out-of-plane direction for all slab geometries. For the molecules, all components of the dipole enter the loss. For the fixed-point model in direct training mode, the models' prediction of the total charge is only constrained by the fact that the real (DFT) atomic multipole moments, in the loss, sum up to the right total charge (see equation \eqref{eq:fixedpoint_direct_loss_equation}). We found that it is always helpful to include a total charge error in the loss, of the form:
\begin{align*}
    w_Q  \left(\frac{Q_\text{total} - \tilde{Q}_\text{total}}{N_\text{atoms}}\right)^2
\end{align*}
Since the model is evaluated by iterating to self-consistency, the total charge during inference is always correct regardless of the choice of loss function. The loss weights for this test system ($w_E$, $w_D$, ...), unless otherwise stated, are given in Table \ref{table:scf:loss_weights}. The relative size of the total dipole and total charge weight reflects the fact that the total dipole per atom is generally much smaller than the other properties in our unit system of (eV,e,\AA). All the fits in this paper use the loss in equation \eqref{eq:scf:loss}, but we note that using an extensive total charge and total dipole loss (not per atom) may be a better choice. This would also make the loss weights for these terms closer to 1 in magnitude.

Another special case is direct training of the energy functional model. As can be seen in equation \ref{eq:energy_direct_loss}, it is natural to include in the loss the gradient of the energy with respect to the charge density coefficients, $\|\nabla_{\mathbf{p}} \mathcal{E}_\text{ML}(\bm{\theta}, \mathbf{p})\|_{\mathbf{p}_\text{DFT}}^2$. This was also used in the some of the fits with a weight of 1,000, but we will show that direct training is impractical for this model. 

One must be careful when training on both the total dipole and partial atomic multipoles, since it is possible that these properties are inconsistent. The DFT atomic multipoles in this dataset are consistent with the total dipole so long as both partial charges and partial dipoles are included. For this reason, we reduce the atomic multipole component of the loss to $1$ for any models which do not include partial dipoles. Relevant details of the training method and number of epochs in each stage will be mentioned next to the individual results, and complete tables of the loss weights and training schedule for all fits in this paper can be found in the supplementary information.

All models used the schedule-free optimizer for training \cite{Defazio2024TheScheduled}.

\begin{table}[t]
\caption{Water--metal dataset composition.}
\label{table:dataset_summary}
\centering
\begin{tabular}{@{}l r@{}}
    \toprule
    \textbf{Configuration} & \textbf{Number} \\
    \midrule
    Aluminium slab with field & 308 \\
    Water slab with field     & 438 \\
    Water slab                & 644 \\
    Water cluster (Hydronium) & 600 \\
    Water cluster (Neutral)   & 600 \\
    \midrule
    \textbf{Total}            & \textbf{2590} \\
    \bottomrule
\end{tabular}
\end{table}

\begin{table}[t]
\caption{Hyperparameters for the local component of all models.}
\label{table:scf:model_hypers}
\centering
\begin{tabular}{@{}l c@{}}
    \toprule
    \textbf{Hyperparameter} & \textbf{Value} \\
    \midrule
    Number of chemical channels & 128 \\
    Maximum equivariance order $L$ & 1 \\
    Single layer cutoff radius (\AA) & 6 \\
    \bottomrule
\end{tabular}
\end{table}

\begin{table}[t]
\caption{Loss weights for all water--metal fits.}
\label{table:scf:loss_weights}
\centering
\begin{tabular}{@{}l r@{}}
    \toprule
    \textbf{Property} & \textbf{Loss weight $w$} \\
    \midrule
    Energy                     & 100 \\
    Force                      & 100 \\
    Atomic multipoles ($l=1$)  & 100 \\
    Total dipole               & 1\,000\,000 \\
    Total charge               & 10\,000 \\
    \bottomrule
\end{tabular}
\end{table}
\subsection{Training Dynamics}
\label{sec:scf:res:training_dynamics}

\subsubsection{Energy Functional}


\begin{figure*}
    \centering
    \includegraphics[width=\linewidth]{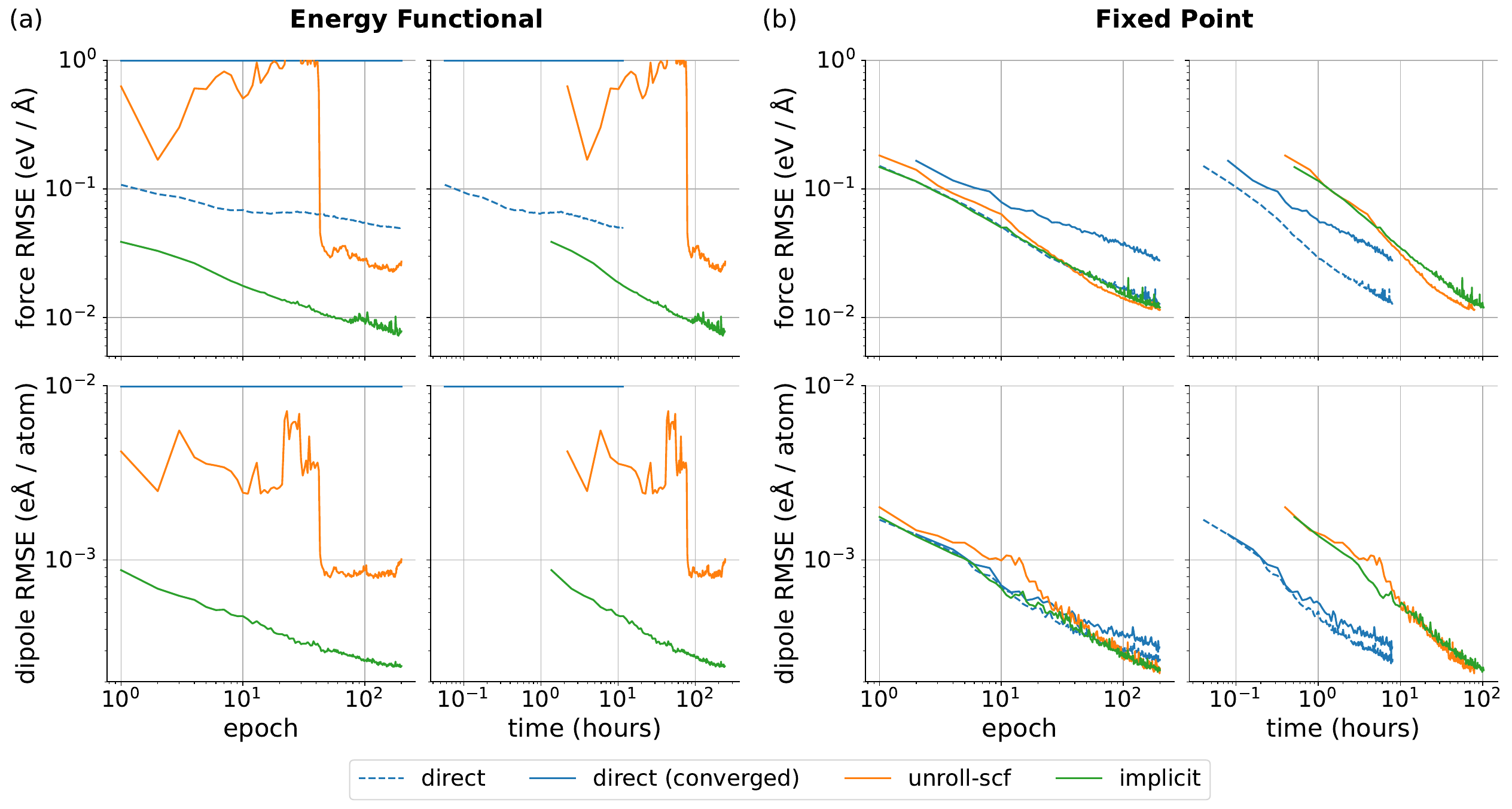}
    \caption{Training characteristics of the energy functional (a) and fixed-point models (b) for the metal--water slab example. The difference between direct and direct (converged) is explained in the text.}
    \label{fig:scf:model_training_dynamics}
\end{figure*}

We first investigate how each model architecture interacts with the different training methods outlined in section \ref{sec:scf:training}. For the energy functional approach, using a one-body quadratic functional layer, the training dynamics for different training methods are shown in Figure \ref{fig:scf:model_training_dynamics}a.

In the direct training method, the model is trained at the DFT set of atomic multipoles. This means that the loss \eqref{eq:scf:loss} contains just the energy and force terms, as well as the gradient term $\|\nabla_\mathbf{p}\mathcal{E}(\mathbf{p})\|^2$. After training in direct mode, if one wants to use the model, the model needs to be converged to its own minimum energy, which will not in general be the DFT solution. There are therefore two `direct' lines on the graphs in Figure \ref{fig:scf:model_training_dynamics}a: The line labelled `direct' is the error of the model when evaluated with the DFT atomic multipoles, whereas `direct (converged)' is the error if one instead tries to minimize the model to its own solution for all configurations in the validation set. During direct training the atomic multipoles are inputs, not outputs, of the model and hence there is no recorded dipole loss for direct training in panel (a).

In this case, we found that the simple direct method of training \textit{never} resulted in a model which behaved properly for all validation set configurations. The error is therefore shown as a line across the top if the plot indicating a failure to converge. Other studies of energy functional based models have always included some amount of bias to enhance the model's stability \cite{scfnn, eMLP}. 

By contrast, training using implicit differentiation is extremely well behaved. The model remains stable at all points and both metrics steadily improve. One can note, however, the difference in time to train. For the same number of epochs, the direct training mode takes just over 10 hours, whereas the implicit training takes 9 days. Training by unrolling the self-consistency loop is not stable in this example; we have observed that differentiating through a minimization process can be made to work, but was unreliable.

The conclusion is that for this implementation of an energy functional model with atomic multipoles describing the charge density, implicit differentiation is greatly superior when measured in improvement per epoch, but is extremely expensive. In this case, implicit differentiation also gives a model which is well behaved, in that the energy minimization is always successful, without any need for data augmentation, constrained DFT, or classical baselines. 

\subsubsection{Fixed-Point}


A similar experiment performed for the fixed-point architecture with a linear update layer is shown in Figure \ref{fig:scf:model_training_dynamics}b.

Some differences in methodology separate these result from those of the energy model. In order to evaluate the fixed-point model at a specified total charge, one must adjust the Fermi level during the SC loop to achieve the right total charge (section \ref{sec:scf:enforcing_total_charge}). In direct training mode, this concept does not apply, since the model is only making one prediction of the charge density, based on the DFT electric potential and DFT Fermi level (equation \eqref{eq:fixed_point_direct_loss}). 

In the other training modes where the model is converged to its solution, one can naturally choose to evaluate the model in two ways: Firstly, at constant charge, by optimizing the Fermi level during the SC loop to achieve the right total charge, and secondly in `constant potential mode' where the Fermi level is kept fixed at the DFT value. We found the latter to be more well behaved with untrained, randomly initialized, models and therefore the results in Figure \ref{fig:scf:model_training_dynamics}b correspond to the constant potential approach. 

One can see immediately that for the fixed-point model on this dataset, the behaviour of all training methods is quite similar. Most importantly, the direct method works well. The errors reduce in a predictable manner, and the model is always stable when it is re-evaluated to convergence on the validation set. One can see that when re-evaluating, the force errors increase as one might expect. The dipole error on the other hand is almost unchanged.

Both implicit differentiation and unrolling of the SC loop also perform similarly on a per-epoch (per gradient-update) basis. Again, when measured per unit time, despite the fact that the gradient of learning for direct is slightly lower, direct training still seems to more efficient for learning the total dipole. For learning the atomic forces, it is more balanced. 

When unrolling the SCF loop, the behaviour of differentiation through the SC cycle depends on what scheme is used to converge the SC cycle. As discussed in section \ref{sec:scf:methods:scfloop_details}, a simple linear update is used for the SC loop in the fixed-point models. For the result in Figure \ref{fig:scf:model_training_dynamics}, the mixing parameter was 0.2, and a cap on the maximum number of SCF steps was also imposed at 40 steps.

\subsubsection{Fast Training of fixed-point models}
\label{sec:shortcut_scf}

\begin{figure}
    \centering
    \includegraphics[width=\linewidth]{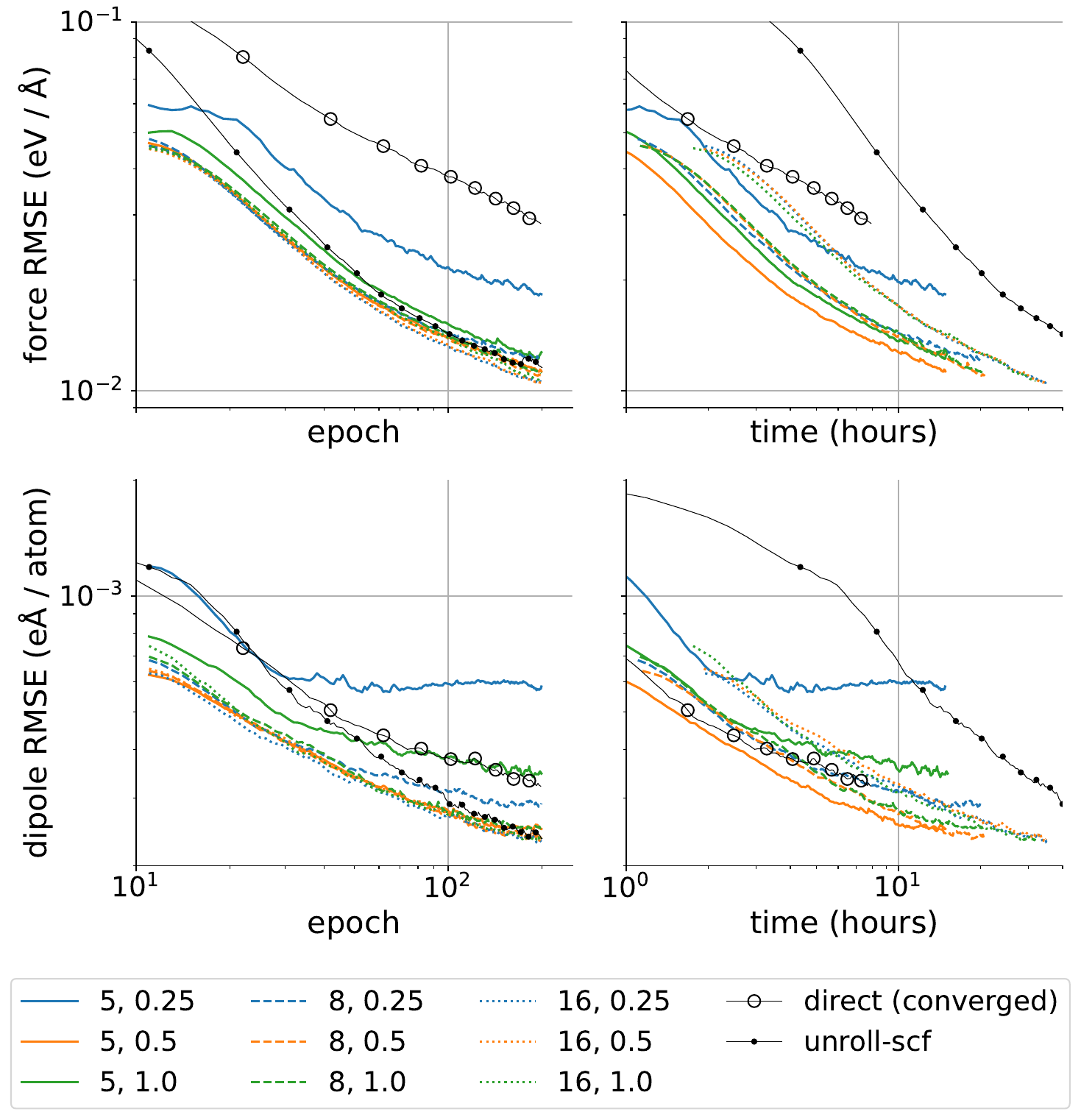}
    \caption{Training a fixed-point model with a truncated SC-cycle. 9 different runs are shown with different choices of the number of SC-steps to perform and the mixing parameter. The legend shows this as ``number-of-steps, mixing-parameter''.}
    \label{fig:scf:shortcut_scf_ema}
\end{figure}

We note one efficient way to train the fixed-point model, which greatly increases the practicality of training self-consistent MLIPs. 

The fixed-point SC-cycle is a series where one expects the differences between terms to rapidly become small. It is therefore natural to try truncating the SC cycle after a fixed (and small) number of steps, and treat this model like a message passing neural network. For some systems, this might give an accurate model which is not truly a self-consistent model but which is close to one in parameter space. Then one can converge the model and the error may not increase as much as in direct training, or one can finish the training process by switching to implicit differentiation or increasing the number of steps so that the model fully converges during training.

We find that for appropriate settings this is an efficient way to train. Figure \ref{fig:scf:shortcut_scf_ema} shows how different variants of this scheme perform. Specifically, we tested truncating the SC-loop after 5, 8 and 16 steps, and with mixing parameters of 0.2, 0.5 and 1.0. Figure \ref{fig:scf:shortcut_scf_ema} presents the error throughout training of the models when re-evaluated by iterating to self-consistency. This means that even though only a few steps are using during training, we treat the resulting models as SC models and converge the SC loop. The lines in the figure have been smoothed with an exponential moving average filter to make the plot readable. A larger version without the smoothing is presented in the supplementary information.

In all cases, the initial guess for the charge density is the local part of the model's charge density prediction (see  section \ref{sec:scf:fixedpoint_update_layer_details}). In these experiments the Fermi level of the model is updated during the SC cycle to reach the right total charge. The initial guess for this Fermi level during training is the DFT Fermi level, and at inference the initial guess is zero. 

One can see that by truncating the SC loop, it is possible to interpolate between two limits: direct training and unrolling a fully converged SC-cycle, which may take many tens of steps. The shortcut-SCF training is clearly a large improvement over direct training, while still being faster than implicit differentiation or differentiating through a fully converged SC-cycle. As one can see in the plot, most choices for the number of SC steps and mixing parameters behave well, but the performance with only 5 steps is very variable. This suggests that when training with only 5 steps, the model does not tend to learn something close to a real fixed-point function. 

In the remaining sections of this paper, this training method is used for some results. When doing this, we adopted 8 SC steps and a mixing parameter of 0.5 for most of the training process, and then switch to either implicit differentiation or a fully unrolled SC-loop, for the final part of training. This means that the final model is a truly self-consistent model.

\subsection{Comparing Charge-Aware Models}
\label{sec:scf:res:comparing_charge_models}

\begin{figure*}
    \centering
    \includegraphics[width=0.75\linewidth]{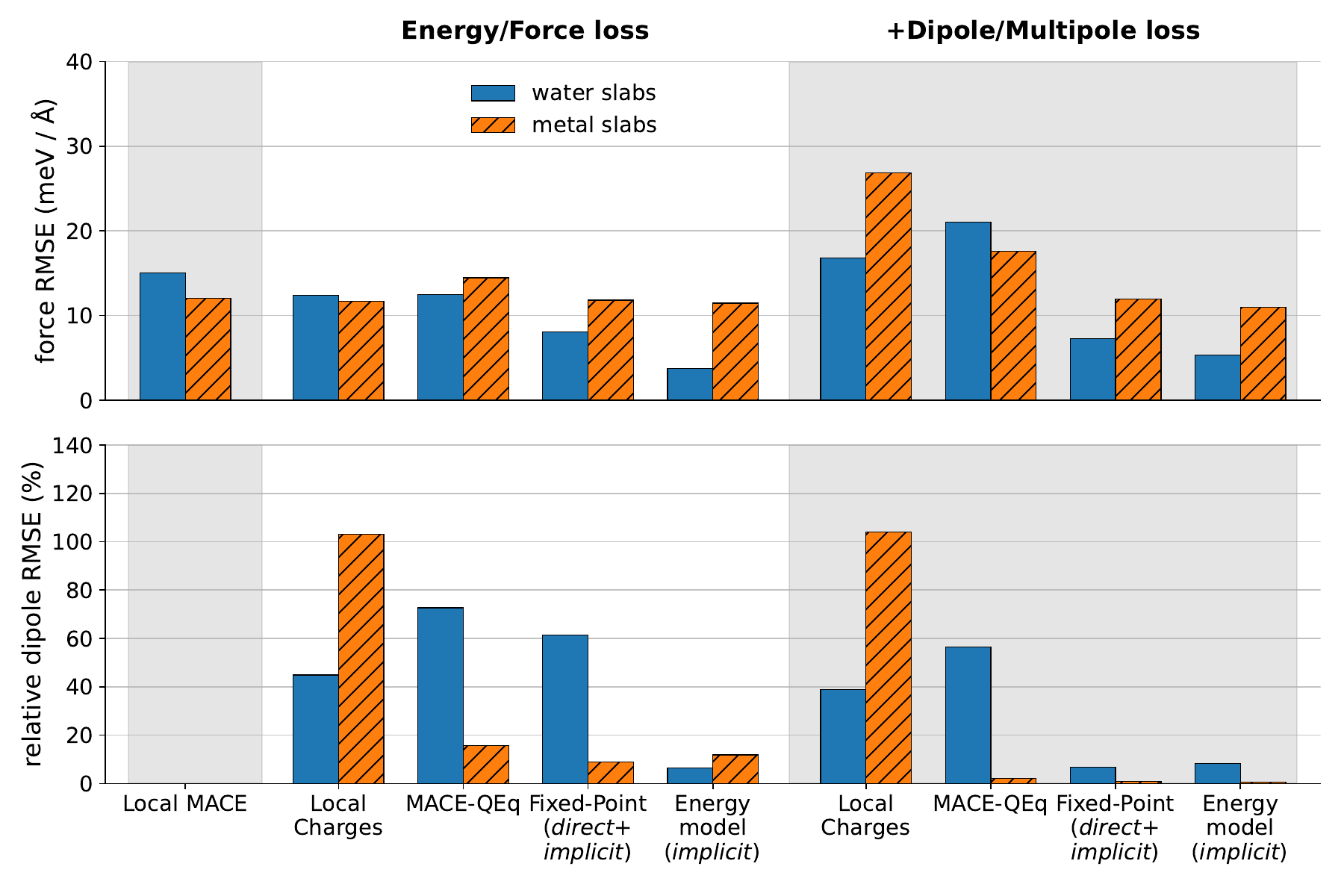}
    \caption{Comparing the performance of charge-aware models, with and without including multipoles in the loss.}
    \label{fig:scf:effect_of_scf_and_loss}
\end{figure*}

The dataset introduced in section \ref{sec:scf:dataset} exhibits varied physics which probes the limits of different architectures. This is summarised in Figure \ref{fig:scf:effect_of_scf_and_loss} in which the baseline models and the newly introduced self-consistent MACE models are compared. We show the test error in total dipole and atomic force. Furthermore, all models were trained both with the combined loss in equation \eqref{eq:scf:loss} as well as simple loss of only energy per atom and force (with the same weights as in Table \ref{table:scf:loss_weights}). For the fixed-point and energy functional models, the term in parentheses is the training method: The energy model was trained entirely with implicit differentiation for 200 epochs, while the fixed-point model was trained with 200 epochs of direct training, followed by 30 epochs of implicit differentiation. The supplementary information contains full details of the training schedules for all fits. The fixed-point model in this figure uses the one-body linear update function, and the energy functional model uses a one-body quadratic functional. 

In this figure, all models use the same electrostatics implementation with atomic charges and dipoles, with the exception of the local MACE model (extreme left of Figure \ref{fig:scf:effect_of_scf_and_loss}) which has no charge density or Coulomb energy terms, and MACE-QEq which has only atomic charges.

\subsubsection*{Training on energy and forces only}

We first train all models, both baselines and self-consistent models, on just the total energy and atomic forces. In the materials MLIP community, the vast majority of research is done using short ranged models with no electrostatics. This is partly because it is difficult to find examples where modern short range MLIPs are unable to accurately regress the energy and forces. Most researchers therefore conclude that short range models are often good enough. Here, at first sight it seems that this is the case: The force errors of the MACE-LocalCharges model and the MACE-QEq models barely improve upon the errors of a normal, local MACE model. There is, however, improvement to be made when using a more flexible, self-consistent model. The fixed-point model with a linear update function and the quadratic energy functional model improve significantly over the other models in force error. 

One can look at the dipole error of these models as well, even though they are not trained on any charge related quantity. In this case the MACE-LocalCharges model explains about 50\% of the variation in dipole of the water slabs, but makes no progress on the metal slabs. This is expected since the metal dipole only depends on the external field applied, and this model has no knowledge of the field. For the MACE-QEq model the situation is reversed, since its structure is ideal for modelling the charge sloshing in metal. The fixed-point model has similar dipole predictions, whereas the energy functional achieves a much lower dipole error than the other models, which will be discussed below.

\subsubsection*{Training on All Properties}

The right-hand panel of Figure \ref{fig:scf:effect_of_scf_and_loss} shows the same results when training on all properties. Note that as discussed above, the atomic multipole component of the loss is not included when training MACE-QEq, since this would be inconsistent with the total dipole. When the dipole and atomic multipoles are added to the loss, both baseline models have worse force predictions. For the MACE-LocalCharges model, this does not come with an improvement in the total dipole. The remaining 40\% error is the best error that can be obtained with this architecture for this geometry, even when atomic dipoles and charges are used, and when the loss is dominated by the dipole error. 
This error represents the induced component of the dipole from the external fields and internal depolarizing fields (which are significant in a slab geometry). 

The self-consistent models introduced here perform significantly better. Firstly, both architectures do not loose accuracy on the forces when the dipole is added to the loss. Secondly, with a well conditioned self-consistent model, it is possible to explain almost all of the variation in total dipole for both water and metal slab subsets. 

The fact that the energy functional and fixed-point methods achieve the same error when trained on all properties, but perform very differently when trained on only energies and forces, may be because only the former is variational. The force and dipole predictions of the energy functional model are therefore more restrictively linked, which in this case allows for greater information transfer between the observables.

\subsection{Energy Functional and Fixed Point Approaches can be Equally Accurate}
\label{sec:scf:res:energy_vs_field}

\begin{figure*}
    \centering
    \includegraphics[width=0.75\linewidth]{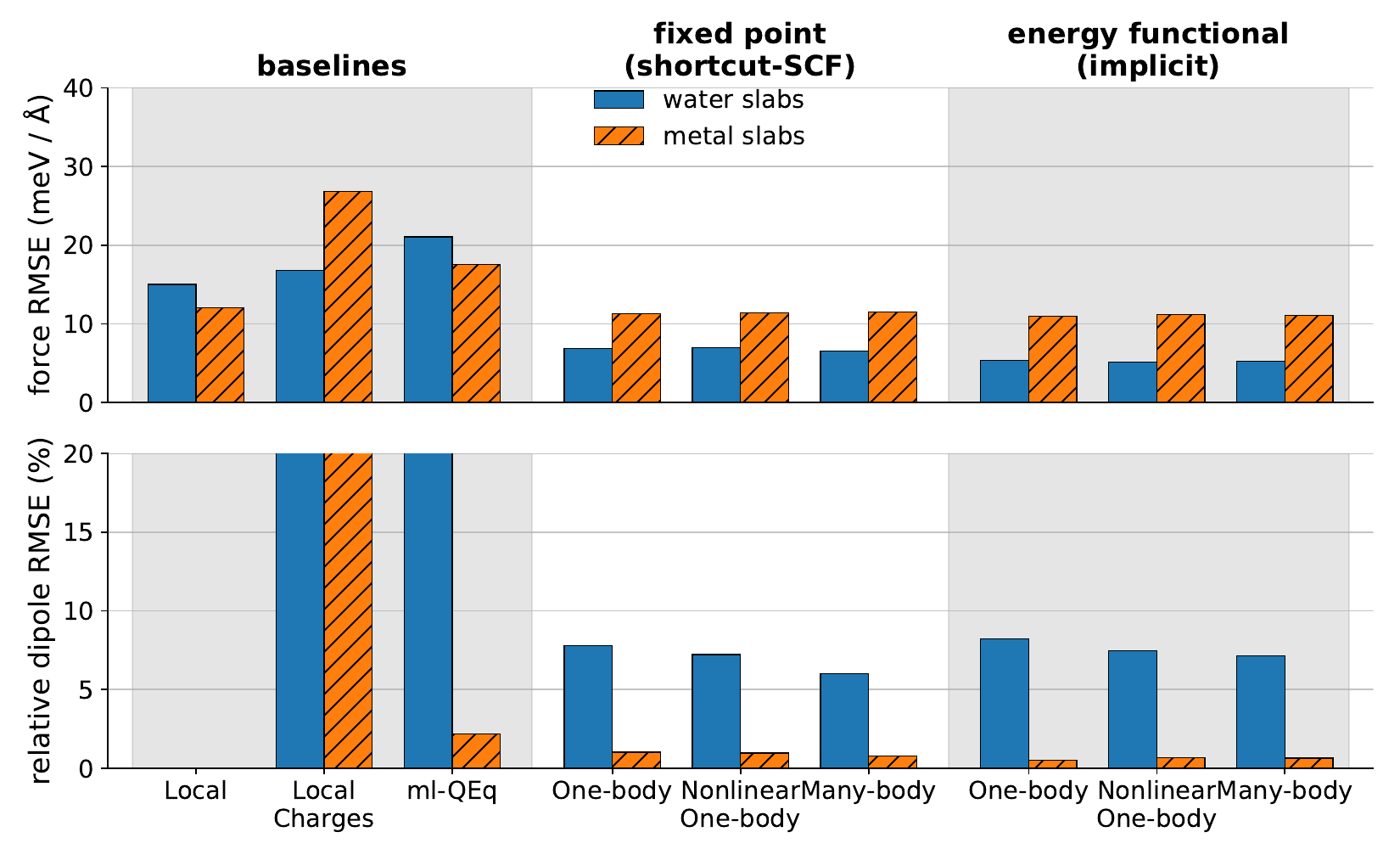}
    \caption{Comparing the energy functional and fixed-point models using different architectures.}
    \label{fig:scf:energy_vs_field}
\end{figure*}

We now compare the two approaches to self-consistency. Figure \ref{fig:scf:energy_vs_field} shows the error of atomic forces and total dipole for the baseline models and for all three variants of the fixed-point and energy functional models. While we have tried to make a fair comparison between the two approaches, some small differences separate the models. Most obviously, for the energy functional model all fits were performed with implicit differentiation for 200 epochs since we have found this to be the only way to reliably make accurate and stable models. The fixed-point models are instead trained using the shortcut-SCF method for 200 epochs, followed by 30 epochs of unrolling the SCF loop. When unrolling the SCF loop, we capped the maximum number of steps SCF at 40. For training the fixed-point models, we also ran this experiment with 200 epochs of direct training followed by 30 epochs of implicit differentiation. The results are quite similar and are presented in the supplementary information.

One can see that both architectures greatly outperform the baselines, and that there are minor differences between the approaches. The energy functional approach, in our implementation, is slightly more accurate on forces while the fixed-point model is more accurate at water dipole predictions. 

For each model, there are small improvements to be made for the dipole moment predictions by going to more expressive functional forms, but the improvement is negligible compared to the difference between our architectures and the baseline models. The simplest variants of the energy functional and fixed-point models are intentionally close in expressivity to classical polarizable force fields and to models like MACE-QEq. In fact, the quadratic energy model has formally very similar expressivity to MACE-QEq, yet the two achieve quite different errors. Sections \ref{sec:results:hessian} and \ref{sec:scf:res:l0_l1} will show where this difference comes from. 

\subsection{Physicality and Extrapolation}
\label{sec:scf:res:physicality}

The physics we want to probe with this test system is (i) quantization of charge on clusters and (ii) dielectric screening in metals and insulators. These are both failure modes of classical charge equilibration models and of MLIPs which predict charges only from local geometry.

\begin{figure*}
    \centering
    \includegraphics[width=0.85\linewidth]{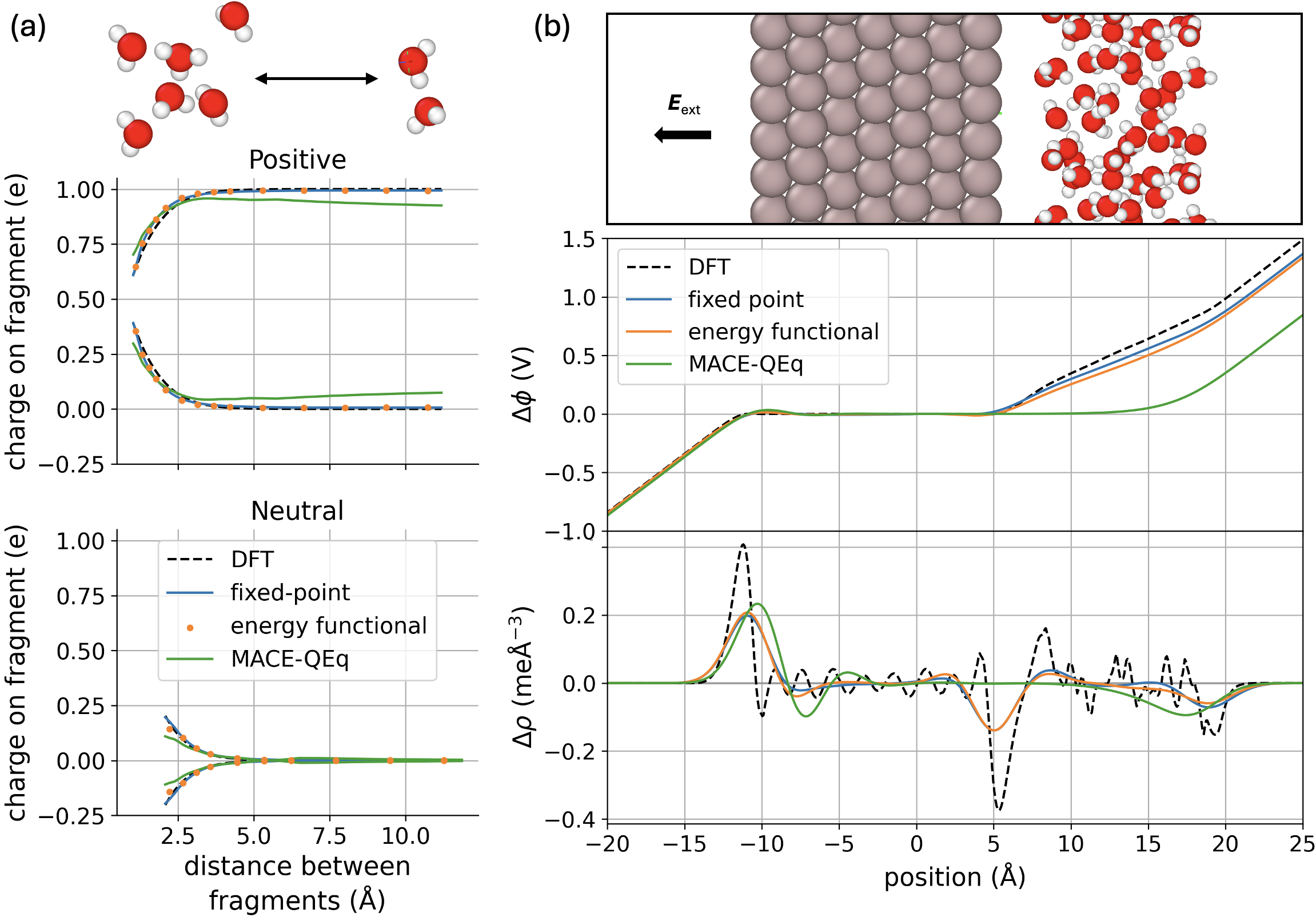}
    \caption{Physicality of the trained models. (a): fragmentation of small water clusters. The top panel shows separation of a cluster including a hydronium ion, whereas the lower plot shows fragmentation of a neutral cluster. (b): finite field response of a combined metal and water system.}
    \label{fig:scf:physicality}
\end{figure*}

\subsubsection*{Fragmentation of clusters}

We first test the physicality of various models at predicting quantization of charge during fragmentation. Figure \ref{fig:scf:physicality}a shows a cluster of water molecules being separated into two fragments, where one fragment contains a hydronium ion. Also shown in the plot labelled `positive' is the total charge on each fragment as a function of distance between the fragments. The charge on each piece is simply defined as the sum of the partial charges. 

One can see that MACE-QEq fails to predict the right separation, and furthermore the model appears to be quite non-smooth. This reinforces previous results which showed that when charge equilibration schemes with machine learning-derived parameters are stretched to fit things which are challenging for the quadratic functional form, unsatisfactory behaviour emerges \cite{Vondrak2025PushingLimits}.

Surprisingly, both the fixed-point and energy models perform well on this test, with the charge on each fragment smoothly approaching an integer. This is certainly not expected since there is nothing in either model's architecture which guarantees this behaviour. However, in both cases the models learn to separate the permanent and induced part of the charge density in a sensible manor, such that the fragmentation is modelled very accurately. Interestingly, the models do learn the correct charge transfer at short distances, and approximately follow the DFT reference. 

It is important to realize that the behaviour is only approximate: At the largest separation of the positive cluster, the energy functional model predicts 0.007 electrons of charge transfer from the positive to neutral fragment and the fixed-point model predicts 0.004 electrons. For comparison, DFT predicts only $3\times10^{-6}$ electrons of transfer. The lowest panel of Figure \ref{fig:scf:physicality}a shows the same test, but when splitting a cluster which is neutral, rather than positively charged.

\subsubsection*{Metallic and Dielectric Behaviour at an Interface}

For another test of physicality, we construct a simple model of a metal--water interface, by placing a slab of aluminium truncated on the 111 plane next to a slab of liquid water taken from the test set. The mixed system is shown at the top of Figure \ref{fig:scf:physicality}b. The two components are separated by 5\AA.

One can examine whether the models have correctly learned insulating and conducting behaviour by checking the response to a small, 0.1 V$/\text{\AA}$ electric field in the out of plane direction. The lower panels of \ref{fig:scf:physicality}b show the difference in electric potential and the difference in charge density respectively between the with-field and without-field cases. For the lines labelled `DFT', the full Hartree potential and electron density was extracted from DFT calculations on a grid. For the ML models, the smooth charge density is defined from the atomic multipoles as described in equation \eqref{eq:atomic_multipole_expansion}. 

Clearly both the fixed-point and energy functional methods perform well in this test and capture both the conducting and dielectric behaviour. By contrast, the MACE-QEq model correctly describes conduction as might be expected, but fails to describe any kind of dielectric response for the water portion. 

This test is difficult not only because the model must predict different responses for each part of the system, but also because there must be no charge transfer between the metal and water due to the applied field. Also, in the training dataset, there are no examples of configurations with both metal and water, so this is a true test of well behaved extrapolation.

We can tie these results back to the theory in section \ref{sec:scf:theory:equivalence}. Specifically, it was shown that a simple energy functional method is capable of modelling perfect conduction, but cannot model systems with zero susceptibility. Roughly speaking this means that it cannot formally model the case where a change in potential gives no change in charge density. By contrast, a simple implementation of the fixed-point approach should not be able to model perfect conduction. The above experiment shows that with a sufficiently flexible machine learning models, both in practice can show the correct behaviour for a real combination of conducting and insulating materials. It should be noted, however, that these properties are learned (rather than being built into the architecture) and hence are not exact. For instance,  the limit of charge transfer between the two fragments in Figure \ref{fig:scf:physicality}a at infinite separation is only approximately an integer.

\subsection{How does each model represent insulating and conducting behaviour?}
\label{sec:results:hessian}

\begin{figure}[h!]
    \centering
    \includegraphics[width=\linewidth]{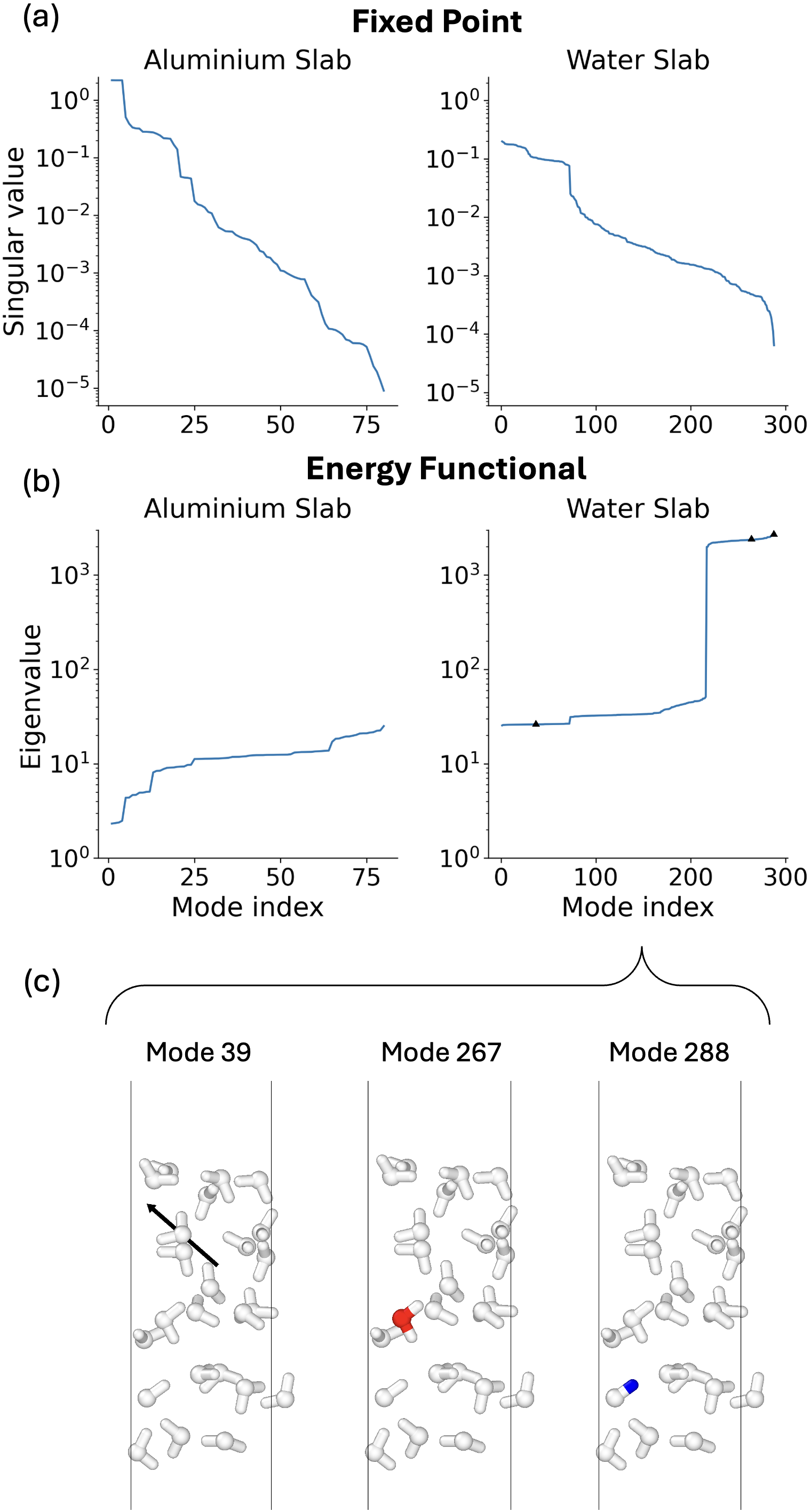}
    \caption{Quantifying the smoothness of the landscape of the self-consistent functions of each model; details are in the main text. In (a) singular values of the Jacobian of the update function of the water-metal slab fixed-point model. In (b) the eigenvalues of the Hessian of the water-metal energy model. In (c): each panel corresponds to a single eigenfunction of the Hessian of $G_\text{ML}[\mathbf{p}]$. The eigenfunctions are drawn by visualising atomic dipoles as arrows, and colouring the atoms according to the charge. Since large eigenvalues in the Hessian of the energy model correspond to small singular values of fixed-point model's update function, we have plotted the eigenvalues in increasing order and the singular values in decreasing order.}
    \label{fig:hessian_figure}
\end{figure}

We have seen that the energy and fixed-point models can be made to perform quite similarly. They have close to the same validation errors, and on the interface test they both have the right screening for both the metal and liquid segment, and the right fragmentation of charge. This is not the case for MACE-QEq. On close inspection however, one can show that functions learned by the two SC models are very different in terms of smoothness, which also explains the stark differences in training dynamics.


In section \ref{sec:theory} we explained that the energy functional model is essentially approximating the hardness kernels. We can probe the learned hardness kernels of the model by examining the Hessian of the learned functional $G_\text{ML}$. To simplify notation, let $k=(ilm)$ as in section \ref{sec:theory}, then define:

\begin{align*}
    H_{kj} = \frac{\partial^2 G_\text{ML}[\mathbf{p}]}{\partial p_k \partial p_j}
\end{align*}

One can compute this Hessian for a trained model at the lowest energy charge density from the model and examine its eigenvalues to quantify the smoothness of the learned functional. Large eigenvalues correspond to a high curvature and a `stiff' charge-energy landscape. In figure \ref{fig:hessian_figure}b, we have plotted the eigenvalues of the Hessian for the trained one-body quadratic energy functional model on two example structures. Specifically, we took one metal slab and one water slab and examined the Hessian after determining the charges self-consistently with the model. The multipoles going into the energy functional $G_\text{ML}$ are the actual atomic multipoles (which are typically $O(0.1)$ in units of $e$ and \AA) so the resulting Hessian eigenvalues can be compared to the hardnesses in QEq in units of eV/e$^2$. 

One can see several clear features. Most importantly, for the liquid water slab most eigenvalues are around 50, but 72 eigenvalues are $\sim$3000. This is related to the case of infinite vs zero susceptibility as discussed in section \ref{sec:scf:theory:equivalence}. We showed that systems which are stiff, in the sense some degree of freedom of the charge does not respond to a change in the relevant potential, implies that the curvature of the energy functional must go to infinity. In this case, the response of water is stiff because it is an insulator and net charge should never build up in response to a small change in potential difference. The model has fitted this landscape by forcing the charge of all 72 atoms to be almost fixed (in that the charge does not respond to the potential). This can be seen by examining the eigenvectors corresponding to the different modes. The very stiff modes all correspond to only the charge on an atom changing (see figure \ref{fig:hessian_figure}c). Because they are stiff, a change in potential induces very little change in the charge which is necessary to get the linear dielectric behaviour of a configuration or water. By contrast, the softer modes all correspond to dipole polarizabilities in which atomic dipoles respond to fields. These partial dipoles are therefore responsible for the screening. In order to fit this system, the energy functional has had to create extremely large eigenvalues. This also explains why MACE-QEq does not perform well, since in that model the actual hardnesses never approach such large values, and are confined to be $\sim O(10)$.

For the aluminium slab, the case is very different. Most of the eigenvalues are quite similar in magnitude, and are $O(10)$. As discussed in section \ref{sec:scf:theory:equivalence}, in a classical perfect conductor with zero screening length, the curvature of the non-Coulomb energy is zero. In a real conductor, however, only the self-consistent response at long wavelengths needs to diverge, which can occur without zero curvature of the learned energy functional, similar to the case of Thomas-Fermi theory. Hence, to get the right screening in the aluminium slab, the curvature of the learned functional does not need to go to zero, and values of $O(1)-O(10)$ are sufficiently small. 

We can make a similar analysis for the fixed-point model. This time, instead of computing a hessian, we can look at the fixed-point update function $\mathbf{p} = F_\text{ML}(\mathbf{v})$ from equation \eqref{eq:prediction_of_multipoles}, and consider its derivative:
\begin{equation*}
    A_{kn} = \frac{\partial p_k}{\partial v_{n}}
\end{equation*}
These terms describe the expansion of the density in terms of the effective potential in the model, and are conceptually similar to the softness kernels. One can look at the singular value decomposition of the matrix $A$, and plot the magnitude of the singular values. This is done for the same two example structures in figure \ref{fig:hessian_figure}a, using the linear update fixed-point model. Since there may be more electric field features than multipole coefficients, we have only plotted the non-zero singular values. The results show that many singular values, for both the aluminium and water system, are very small. This indicates that there are many modes which have very low susceptibility and which do not take part in the SC procedure (since they barely respond to the electric potential features). For a perfectly stiff system where the charge density does not react to any field, these values would all be 0, whereas the eigenvalues of the energy model Hessian would be infinite. Again, we can see the difference between the insulating and conducting systems. The aluminium slab has some singular values around 2.0 which are significantly larger than most other values in aluminium, and all values in water. This is the manifestation of how fixed-point models will struggle, in principle, to fit perfect conductors as explained in section \ref{sec:scf:theory:equivalence}. In practice, the fact that the largest singular values are still $O(1)$ shows that in our implementation, this slab of aluminium can easily be fitted without the model becoming ill conditioned. From the discussions in section \ref{sec:theory:fixedpoint:theory} and \ref{sec:scf:theory:equivalence}, the hardness kernels are, in a sense, the inverse of the softness kernels. This explains why the fixed-point model is able to use an update function with small singular values for the water slab, while the energy functional needs to create very large hardnesses to model the same behaviour. 

We strongly suspect that the relative sharpness of the functional in the energy functional model is why direct training does not work for this model. Since the curvature is large, evaluating the energy for any set of charges which is not the minimizing charge density results in a huge energy prediction. The DFT charge density will never be quite the minimizing density for the model, therefore when training with gradient descent at the DFT charge density, one gets very large and noisy gradient updates, making the optimization impossible. When using implicit differentiation, the model and its derivatives are only evaluated at its own minimum, avoiding this problem. By contrast, in the fixed-point model, many of the singular values are very small. This means that the model is naturally robust to changes in the corresponding electric field features, and variations in the input electric potential features can be tolerated. Therefore the output, and training gradients, are quite smooth even in direct training when one evaluates the model at the DFT charge density rather than its own solution.

\subsection{Effect of Higher-Order Atomic Multipoles}
\label{sec:scf:res:l0_l1}

\begin{figure*}
    \centering
    \includegraphics[width=0.75\linewidth]{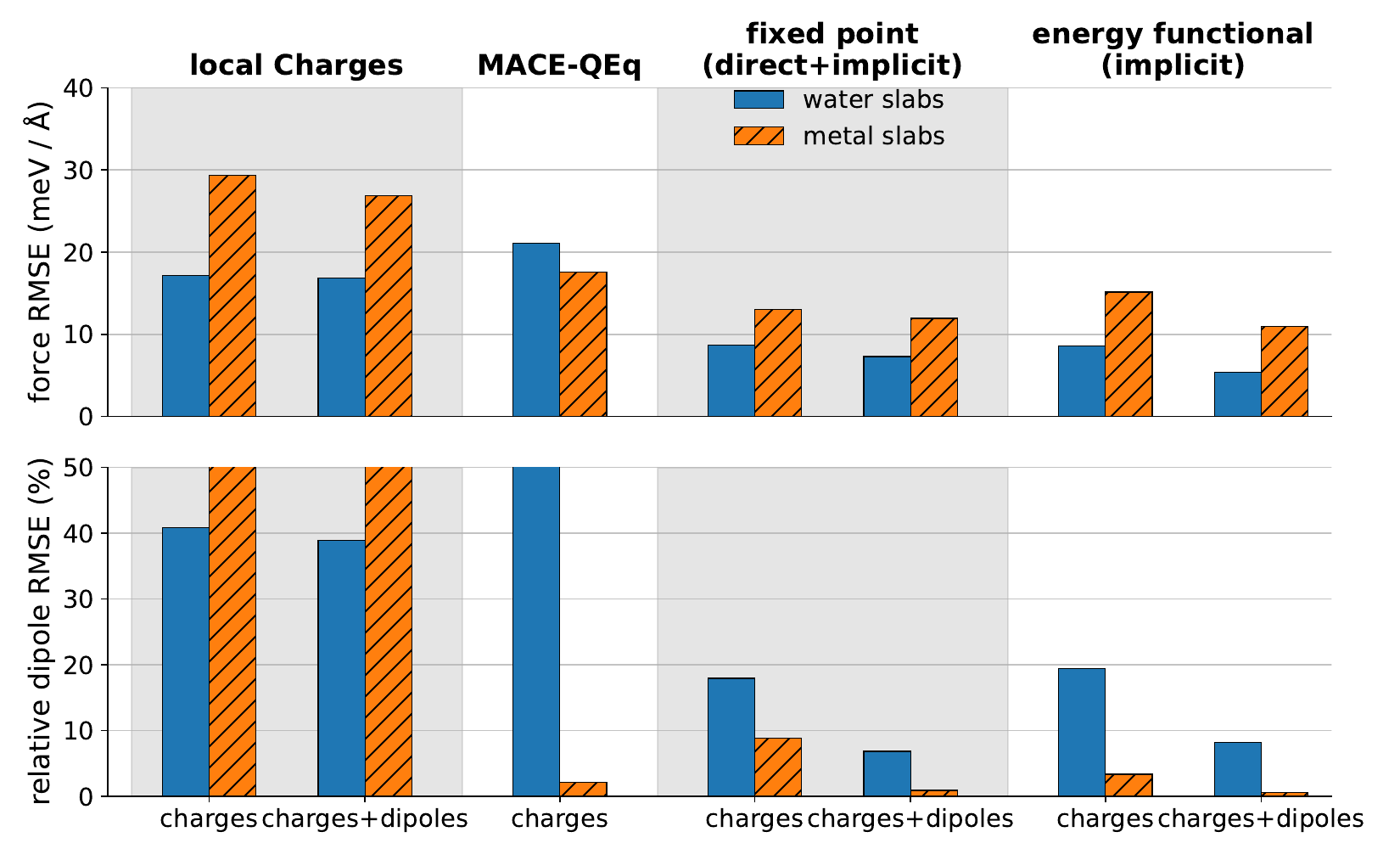}
    \caption{Comparing the effect of including atomic dipoles and vector features of the electric potential.}
    \label{fig:scf:l0_vs_l1}
\end{figure*}

Finally, a glaring difference between our implementations and other electrostatic MLIPs is the inclusion of higher-order atomic multipoles. To assess the impact of atomic dipoles, Figure \ref{fig:scf:l0_vs_l1} shows the performance of the MACE-LocalCharges model, the one-body linear fixed-point model, and the one-body quadratic energy model when using only partial charges on atoms. For the fixed-point model with partial charges only, the number of electric potential features was also reduced to just one scalar feature, corresponding to the most minimal version of this architecture. 

One can note several points: The MACE-LocalCharges model shows \textit{no improvement} when including atomic dipoles---its accuracy is entirely limited by the physics it can simulate. In contrast, both the fixed-point and energy functional models are considerably more accurate when using also atomic dipoles. Finally, the charges-only variants of both approaches outperform MACE-QEq and MACE-LocalCharges on almost all metrics. One can note that, as one would expect from the previous section, the water slab dipole prediction is most severely impacted by removing atomic dipoles from both SC models, although the performance is still better than MACE-QEq. 

In summary, the inclusion of atomic dipole moments in local charge prediction models does not seem to be at all helpful. In contrast, it is highly important for SC models since it allows for more natural ways to describe electronic screening.

\subsection{Conclusions from the metal--water System}

An example dataset was constructed which demonstrates some simple electrostatic phenomena, such as dielectric or metallic screening and fragmentation of clusters into pieces with integer charge. On this dataset, several conclusions can be drawn:
\begin{enumerate}
    \item In our implementation of both the energy functional approach and the fixed-point approach, one can train useful models without the need for data describing non-ground state electronic states (from either constrained DFT calculations or ad-hock data augmentation). This is relevant because it shows that the functional forms we have proposed---combined with the right training methods---naturally learn sensible, physical functions from only ground state data. Our trained models naturally learn to separate clusters into near integer-charge fragments and show the correct screening behaviour for both metals and insulators. This is without any bias towards this behaviour in the model architecture, and was shown in extrapolative tests: Neither fragmentation, nor a metal--water interface, were present in the training set of the models. These phenomena are not  typically described correctly by classical charge models, and the MACE-QEq results support this conclusion even in machine learned QEq models. The electronic screening we have discussed is also impossible in local models of the charge density, such as MACE-LocalCharges.
    \item This test system, which probes two extremes of electronic response, shows a key difference between the two methods. Learning an energy functional necessarily requires a very non-smooth energy landscape in order to treat electronically stiff systems. This is expected to be a common problem as insulators will always include stiff modes, since the total charge in a region should not respond to a shift in the electric potential. For an energy functional model to have this behaviour, the learned energy must have a very large curvature as a function of total charge. On the other hand, the case of metallic systems does not cause a problem for a fixed-point model. This is because in practice one can describe electronic screening in metals with a reasonably smooth fixed-point function.
    \item Finally, the most important result from a model development perspective is that training dynamics is extremely important in self-consistent models. We showed that even though both approaches can reach similar accuracy, the training methods differ widely. In both approaches, one can train accurate models with a simple functional form, but at the cost of expensive training protocols such as implicit differentiation. This is also the main differentiating factor that we have found between the approaches. The iterative scheme of the fixed-point model lends its self to direct training or training by unrolling the SC-loop, which we have not found to be the case for the energy functional approach. It is expected that developments in the efficiency of implicit differentiation may have an impact on the optimal strategy for training large self-consistent models. We have also argued that the difficulty in training the energy functional model is directly related to the non-smoothness of the learned functional.
\end{enumerate}

\section{Results: Charged Oxygen Vacancy in Silicon Dioxide}
\label{sec:results:sio2}

We now consider a second test system which probes a different set of physical effects. Specifically, this section examines charged oxygen vacancies in $\alpha$-quartz \ce{SiO2}. 

\begin{figure}
    \centering
    \includegraphics[width=\linewidth]{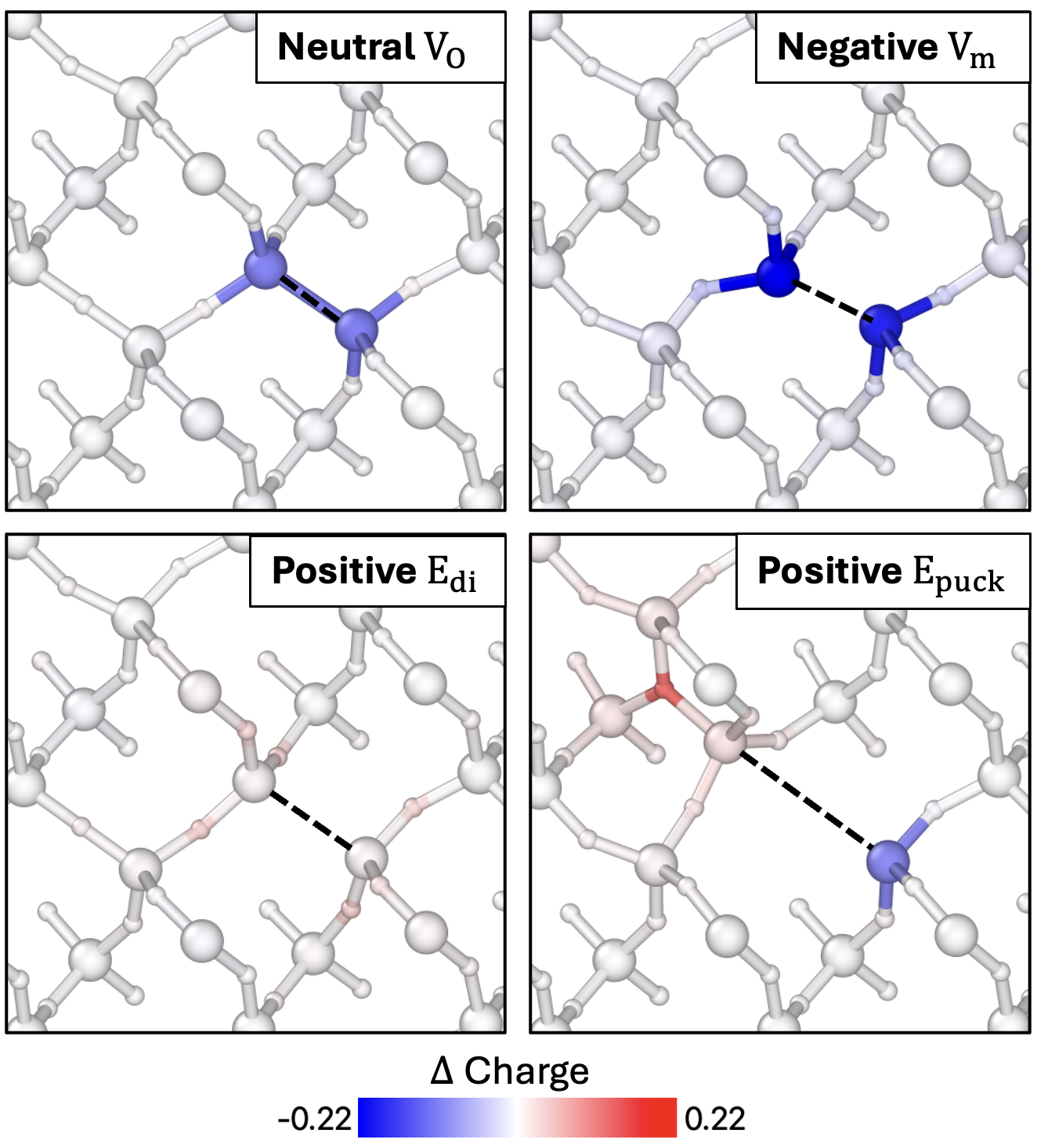}
    \caption{Oxygen vacancies in $\alpha$-quartz silicon dioxide with the atoms coloured according to the change in Hirshfeld charge relative to atoms in a unit cell of the material.}
    \label{fig:sio2_dataset}
\end{figure}

An oxygen vacancy in \ce{SiO2} can exist in three different charge states (with the lowest energy state depending on the Fermi level) and will have different geometries in each charge state \cite{Milton2023DifferenceSiO2}. When the vacancy is neutral, denoted $V_\text{O}$, the two silicon atoms to either side of the vacancy move together to form a bond, as in figure \ref{fig:sio2_dataset}. When the vacancy is negatively charged, denoted $V_\text{m}$, the two silicon atoms also come together, but with an offset perpendicular to the bond (see also figure \ref{fig:sio2_dataset}). When the vacancy is positive, three different stable geometries exist. In this paper we will consider only two. Firstly, the positive `dimer' configuration (E$_{\text{di}}$) has a geometry resembling a long bond between the two silicon atoms. Secondly, the `puckered' configuration (E$_{\text{puck}}$) involves one silicon atom moving backwards through the oxygen atoms behind it, forming a 3-fold coordinated oxygen atom behind the vacancy. These positive defects are also shown in figure \ref{fig:sio2_dataset}, where the atoms are coloured according to the change in Hirshfeld charge relative to each atom in a pristine $\alpha$-quartz \ce{SiO2} lattice.

In this section we examine whether various models can properly distinguish between these defect geometries, and whether they can extrapolate to configurations of multiple defects.

\subsection{Dataset Description}

A training dataset was created in a natural way for applying MLIPs to this problem: A foundation model (MACE-MATPES) was used to run MD simulations of cells of \ce{SiO2} containing a single oxygen vacancy at 400 K, and with either 48 unit cells (143 atoms) or 81 unit cells (242 atoms) of \ce{SiO2}. The foundation model cannot tackle systems with variable charge and so the different geometries of $V_\text{m}$, $V_\text{O}$ and $E_\text{di}$ do not naturally occur in the simulation. Instead, the defect geometry in MD is always close to the neutral vacancy $V_\text{O}$. Therefore, we applied a Hookean force constraint to force the Si--Si distance in the defect to vary over a range from 2.4 \AA \ to 4.5 \AA \ (which is representative of the Si--Si distance found in the neutral and puckered vacancy geometries). Further details are presented in the supplementary information. With this procedure, we sampled 300 configurations, which roughly resembled the dimer geometry in $V_\text{m}$, $V_\text{O}$ and $E_\text{di}$. Each of these structures was labelled with DFT in three different charge states (-1, 0, +1). We also sampled 374 structures with larger Si--Si distances, all of which adopted the $E_\text{puck}$ geometry, with a 3-fold coordinated oxygen. These structures were labelled with DFT using a total charge of +1. The DFT exchange-correlation functional was PBE \cite{PBE} which has been shown to stabilise all the defect geometries we are dealing with in this paper \cite{Milton2023DifferenceSiO2}. Other DFT calculation settings are given in section \ref{sec:scf:methods:dft_settings}. 

Even using the above mentioned constraint, the geometry of the negative defect very rarely occurs in foundation model MD. For this reason, we also added two copies of each DFT relaxed geometry into the training set, in both the 143 and 242 atom supercells. 

\subsection{Single Point Error Performance}

We now compare the performance of the self-consistent models and baselines on this dataset. For all models, the loss function included energy per atom, force and atomic multipoles as in equation \eqref{eq:scf:loss}. We note one important difference for this example was the inclusion of the model Fermi level in the fixed-point loss. When running the fixed-point model in constant charge mode, the Fermi level becomes an output of the model and can thus be included in the loss. All the fixed-point models were therefore trained with the additional loss term:
\begin{align*}
    L_\mu = w_\mu \left(\frac{|\mu - \mu_{\text{DFT}}|}{N_\text{atoms}}\right)^2
\end{align*}
In which $w_\mu=1000$. When training the fixed-point models, we used the shortcut-SCF method as in section \ref{sec:shortcut_scf}, with 15 steps and a mixing parameter of 0.5 for 250 epochs, followed by 50 epochs of unrolled SCF. For training the energy functional models, implicit differentiation was found to perform best, but was simply too expensive for multiple training runs. Therefore, we used an unrolled SCF loop and the models were trained for 500 epochs. Full details of the losses and training schedules for all models are presented in the supplementary information. 

\begin{figure*}
    \centering
    \includegraphics[width=\linewidth]{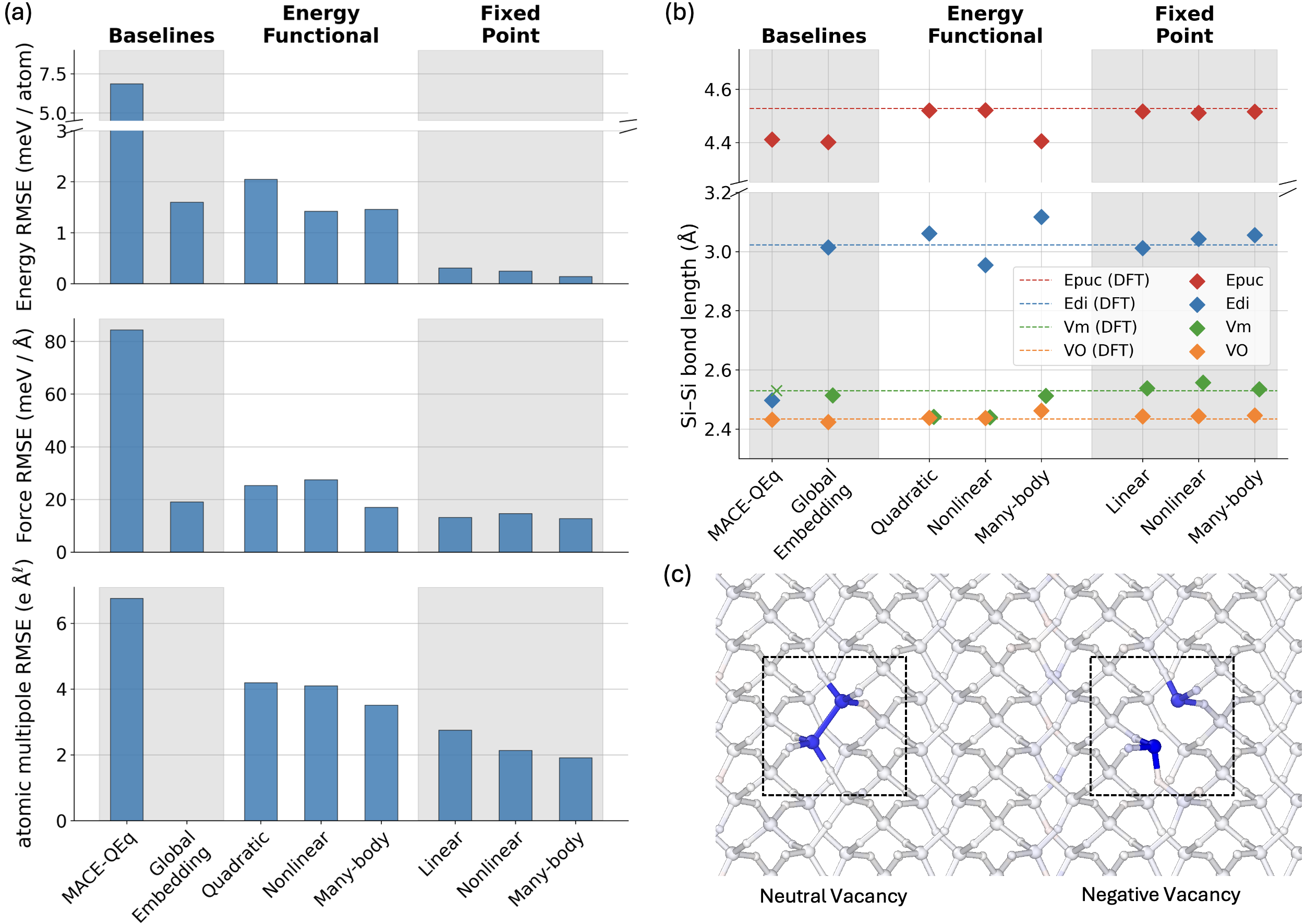}
    \caption{(a): Validation errors for the different models on the \ce{SiO2} training set. The global embedding model is a local model and therefore does not have a prediction for the atomic multipoles. (b): Si--Si bond length after relaxing each of the four defect types in a supercell. (c): Supercell containing one neutral oxygen vacancy and one negative oxygen vacancy, to illustrate the double defect relaxation test. Atoms are coloured according to the difference in Hirshfeld charge relative to pristine $\alpha$-quartz.}
    \label{fig:compound_sio2_results}
\end{figure*}

As well as the SC models, we trained two baselines to compare against. Firstly, the MACE-QEq model and secondly a local MACE model with a global embedding of the total charge, as described in section \ref{sec:scf:baseline_models}. Figure \ref{fig:compound_sio2_results}a presents the validation errors for energy per atom, force and atomic multipole coefficients for all the models we trained. The global embedding MACE model does not predict atomic multipoles and thus does not have an error on this quantity.

There is a clear story from the errors: MACE-QEq struggles to achieve reasonable energy and force errors, while the global embedding model does very well. The energy and force errors for the global embedding are in the range one would generally hope for, and are comparable to the self-consistent models. This illustrates that the relevant physics---at least for achieving a low validation error on single vacancies---is easily captured by simply telling the model the correct overall charge. At the same time, however, MACE-QEq is not able to properly separate the PESs corresponding to different charges. We also trained a MACE-QEq model with only the energy and force components of the loss. This model improved on the single point errors, but not on the key observable we are examining in this section, which is geometry relaxations of multiple charged defects.

We found that the fixed-point models trained with the shortcut-SCF method generally perform quite well. As can be seen in the figure, the fixed-point models all outperform the baselines. One can also see that the many-body update generally performs better than the simpler updates, in contrast to the metal--water interface system. The energy functional models also perform well, but are expensive to train.

\subsection{Relaxing Charged Defects}

The main task for these models is to relax an oxygen vacancy to the correct geometry based on a reasonable starting guess and the total charge. In order to assess whether a relaxed defect structure has the correct geometry, we will consider the Si--Si distance between the two atoms on either side of the vacancy, represented by the dashed line in figure \ref{fig:sio2_dataset}. This is not a perfect metric, since the difference between the negative and neutral geometries is quite subtle, but for the following results it is sufficient to tell the defects apart. 

To determine which models have learned the correct PES, we first verify whether each defect geometry is a local minimum at the correct charge. This was done by taking a copy of the negative ($V_\text{m}$), neutral ($V_\text{O}$), positive dimer ($E_\text{di}$) and positive puckered ($E_\text{puck}$) in a 242 atom cell, and relaxing the geometry to the nearest minimum at the corresponding charge. The results of this experiment are shown in figure \ref{fig:compound_sio2_results}b. In the plot, each point represents the Si--Si distance from a given model when relaxing a given defect, while the dashed lines are the values from DFT relaxations, which are in agreement with other values in the literature \cite{Milton2023DifferenceSiO2}. The `$\times$' represents a failed geometry relaxation by a model, which occurred when relaxing the negative defect with MACE-QEq. 

The performance of the baseline models mirrors that seen in single point errors. The MACE-QEq model is unable to relax the negative or positive structures correctly. The negative structure simply failed to relax using l-BFGS (the forces diverged), and the positive structure adopted the geometry of the neutral defect. By contrast, the simple global embedding correctly relaxed all 4 defects to the right structure. 

For the SC models, one can see some clear trends. Overall, the fixed-point models can reliably recover the right geometry at all charges. The energy models, however, do not. The negative defect incorrectly adopts the geometry of the neutral structure for both the one-body quadratic and one-body nonlinear model. Only the many-body model correctly relaxes this structure. In general, we found that recovering the geometry of the negative defect from this training set is a difficult task. 

\subsubsection*{Relaxing Multiple Defects}

So far, the global embedding model has performed well, in both errors and when relaxing single defects, without using any long-range interactions. We now examine the more challenging case of multiple defects in a single supercell. To do this, we consider a supercell of 484 atoms containing two oxygen vacancies, 16~\AA~apart. An example is shown in figure \ref{fig:compound_sio2_results}c. To assess the models, we relax this supercell, which initially contains one copy of the negative defect geometry and one copy of the neutral defect geometry, with relaxations performed with various different total charges. After each relaxation, we determined which defect each vacancy most closely resembled based on both the Si--Si bond length, and by inspecting for the visually apparent offset which is present in the negative vacancy. 

We expect that the charge on each defect should be quantised, with each vacancy relaxing to one of the three charge states, rather than adopting geometries which are a mixture. One therefore expects that when relaxing this supercell with a total charge of +1, we should get one neutral vacancy and one positive vacancy. When relaxing with +2 total charge, we should find the geometries of two positive vacancies, and similarly for total charges of 0, -1 and -2. The results of this experiment for all models are presented in table \ref{table:double_defects}. In the table, the notation (a, b), where a and b can be -1, 0, or +1, is used as shorthand for the geometry of defects in a supercell. (0, +1) indicates that one defect relaxed to the neutral vacancy $V_\text{O}$, and the other to the positive dimer $E_\text{di}$. This time, the global embedding model does not predict the correct structures. It is easy to see why this is: When the total charge is set to +1 in the global embedding model, each vacancy will relax to the positive dimer geometry, unaware of the other vacancy. Furthermore, when trying to evaluate at +2 or -2 charge, the model simply fails to relax, since this embedding value has never been seen by the model. This is a failure of extensivity. 


MACE-QEq does have the flexibility in principle to represent the different combinations of charges. In this test however, the model always relaxes both defects to approximately the same structure, which most resembles the neutral defect. 

The SC models perform better. The many-body fixed-point model successfully relaxes all charge states, starting from the (0, -1) structure, to the expected pair of defects. The -2 case is marked as a question mark, since the real DFT solution is actually not the expected (-1, -1), which is discussed in detail below. The other fixed-point models make some errors for either the -1 or +1 case. 

The energy functional models struggle more on this test. The quadratic model gives the same result as MACE-QEq (even though the single point errors are considerably better). Only the one-body nonlinear model performs better, relaxing the doubly positive and doubly negative structures to reasonable geometries. 

The total charge of -2 is marked by a question mark when the model relaxes to (0, -1). This is because although one would expect to get two negative defects (-1, -1), a DFT relaxation of this system from this starting structure also gives the (0, -1) structure. The models are therefore correctly mimicking the level of theory they are trained on, which is PBE. To explore this further, we performed DFT relaxations of the (0, -1) structure and the (-1, -1) structure, each with a total charge of -2 and with different constraints on the projected total spin. Table \ref{table:-2_defects} shows the resulting geometries, as well as the energy difference in the final structure relative to the (-1,-1) case. In all cases, the resulting geometry is the same as the initial geometry, meaning that both the (0, -1) and (-1, -1) geometries are local minima for a total charge of -2, even when one forces the system to have two unpaired electrons. We suspect that this behaviour may not transfer to hybrid DFT functionals. 

Nevertheless, at this level of theory we can compare the energy of the different structures in the table. This reveals that when one forces the system to have two unpaired electrons, the (-1, -1) structure is more stable by 0.107 eV. If one instead constrains the projected total spin to be 0, the (0,-1) structure is more stable by 0.363 eV. When relaxing both structures with the many-body fixed-point model (which is not aware of spin) we found that it also stabilises both structures, and the energy difference is in between the two DFT predictions for $N_\uparrow-N_\downarrow=0$ and $N_\uparrow-N_\downarrow=2$, with the (0, -1) case being slightly more stable.

\begin{table}[]
    \centering
    \renewcommand{\arraystretch}{1.2}
    \begin{ruledtabular}
    \begin{tabular}{lccc}
    \multirow{2}{*}{\textbf{Model}} & \textbf{Total} & \textbf{Resulting} & \textbf{Expected?} \\
     & \textbf{Charge} & \textbf{Defects} & \\
    \colrule
    \multirow{5}{*}{\parbox{2.5cm}{\raggedright Global Embedding}} 
     & $-2$ & relaxation fails & \xmark \\
     & $-1$ & $0,\,-1$ & \cmark \\
     & $0$ & $0,\,\ \ 0$ & \cmark \\
     & $1$ & $1,\,\ \ 1$ & \xmark \\
     & $2$ & relaxation fails & \xmark \\
    \colrule
    \multirow{5}{*}{\parbox{2.5cm}{\raggedright MACE-QEq}} 
     & $-2$ & $0,\,\ \ 0$ & \xmark \\
     & $-1$ & $0,\,\ \ 0$ & \xmark \\
     & $0$ & $0,\,\ \ 0$ & \cmark \\
     & $1$ & $0,\,\ \ 0$ & \xmark \\
     & $2$ & $0,\,\ \ 0$ & \xmark \\
    \specialrule{1pt}{1pt}{1pt}
    \multirow{5}{*}{\parbox{2.5cm}{\raggedright Fixed-Point Linear}} 
     & $-2$ & $0,\,-1$ & ? \\
     & $-1$ & $0,\,\ \ 0$ & \xmark \\
     & $0$ & $0,\,\ \ 0$ & \cmark \\
     & $1$ & $0,\,\ \ 1$ & \cmark \\
     & $2$ & $1,\,\ \ 1$ & \cmark \\
    \colrule
    \multirow{5}{*}{\parbox{2.5cm}{\raggedright Fixed-Point Nonlinear}} 
     & $-2$ & $0,\,-1$ & ? \\
     & $-1$ & $0,\,-1$ & \cmark \\
     & $0$ & $0,\,\ \ 0$ & \cmark \\
     & $1$ & $0,\,\ \ 0$ & \xmark \\
     & $2$ & $1,\,\ \ 1$ & \cmark \\
    \colrule
    \multirow{5}{*}{\parbox{2.5cm}{\raggedright Fixed-Point Many-Body}} 
     & $-2$ & $0,\,-1$ & ? \\
     & $-1$ & $0,\,-1$ & \cmark \\
     & $0$ & $0,\,\ \ 0$ & \cmark \\
     & $1$ & $0,\,\ \ 1$ & \cmark \\
     & $2$ & $1,\,\ \ 1$ & \cmark \\
    \specialrule{1pt}{1pt}{1pt}
    \multirow{5}{*}{\parbox{2.5cm}{\raggedright Energy Functional Quadratic}} 
     & $-2$ & $0,\,\ \ 0$ & \xmark \\
     & $-1$ & $0,\,\ \ 0$ & \xmark \\
     & $0$ & $0,\,\ \ 0$ & \cmark \\
     & $1$ & $0,\,\ \ 0$ & \xmark \\
     & $2$ & relaxation fails & \xmark \\
     \colrule
    \multirow{5}{*}{\parbox{2.5cm}{\raggedright Energy Functional Nonlinear}} 
     & $-2$ & $0,\,-1$ & ? \\
     & $-1$ & $0,\, \ \ 0$ & \xmark \\
     & $0$ & $0,\,\ \ 0$ & \cmark \\
     & $1$ & $0,\,\ \ 0$ & \xmark \\
     & $2$ & $1,\,\ \ 1$ & \cmark \\
     \colrule
    \multirow{5}{*}{\parbox{2.5cm}{\raggedright Energy Functional Many-Body}} 
     & $-2$ & $0,\,-1$ & ? \\
     & $-1$ & $0,\, \ \ 0$ & \xmark \\
     & $0$ & $0,\,\ \ 0$ & \cmark \\
     & $1$ & $0,\,\ \ 0$ & \xmark \\
     & $2$ & relaxation fails & \xmark \\
    \end{tabular}
    
    \end{ruledtabular}
    \caption{Relaxing pairs of defects in different charge states. In all cases, the initial geometry included one copy of the neutral defect and one copy of the negative defect. The `expected' entry is omitted when models relax to the $(0, -1)$ structure as discussed in the text and table \ref{table:-2_defects}.}
    \label{table:double_defects}
\end{table}

\begin{table}[t]
\renewcommand{\arraystretch}{1.2}
\begin{ruledtabular}
\begin{tabular}{cccc}
\multirow{2}{*}{\textbf{Model}} & \textbf{Initial} & \textbf{Resulting} & \textbf{Relative} \\
 & \textbf{Defects} & \textbf{Defects} & \textbf{Energy (eV)} \\
\colrule
\multirow{2}{*}{\parbox{2.5cm}{\centering DFT ($N_\uparrow - N_\downarrow=0$)}} & $(0, -1)$ & $(0, -1)$ & $-0.363$ \\
 & $(-1, -1)$ & $(-1, -1)$ & $0.000$ \\
\colrule
\multirow{2}{*}{\parbox{2.5cm}{\centering DFT ($N_\uparrow - N_\downarrow=2$)}} & $(0, -1)$ & $(0, -1)$ & $0.107$ \\
 & $(-1, -1)$ & $(-1, -1)$ & $0.000$ \\
\colrule
\multirow{2}{*}{\parbox{2.5cm}{\centering Fixed Point}} & $(0, -1)$ & $(0, -1)$ & $-0.084$ \\
 & $(-1, -1)$ & $(-1, -1)$ & $0.000$ \\
\end{tabular}
\end{ruledtabular}
\caption{Defect pairs with -2 total charge do not always relax to the expected structure. Two defects are initialised in a supercell where either (i) both defects have the geometry of the isolated negative vacancy, denoted (-1, -1), or (ii) one defect has the geometry of the negative vacancy while the other has the geometry of the neutral vacancy, denoted (0, -1). For each model in the table, a defect pair with each of these initializations is relaxed to the nearest local minima with a total charge of -2 and the final energy is reported relative to the (-1,-1) case. $N_\uparrow - N_\downarrow$ refers to the constraint on total projected spin.}
\label{table:-2_defects}
\end{table}

\subsection{Conclusions from Silicon dioxide}

This test system behaves quite differently to the metal--water system. Induced polarization over long distances (which gives rise to screening in the metal--water example) does not appear to be important for getting the right properties, since a global embedding model can relax single defects to the correct geometry depending on the charge. The difficult physics is that of correctly assigning a limited total charge across multiple defect sites. This was an extrapolation test since no multiple defect configurations were in the training set, and thus it is not a true test of the expressiveness of different models, but the results are quite stark. 

Our implementation of MACE-QEq cannot correctly extrapolate to multiple defects at all, even though in principle one can evaluate the model in higher charge states than what it was trained on. In fact, that model cannot relax the single defects, suggesting that the PES as a function of geometry and atomic charge is simply a difficult function to fit. 

The self-consistent models we trained attained broadly similar errors to one another, but we found considerable differences when relaxing oxygen vacancies. It seems that the energy functional models were generally more difficult to train. All three energy functional models were able to relax most defects correctly, but only the most complicated, many-body variant could relax the negative vacancy. When relaxing pairs of defects, in this extrapolative test the models struggled to produce the right geometries, and sometimes exhibited unphysical behaviour such as failed geometry relaxations. The fixed-point models performed quite differently, and the most complicated architecture was able to relax all charge states of the double defect structure to the correct geometry. We stress, however, that this is a considerable extrapolation and in practical use of these models, one would almost certainly add multiple defect examples into the training set. It is not immediately clear how this would change the results. 

Finally, it is worth pointing out the practical limitations of training these models. The training cost in this example was very large for the energy functional models, since one cannot use the direct or shortcut-SCF methods. Training and using the many-body fixed-point model is also very expensive, and further training of the one-body variants (adding multiple configurations with defect pairs into the training set) would be preferable in practice.  

\section{Conclusions}

\label{sec:scf:conclusions}

We have presented a framework for understanding many existing electrostatically self-consistent MLIP architectures. This perspective is useful because it makes explicit which aspects of the underlying electronic structure theory are being approximated by a given model. In addition, we have shown that the two self-consistent approaches outlined in this work can, under certain conditions, be equivalent to one another. Broadly speaking, a machine-learned local energy functional that depends on partial charges can be reformulated as a machine-learned local fixed-point function. The differences between learning an energy functional and learning a fixed-point function are twofold. First, when the energy functional is genuinely many-body, the equivalent fixed-point operation may be global rather than local. Second, when certain degrees of freedom in the charge density do not respond to changes in fields or potentials, a fixed-point formulation can simply predict no update for those variables, whereas an energy functional must generate a highly curved energy landscape in order to effectively constrain those degrees of freedom. Our results showed that this is not an abstract concern: When modelling the dielectric behaviour of clusters and slabs of water, there are stiff modes in the energy landscape which make training energy functional models difficult, but do not pose a problem for fixed-point style models. 

We have also investigated several natural strategies for training self-consistent ML models and discussed how these approaches apply to both of the architectures considered here. These training methods are critically important for the practical usability and scalability of self-consistent electrostatic MLIPs.

To understand which architectural and training decisions are most important in practice, we implemented a range of self-consistent models, together with representative baselines, within a single common framework. This allowed the different approaches to be compared on an even footing, since the representation of the charge density and the local components of the models were held fixed.

The test systems used in this study are quite simple, but have revealed the key issues faced by non-self-consistent models, as well as the differences between different self-consistent model architectures. We showed that our self-consistent models naturally reproduce the correct physics of a metal--water interface (a key ingredient in applications such as heterogeneous catalysis and energy storage) using only ground-state training data. These models also extrapolate correctly to cluster fragmentation and capture the high-frequency response of a metal–water interface, behaviour that was not observed for the baseline models. In particular, MACE--QEq exhibited the same qualitative failures previously documented for classical charge equilibration models.

A similar picture emerged for charged vacancies in silicon dioxide, but in this case the distinction between fixed-point and energy-functional approaches was more pronounced. The most sophisticated fixed-point model we trained successfully extrapolated to configurations containing multiple defects, whereas the energy-functional models did not. 

Finally, we emphasize that for self-consistent MLIPs, the training procedure is often more critical to performance than the expressiveness of the model itself. In our experience, accurate fixed-point models are generally easier to train, since unrolled or truncated self-consistent field procedures provide a controlled and stable training strategy. In the framework we have proposed, we think that part of this is an inherent property of parametrising smooth energy functionals for insulating systems. 

While the models presented here are physically motivated and accurate, their architectural, training, and inference complexity is substantially greater than that of simpler MLIP approaches. We expect that a promising direction for future research will be to retain the essential physical ingredients identified in this work while developing training and inference strategies that avoid the need for sophisticated training procedures such as implicit differentiation. 

Finally, given the direction of the wider research field, a key question is whether one should pursue the development of self-consistent foundation models, and, if so, what are the remaining challenges? In this work we have introduced considerably more physics into MLIPs, but at the cost of more expensive inference and training protocols. For building foundation models, it is not yet clear whether substantially more physically motivated models will prevail over the traditional machine learning strategy of ``more data, larger model''. If self-consistent models are the way forward, the key bottleneck for now is the complexity of training. This is particularly important for foundation models, since with large diverse datasets which contain varied chemistry and potentially some outliers, a reliable and efficient training method is essential. A related open question is what kind of data is needed in foundation datasets to train self-consistent models. Most existing large datasets for materials science include only total energies and forces, but in this work we also trained on properties like the total dipole and Fermi level. Whether this additional information is still needed when training foundation models will be a key factor determining how quickly and easily self-consistent foundation models can be made. 

\section{Acknowledgements}

W.J.B. thanks the Cambridge University Engineering Department for funding through the Ashby postdoctoral fellowship. I.B. was supported by the Harding Distinguished Postgraduate Scholarship. The authors also acknowledge helpful discussions with Jack Thomas and Christoph Ortner about self-consistent model development. 

This work used computational resources from the ARCHER2 UK National Supercomputing Service (http://www.archer2.ac.uk) which is funded by EPSRC via our membership of the UK Car–Parrinello Consortium. We also used the Cambridge Service for Data Driven Discovery (CSD3), with some computing hours supplied by the Lennard-Jones Center.  
The authors acknowledge the use of resources provided by the Isambard-AI National AI Research Resource (AIRR)~\cite{bristolai}. Isambard-AI is operated by the University of Bristol and is funded by the UK Government’s Department for Science, Innovation and Technology (DSIT) via UK Research and Innovation; and the Science and Technology Facilities Council [ST/AIRR/I-A-I/1023].
We would like to thank UK Sovereign AI for providing compute on Isambard-AI.
We acknowledge the Jean Zay cluster access to compute as part of the Grand Challenge: GC010815458 (Grand Challenge Jean Zay H100).

\section{Data Availability Statement}

The datasets and all models created in this study are available in a Zenodo repository at \url{10.5281/zenodo.21610371}. An implementation of the models is available at \url{https://github.com/ACEsuit/mace-scf}.

\section{Conflicts of Interest}

G.C. is a partner in Symmetric Group LLP that licenses force fields commercially and also has equity interest in Ångström AI.

\bibliographystyle{IEEEtran}
\bibliography{refs}

\section{Methods}

\subsection{fixed-point model Implementation Details}
\label{methods:implementation_fixed_point}

The fixed-point SC architecture is based around a mapping from the effective potential to the charge density. As in the main text, this is decomposed into a local component dependent only on the local node features, and a non-local component depending on the electric potential.

\begin{align}
    \mathbf{p}_{i} &= F_\text{ML}(\tilde{\mathbf{h}}, \mathbf{v}) = \mathbf{p}_{i}^{(\text{local})} + \mathcal{F}(\tilde{\mathbf{h}}, \mathbf{v})
\end{align}

The local piece is a linear function of the local node features, as described in the main text. The three choices for $\mathcal{F}$ are described below.

\subsubsection{One-Body Linear Update}

In the simplest kind of update, the atomic multipoles $p_{i,lm}$ at step $t+1$ are predicted as an affine function of the electric potential descriptors $v_{i,nlm}$ at step $t$. This is equivalent in spirit to the implementation of SCFNN and to polarizable force fields. The implementation involves first embedding the normalized electrostatic feature vector $v_{i,nlm}$ into a wider vector:
\begin{align}
    V_{i,klm} &= \text{LinearEmbedding}(v_i, \tilde{h}_i) \nonumber \\
    &=\sum_{n}^{N_{radial}} W_{z_i l, kn}v_{i,nlm}+ \sum_{k'}^{K} W'_{lkk'} \tilde{h}_{i,k'lm}
    \label{eq:linear_biased_embedding}
\end{align}
The embedding dimension $K$ is the number of channels of the MACE feature vector. In this work, the normalization constants $\mathcal{N}_{NL}$ were found by taking the average over absolute the values of $v_{i,nlm}$ for each $n,l$ combination across the dataset. If DFT atomic multipoles are available in the training set, one can compute the typical size of $v_{i,nlm}$ before training. The purely local features $\tilde{\mathbf{h}}$ enter the embedding as a bias term. Following this, atomic multipoles are predicted by first forming invariants $I_{i,k}$, and then again multiplying the local features:
\begin{align}
    I_{i,k} &= \sum_{k'lm} W_{lkk'}  \tilde{h}_{i,klm} V_{i,k'lm} \label{eq:fp_outout_calc_Isa} \\
    p_{i,lm}^{\text{(nonlocal)}} &= \sum_{kk'} W'_{z_i,lkk'} I_{i,k} \tilde{h}_{i,k'lm}\label{eq:fp_outout_calc_Isb}
\end{align}
$p^\text{(nonlocal)}$ is the output of $\mathcal{F}$, and is combined with the local contribution to give the full set of multipoles as in \eqref{eq:prediction_of_multipoles:sum}. $W$ and $W'$ are matrices of weights. 

\subsubsection{One-Body Nonlinear Update}
To allow for nonlinear dependence of the atomic multipoles on the electric potential features, one can repeat exactly the same process as above, but adding a potentially deep nonlinearity to the rotationally invariant quantities $I_{i,k}$. Specifically, we present results corresponding to using the same logic as in equations \eqref{eq:linear_biased_embedding}--\eqref{eq:fp_outout_calc_Isb}, but applying a nonlinearity before predicting the multipoles:
\begin{align}
I_{i,k} &= \sum_{k'lm} W_{lkk'}  \tilde{h}_{i,k'lm} V_{i,k'lm} \label{eq:methods_dot_products} \\
    I'_{i,k} &= \text{MLP}(I_{i,k}) \label{eq:methods_mlp} \\
    p_{i,lm}^{\text{(nonlocal)}} &= \sum_{kk'} W'_{z_i,lkk'} I'_{i,k} \tilde{h}_{i,k'lm} \label{eq:methods_readout}
\end{align}
In which ``MLP'' refers to a three layer multi-layer perceptron with 64 hidden neurons. 

\subsubsection{Many-body Update}
\label{sec:scf:fixedpoint_manybody_arch}
For a many-body update function, we make use of the MACE layer with some modifications which were found to be important when working with electric potentials and charges on atoms. Again, projections of the electric potential are embedded as in equation \eqref{eq:linear_biased_embedding}. This information is then shared between neighbouring atoms through first forming the following descriptors, which are analogous to the MACE 1-particle basis \cite{Kovacs2023EvaluationScience}:
\begin{align*}
    \psi_{ij,k\eta l_3 m_3} = \sum_{l_1m_1l_2m_2} C^{l_3m_3}_{\eta l_1m_1l_2m_2} R_{k\eta l_1l_2l_3} Y_{l_1}^{m_1} (\tilde{\mathbf{r}}_{ij})  \\ \times \left( V_{j,kl_2m_2} + \sum_{k'} w_{l_2kk'} V_{i,k'l_2m_2} \right)
\end{align*}
This differs from the MACE 1-particle basis in that the node features of neighbouring atoms ($j$) are replaced with a combination of the neighbour and the central ($i$) atoms' embedded electric potential features $\tilde{V}$. Following this, the neighbourhood information is aggregated by summing over neighbours, where the sum is normalized by a smooth learnable function of the number of neighbours ($f(N)$):
\begin{align}
    A_{i,kl_3m_3} = f(N) \sum_{k'\eta} W_{kk' \eta l_3} \sum_{j \in \mathcal{N}(i)} \psi_{ij,k'\eta l_3 m_3}
\end{align}
The learnable normalization is the same as that used in the MACE-MP-0 foundation models \cite{MP0}:
\begin{align}
    &f(N) = \nonumber \\
    & \quad \Biggr( 1+\sum_{j\in \mathcal{N}(i)} \text{tanh}\left(\text{SiLU}\left(\sum_k W_k B_k(r_{ij})\right)^2\right)\Biggr)^{-1} \nonumber
\end{align}
Where $W_k$ are weights and $B_k$ are Bessel functions which serve as the radial embedding in a MACE model. To create high body-order features, equivariant products of $A$ are formed with correlation order $\xi=1,...,\nu$ , to produce body ordered features $B^\nu$, through exactly the equations used in the MACE product operation. This allows arbitrary body order interactions, but for all experiments shown in this paper, we found that correlation $\nu=1$ (two-body information) was the highest practical value of $\nu$ due to the fact that one must iteratively evaluate this part of the model to converge the SC loop.

Finally, a linear contribution from the central atom only is added, to create a new set of atomic features which contain body ordered information about the geometry and electric potential around atom $i$:
\begin{align}
    T_{i,klm}=\sum_{k'} W_{kk'l} V_{i,k'lm} + \sum_{\nu k'} W_{kk'\nu} B_{i,k'lm}^\nu
\end{align}
The nonlocal part of the atomic multipoles are derived from these features exactly as in equation \eqref{eq:methods_dot_products}--\eqref{eq:methods_readout}, using $T_{i,klm}$ in place of $V_{i,klm}$. 

\subsubsection{Nonlocal Energy Readout}

For the nonlocal energy readout, all models use a flexible manybody function based on the MACE layer. Firstly, the final charge density $p_{i,lm}$ and electric potential features $v_{i,nlm}$ are embedded using the same linear embedding as in equation \ref{eq:linear_biased_embedding} to create mixed features $X_{i,klm}$ as follows:
\begin{align}
    X_{i,klm} =& \text{LinearEmbedding}(v_i, \tilde{h}_i) \nonumber \\
    &+ \text{LinearEmbedding}(p_i, \tilde{h}_i) \nonumber
\end{align}
This is then used as the input features to a MACE layer. As in MACE, one first creates the one-particle basis:
\begin{align}
    \phi_{ij,kl_3m_3} = \sum_{l_1m_1l_2m_2} \mathcal{C}^{l_3m_3}_{l_1m_1,l_2m_2} &R_{kl_1l_2l_3}(r_{ij}) \times \nonumber \\ &\times Y_{l_1m_1}(\hat{\mathbf{r}}_{ij}) X_{j,kl_2m_2} \nonumber
\end{align}
The hyperparameters of this operation are the same as the local part of the model. Then, information is aggregated over neighbours:
\begin{align}
    A_{i,klm} = \sum_{k'} W_{lkk'} X_{i,k'lm}+ \frac{1}{N}\sum_j \phi_{ij,klm}
\end{align}
Where $N$ is the average number of neighbours. In this case, we also add a contribution from the central atom $i$, which is the first term in the above equation, in contrast to a normal MACE layer where this is absent. The MACE product is then used to create body ordered features from $A_{i,klm}$ of order $\nu$, denoted $B^\nu$ \cite{MACE, Kovacs2023EvaluationScience}. The correlation of this term in all our models is 2. Finally, we add a skip connection to create body ordered equivariant features of the potential and density:
\begin{align*}
    T_{i,klm}=\sum_{k'} W_{kk'l} X_{i,k'lm} + \sum_{\nu k'} W_{kk'\nu} B_{i,k'lm}^\nu
\end{align*}
Once again, $T_{i,klm}$ is used in place of $V_{i,klm}$ in \eqref{eq:methods_dot_products}--\eqref{eq:methods_readout}. The only modification required to create an energy readout is that the final linear map in \eqref{eq:methods_readout} predicts only scalars.

\subsection{Energy Model Implementation Details}
\label{methods:implementation_energy}

As discussed in the main text, the energy functional in all our models is decomposed into two pieces:
\begin{equation}
    G_{\text{ML},i} = Q_i + U_i
\end{equation}
Where $Q_i$ is a learnable, positive quadratic function of the atomic multipoles (which is the same in all models) and $U_i$ is intended to be a more flexible function of the embedded charge density.

\subsubsection{Positive Quadratic Term} The quadratic term, present in all our energy functional models is defined as
\begin{align}
    H_{i,k} = \text{SoftPlus}\left(\sum_{k'} W_{kk'}\tilde{h}_{i,k'00} + W'_{kz_i}\right) \\
    Q_i = \sum_{kl} \text{SoftPlus}\left(\tilde{W}_{kl}\right) H_{i,k}  \left\| p_{i,l} - p_{i,l}^{(\text{local})} \right\|^2
\end{align}

The $\text{SoftPlus}$ function means that for any value of the learned weights $W$, $W'$ and $\tilde W$, the output is convex in the charges. $\|p_{i,l}\|$ denotes the norm over the $m$ components of a set of atomic multipoles $\{p_{i,lm}\}_{ilm}$. This term is intended to encourage stable models at the beginning of training, but one can see that $Q_i$ can be made arbitrarily small if $\tilde{W}$ takes large negative values. 

\subsubsection{one-body Quadratic}
The simplest function we consider for $U_i$ is formally similar to the quadratic form of QEq. We construct a quadratic form for the embedded charges $V_{i,klm}$ as follows:
\begin{align}
    B_{i,kLM}^{1} &= V_{i,kLM}, \quad B_{i,kLM}^{2} = \sum_{lml'm'} C_{lml'm'}^{LM} V_{i,klm} V_{i,kl'm'} \label{eq:B_basis_funca} \\
    T_{i,kLM} &=  W_{k\tilde{k}} B_{i,\tilde{k}LM}^1 +  W_{k\tilde{k}}' B_{i,\tilde{k}LM}^2
   \label{eq:B_basis_func}
\end{align}
The features $T$ are then combined with the purely local node features in an analogous way to the fixed-point model (equations \eqref{eq:fp_outout_calc_Isa}--\eqref{eq:fp_outout_calc_Isb}):
\begin{align}
    I_{i,k} &= \sum_{k'lm} W_{lkk'}  \tilde{h}_{i,k'lm} T_{i,k'lm} \label{eq:form_invars} \\
    U_i &= \sum_{kk'} W'_{kk'} I_{i,k} \tilde{h}_{i,k'00} 
\label{eq:form_invars_and_readout}
\end{align}
The output $U_i$ is clearly a much more complicated function of the geometry, but can in principle be re-expressed into a quadratic form of the charge density.

\subsubsection{One-body Nonlinear}
To go beyond quadratic while still being one-body in the charge density we apply a non-linearity to the invariant features $I$, exactly as in the fixed-point model.
\begin{equation}
    U_{i} = \text{MLP}(I_{i,k})
    \label{eq:energy-inv-readout}
\end{equation}

\subsubsection{many-body Functional}
Finally, we consider a more flexible functional form which is many-body in the charge descriptors. We have tested several variants that were closely related to the many-body fixed-point layer (section \ref{sec:scf:fixedpoint_manybody_arch}), but found that the computational cost of training such models was extremely large. We therefore present a more simple many-body energy functional by replacing the initial density embedding in equation \eqref{eq:energymodel_rhotilde}, with the following:
\begin{align}
    \tilde{p}_{i,klm}^{\text{neighbours}} = W_{k,z_i} \sum_{j\in \mathcal{N}(i)} p_{j,lm} + W_{k,z_i}' \sum_{j\in \mathcal{N}(i)} p_{j,lm}^{(local)}
\end{align}
Where $\mathcal{N}(i)$ is the set of atoms which are neighbours of atom $i$. This gives the model some information about the charge of neighbouring atoms as well as the central atom's charge. This is then embedded in the same way as \eqref{eq:energymodel_density_embedding} to give a new feature vector denoted $\bar{V}_{i,klm}$:
\begin{align}
    V_{i,kl_3m_3}^{\text{neighbours}} &= \sum_{\tilde{k}} W_{k\tilde{k}} \sum_{l_1m_1l_2m_2} C^{l_3m_3}_{l_1m_1l_2m_2} \tilde{h}_{i,\tilde{k}l_1m_1}\tilde{p}_{i,\tilde{k}l_2m_2}^{\text{neighbours}} \\
    \bar{V}_{i,klm} &= V_{i,klm} + V_{i,klm}^{\text{neighbours}}
\end{align}
In which $V_{i,klm}$ is the original embedding \eqref{eq:energymodel_density_embedding}. This is used in equations \eqref{eq:form_invars}-\eqref{eq:form_invars_and_readout} to give the predicted energy. 

\subsection{Implicit Differentiation}
\label{sec:methods:implicit_diff}

\subsubsection{Fixed-Point Architecture}

Implicit differentiation in the fixed-point model is handled by the TorchOpt package \cite{Ren2023TorchOpt:Optimization}. 

The fixed-point model can naturally be evaluated at constant charge or at constant chemical potential. In constant chemical potential mode, the fixed-point equation that needs to be solved is equation \eqref{eq:fixedpoint_constant_mu_eq}. As explained in the main text, one can compute the Jacobian matrix $  \frac{\partial \mathbf{p}^\star}{\partial \bm{\theta}}$ using equation \eqref{eq:implicit}. A simple fixed-point system such as this is a classic example usage of implicit differentiation, and is easily implemented in the TorchOpt package. 

When operating in constant charge mode, the Fermi level is adjusted to obtain the right charge (see also section \ref{sec:scf:methods:scfloop_details}). This makes the Fermi level one of the implicitly defined parameters, along with the set of atomic multipoles. The new condition that the model is solved is:
\begin{align}
    \bm{f}(\bm{\theta}, \mathbf{p}^\star(\bm{\theta}), \mu^\star(\bm{\theta})) &= \mathbf{0} \label{eq:fixedpoint_eq_costraineda} \\
    \Big(\sum_i p_{i,00}(\bm{\theta})\Big) -Q_{\text{target}}&= 0
\label{eq:fixedpoint_eq_costrainedb}
\end{align}
The first equation is simply the same fixed-point equation, but with the Fermi level as a new parameter, and the second is the charge constraint. Both equations together serve to define $\mathbf{p}^\star$ and $\mu^\star$ for any value $\bm{\theta}$. We can combine these two equations into one. Firstly, let $\bar{\mathbf{p}} = (\mathbf{p}, \mu)$ be a vector which just concatenates the charge density coefficients and the Fermi level. Then define:
\begin{align}
    \bm{f}^\text{(const. Q)}(\bm{\theta}, \bar{\mathbf{p}}^\star(\bm{\theta})) := 
    \begin{pmatrix}
        \bm{f}(\bm{\theta}, \mathbf{p}^\star(\bm{\theta}), \mu^\star(\bm{\theta})) \\[6pt]
        \sum_i p_{i,00}(\bm{\theta}) -Q_{\text{target}}
    \end{pmatrix}
\end{align}
We now simply replace equation \eqref{eq:fixedpoint_constant_mu_eq} with
\begin{align}
    \bm{f}^\text{(const. Q)}(\bm{\theta}, \bar{\mathbf{p}}^\star(\bm{\theta})) = \bm{0},
    \label{eq:fixedpoint_constant_q_eq}
\end{align}
and perform the same algebra as in the main text equations \eqref{eq:fixedpoint_constant_mu_eq}--\eqref{eq:implicit}, but with $\bar{\mathbf{p}}$ replacing $\mathbf{p}$, to get the required Jacobian matrix. 

This is what we do in our implementation, but one could also construct the equations for $\mathbf{p}^\star$ and $\mu^\star$ manually. For instance, starting from \eqref{eq:fixedpoint_eq_costraineda} and \eqref{eq:fixedpoint_eq_costrainedb} which remain solved for all $\bm{\theta}$, one can take the derivative in the same way as equation \eqref{eq:implicit_derivative_1}, and get:
\begin{align}
    0 &= \frac{d}{d \bm{\theta}} f(\bm{\theta}, \mathbf{p}^\star, \mu) = \frac{\partial f}{\partial \bm{\theta}} + \frac{\partial f}{\partial \mathbf{p}^\star}\frac{\partial \mathbf{p}^\star}{\partial \bm{\theta}} + \frac{\partial f}{\partial \mu} \frac{\partial \mu}{\partial \bm{\theta}} \\
    0 &= \frac{d}{d\bm{\theta}} \sum_i p_{i,00}^\star = \sum_i \frac{\partial p_{i,00}^\star}{ \partial \bm{\theta}} = \mathbf{c}^T \frac{\partial \mathbf{p}^\star}{\partial \bm{\theta}}
\end{align}
Where $\mathbf{c}$ is a vector with length equal to the number of charge density coefficients, and which is 1 for coefficients representing partial charges and zero otherwise. The matrix equation then becomes:
\begin{align}
    \begin{pmatrix}
        \frac{\partial \mathbf{p}^\star}{\partial \bm{\theta}} \\[6pt]
        \frac{\partial \mu^\star}{\partial \bm{\theta}}
    \end{pmatrix} = -\begin{pmatrix}
    \frac{\partial f}{\partial \mathbf{p}^\star} & \frac{\partial f}{\partial \mu^\star} \\[6pt]
    \mathbf{c}^T & 0
\end{pmatrix}^{-1} \begin{pmatrix}
        \frac{\partial f}{\partial \bm{\theta}} \\[6pt]
        0
    \end{pmatrix}
\end{align}
This is the same result one gets using the combined equation \eqref{eq:fixedpoint_constant_q_eq}, but equation \eqref{eq:fixedpoint_constant_q_eq} can be implemented more cleanly using TorchOpt.

\subsubsection{Energy Functional Model}

Implicit differentiation for the energy functional is conceptually similar. The solution $\mathbf{p}^\star$ satisfies
\begin{align}
    \bm{f}(\theta, \mathbf{p}^\star) := \frac{\partial}{\partial \mathbf{p}} \Big(E(\bm{\theta}, \mathbf{p})\Big)_{\mathbf{p}=\mathbf{p}^\star} = 0
\end{align}
where $E$ is the model's total energy (local functional contribution and Coulomb energy). One can compute $\bm{f}(\bm{\theta},\mathbf{p})$ easily using backpropagation through $E$. Then, if the total charge is not constrained, the same logic as for the field model (equations \eqref{eq:define_f}--\eqref{eq:implicit}) can be used to derive the equations for computing the gradients of the minimizing charge density with respect to the parameters.

If the total charge is constrained, we again need to modify the equation. It has been shown that the appropriate stationary condition is \cite{NEURIPS2022_228b9279}:
\begin{equation}
    \bm{f}^\text{(energy const. Q)}(\mathbf{p}^\star, \bm{\theta}) := \text{proj}\big(\mathbf{p}^\star - \eta \bm{f}(\theta, \mathbf{p}^\star)\big) - \mathbf{p}^\star = 0
\end{equation}
In which $\eta > 0$ is a step size and $\text{proj}(\cdot)$ is a projection operation that enforces the total charge constraint. One then extracts the gradients through the same matrix algebra as in the main text equations \eqref{eq:define_f}--\eqref{eq:implicit}, using $\bm{f}^\text{(energy const. Q)}$  in place of $\bm{f}$.

\subsection{SC-loop Implementation Details}
\label{sec:scf:methods:scfloop_details}
\subsubsection{Mixing Schemes for Fixed-Point Iterations}

We use $F_\text{ML}$ to denote the map from a vector of electric potential features, to a vector of atomic multipoles, $\mathbf{p} = F_\text{ML}(\mathbf{v})$, and $\mathbf{v}[\mathbf{p}, \mu]$ to be the potential features arising from multipoles $\mathbf{p}$, and the Fermi level $\mu$. Explicitly, $\mathbf{v}[\mathbf{p}, \mu]$ is
\begin{align*}
    \mathbf{p} &\rightarrow \rho(\mathbf{r}) \\
    v_\text{eff}(\mathbf{r}) &= \int \frac{\rho(\mathbf{r}')}{|\mathbf{r} - \mathbf{r}'|}d\mathbf{r}' + v_\text{app}(\mathbf{r}) + \mu \\
    v_\text{eff}(\mathbf{r}) &\rightarrow \mathbf{v} \ .
\end{align*}
The first and last lines refer to equations  \eqref{eq:atomic_multipole_expansion} and \eqref{eq:field_featurization}. The fixed-point cycle can be converged in two ways. Firstly, at constant chemical potential, which means treating $\mu$ as a constant and allowing the total charge to converge to an arbitrary value. Secondly, at constant charge, where $\mu$ is adjusted within the SC cycle to obtain the desired charge $Q_{target}$. In the former case, we use a simple linear update with mixing parameter $\lambda$:
\begin{align}
    \Delta &:= F_{\text{ML}}(\mathbf{v}[\mathbf{p}^{(t)}, \mu]) - \mathbf{p}^{(t)}\\
    \mathbf{p}^{(t+1)} &= \mathbf{p}^{(t)} + \lambda \Delta
\end{align}
Finally, we use a termination criteria based on the average absolute change in the atomic multipole moments between steps with a tolerance $\varepsilon_{\text{MAE}}$. The default value of this is $10^{-6}$. The overall algorithm is shown in Algorithm \ref{alg:scf_fixed_mu}.

For the constant charge case, the Fermi level also needs to be adjusted. Let $Q^{(t)}$ denote the total predicted charge at step $t$. The objective is to keep the number of evaluations of $\mathbf{v}[\mathbf{p}, \mu]$ and $F_\text{ML}$ to a minimum while finding the correct $\mu$ so that $Q^{(t)} = Q_{\text{target}}$. We can leverage automatic differentiation to do this. Firstly, one proceeds as in the constant Fermi level case:
\begin{align}
    \Delta &= F_{\text{ML}}(\mathbf{v}[\mathbf{p}^{(t)}, \mu^{(t)}]) - \mathbf{p}^{(t)} \\
    \mathbf{p}^{(t+1)} &= \mathbf{p}^{(t)} + \lambda \Delta \label{eq:methods_pred_p_plus_1} \\
    Q^{(t+1)} &= \sum_i p^{(t+1)}_{i,00}
\end{align}
Following this, backpropagation is used to find the gradient of each atom's charge with respect to the Fermi level at the previous step. One can compute
\begin{align}
   g_i &:= \frac{\partial p_{i,00}^{(t+1)}}{\partial \mu^{(t)}} \label{eq:methods_def_gi} \\
    \frac{\partial Q^{(t+1)}}{\partial \mu^{(t)}} &= \sum_i g_i, \label{eq:methods_total_q_by_mu}
\end{align}
and then make a first order estimate to the required change in Fermi level:
\begin{align}
    \Delta\mu &= (Q_{\text{target}} - Q^{(t+1)}) / \frac{\partial Q^{(t+1)}}{\partial \mu^{(t)}} \label{eq:methods_delta_mu} \\
    \mu^{(t+1)} &= \mu^{(t)} + \text{clip}(\Delta\mu) \label{eq:methods_update_mu}
\end{align}
Here, $\text{clip}(\Delta\mu)$ equals $\Delta\mu$ when $-1 \leq \Delta\mu \leq 1$, and is clipped to $-1$ or $1$ otherwise.

Finally, we found that convergence is generally easier if one only ever computes the electrostatic features using multipoles which sum to the correct total charge. To this end, when computing a new set of multipoles $\mathbf{p}$ as in \eqref{eq:methods_pred_p_plus_1}, we also compute a normalized version $\mathbf{p}'$ by redistributing excess charge using a scheme inspired by the AIMnet-NSE equilibration method. In AIMnet-NSE, one distributes excess charge between all atoms according to machine learned values $f_i$ on each atom \cite{aimnet_nse}. Here, rather than predicting new values $f_i$, we use the actual sensitivity of the charge on each atom to the Fermi level, which is just $g_i$. Specifically, the normalized atomic multipoles $\mathbf{p}'$ are computed as:
\begin{align}
    p_{i,00}'^{(t+1)} &= p_{i,00}^{(t+1)} + (Q_{target} - Q^{(t+1)} )\frac{g_i}{\sum_j g_j} \\
    p_{i,lm}'^{(t+1)} &= p_{i,lm}^{(t+1)} \quad l,m \neq 0
\end{align}
 We will denote this operation as  $\mathbf{p}'=\mathrm{SpreadCharge}(\mathbf{p}, \mathbf{g})$. This step ensures that at each SC step, even before we have found the correct $\mu$, the sum of the charges in $\mathbf{p}'$ is the correct value. This greatly helps stability especially for models at the start of training, since the electric field features are never computed from nonsensical charges. It does not change the result or meaning of the SC-cycle, since the Fermi level still needs to be found for which $Q_{\text{target}} = Q^{(t)}$, after which this re-distribution of excess charge has no effect.

With this trick included, the overall algorithm for constant charge evaluations is shown in Algorithm \ref{alg:scf_fixed_charge}. The same convergence criterion is used for terminating the SCF loop.

\begin{figure}
\begin{algorithm}[H]
\caption{SCF iteration at fixed chemical potential}
\label{alg:scf_fixed_mu}
\begin{algorithmic}[1]
\Require $\mathbf{p}^{\text{(initial)}}$, $\mu$, $\lambda$, $\varepsilon_\text{MAE}$

\AlgRule
\AlgSection{Initialization}
\AlgRule
\State $\mathbf{p}^{(0)} \gets \mathbf{p}^{\text{(initial)}}$
\State $t \gets 0$
\State $\delta^{(0)} \gets \infty$

\AlgRule
\AlgSection{Self-consistent loop}
\AlgRule
\While{$\delta^{(t)} > \varepsilon_\text{MAE}$}
    \Statex \hspace{\algorithmicindent} \textit{Update with mixing}
    \State $\mathbf{v}^{(t)} \gets \mathbf{v}[\mathbf{p}^{(t)},\mu]$
    \State $\Delta^{(t)} \gets F_\text{ML}(\mathbf{v}^{(t)}) - \mathbf{p}^{(t)}$
    
    \State $\mathbf{p}^{(t+1)} \gets \mathbf{p}^{(t)} + \lambda \Delta^{(t)}$
    \Statex 
    \Statex \hspace{\algorithmicindent} \textit{Check Convergence}
    \State $\delta^{(t+1)} \gets \frac{1}{N_{\mathbf p}} \left|\mathbf{p}^{(t+1)}-\mathbf{p}^{(t)}\right|$
    \State $t \gets t+1$
\EndWhile

\AlgRule
\State \Return $\mathbf{p}^{(t)}$
\end{algorithmic}
\end{algorithm}
\end{figure}

\begin{figure}
\begin{algorithm}[H]
\caption{SCF iteration at fixed total charge}
\label{alg:scf_fixed_charge}
\begin{algorithmic}[1]
\Require $\mathbf{p}^\text{(initial)}$, $\mu^\text{(initial)}$, $Q_\text{target}$, $\lambda$, $\varepsilon_\text{MAE}$

\AlgRule
\AlgSection{Initialization}
\AlgRule
\State $\mathbf{p}^{(0)} \gets \mathbf{p}^{\text{(initial)}}$
\State $\mu^{(0)} \gets \mu^{\text{(initial)}}$
\State $\delta^{(0)} \gets \infty$
\State $t \gets 0$
\Statex
\AlgSubsection{Normalised starting guess for use in $\mathbf{v}$}
\State $\mathbf{v}^{(0)} \gets \mathbf{0} + \mu^{(0)}$
\State $\mathbf{g}^{(0)} \gets \mathrm{BackProp}(F_\text{ML}(\mathbf{v}^{(0)}), \mu^{(0)})$
\State $\mathbf{p}'^{(0)} \gets \mathrm{SpreadCharge}(\mathbf{p}^{(0)}, \mathbf{g}^{(0)})$

\AlgRule
\AlgSection{Self-consistent loop}
\AlgRule
\While{$\delta^{(t)} > \varepsilon_\text{MAE}$}
    \Statex \hspace{\algorithmicindent} \textit{Update with mixing}
    \State $\mathbf{v}^{(t)} \gets \mathbf{v}[\mathbf{p}'^{(t)},\mu^{(t)}]$
    \State $\Delta^{(t)} \gets F_\text{ML}(\mathbf{v}^{(t)}) - \mathbf{p}^{(t)}$
    \State $\mathbf{p}^{(t+1)} \gets \mathbf{p}^{(t)} + \lambda \Delta^{(t)}$
    \State $\Delta Q^{(t+1)} \gets \sum_i p^{(t+1)}_{i,00} - Q_\text{target}$

    \Statex
    \Statex \hspace{\algorithmicindent} \textit{Normalised version of multipoles for use in $\mathbf{v}$}
    \State $\mathbf{g}^{(t+1)} \gets \mathrm{BackProp}(\mathbf{p}^{(t+1)}, \mu^{(t)})$
    \State $\mathbf{p}'^{(t+1)} \gets \mathrm{SpreadCharge}(\mathbf{p}^{(t+1)}, \mathbf{g}^{(t+1)})$

    \Statex
    \Statex \hspace{\algorithmicindent} \textit{Fermi level update}
    \State Compute $\Delta \mu^{(t)}$ from $\Delta Q^{(t+1)}$ and $\mathbf{g}^{(t+1)}$ using Eqs.~\eqref{eq:methods_total_q_by_mu} and \eqref{eq:methods_delta_mu}
    \State $\mu^{(t+1)} \gets \mu^{(t)} + \operatorname{clip}\!\left(\Delta\mu^{(t)}\right)$

    \Statex 
    \Statex \hspace{\algorithmicindent} \textit{Check Convergence}
    \State $\delta^{(t+1)} \gets \frac{1}{N_{\mathbf p}} \left|\mathbf{p}^{(t+1)}-\mathbf{p}^{(t)}\right|$
    \State $t \gets t+1$

\EndWhile
\AlgRule
\State \Return $\mathbf{p}^{(t)}, \mu^{(t)}$
\end{algorithmic}
\end{algorithm}
\end{figure}

\subsubsection{Trust-Region Newton Conjugate Gradient Minimizer}

All energy minimizations were performed with a pytorch reimplementation of the scipy \cite{Virtanen2020SciPyPython} trust-region newton conjugate gradient approach \cite{Nocedal2006NumericalOptimization}.

\subsection{Density Functional Theory Calculations}
\label{sec:scf:methods:dft_settings}

In both the metal--water and \ce{SiO2} examples, we used FHI-aims for DFT calculations \cite{Blum2009AbOrbitals} with the PBE exchange correlation functional \cite{PBE}. Tight basis set and integration grid settings were used throughout.

The metal--water calculations were not spin polarized and used a dipole correction for the slab geometries. An (in-plane) k-point density of 6 /\AA$^{-1}$ was used for the water slabs, and a density of 16 /\AA$^{-1}$ was used for the aluminium slabs. 

For silicon dioxide, spin polarized calculations were performed whenever the total charge was -1 or 1, and non-spin polarized calculations were used for a total charge of 0. For the singly charged calculations, the projected spin was constrained to 1 unpaired electron. For relaxations of the doubly charged supercells, we used spin polarized calculations for both -2 and +2 overall charge, with the total spin constrained to 2 unpaired electrons, with the exception of the relaxations discussed in table \ref{table:-2_defects} where the settings are stated in the caption. A k-point density of 4 /\AA$^{-1}$ was used for all \ce{SiO2} calculations. 

\subsection{Extracting DFT Atomic Multipoles}
\label{sec:scf:methods:aims_multipoles}

Throughout this project, reference atomic multipoles were computed from the DFT reference data. The electron density can be partitioned into atom-centred contributions in the following way:
\begin{align}
    n_i(\mathbf{r}) = n(\mathbf{r}) u_i(\mathbf{r})
\end{align}
In which $n(r)$ is the electron density, $n_i$ is the contribution assigned to atom $i$ and $\{u_i(r)\}_i$ is a set of partition functions which have the property $\sum_i u_i(r) = 1$. This allows one to coarse-grain the electron density $n(\mathbf{r})$ into atomic multipole moments:
\begin{align*}
    p_{i,lm} = Z_i\delta_{l0} - \sqrt{\frac{4\pi}{2l+1}}\int |\mathbf{r}-\mathbf{r}_i|^{l} Y_{lm}(\widehat{\mathbf{r} - \mathbf{r}_i}) \cdot u_{i}(\mathbf{r}) n(\mathbf{r})  \mathrm{d}\mathbf{r}.
\end{align*}
$p_{i,lm}$ is then the multipole moment of the $i$'th atom. All electronic structure calculations in this project used the FHI-aims DFT code. In FHI-aims, the long-range part of the Hartree potential at each SC step is represented using partial atomic multipoles in exactly the way described above for a certain choice of weighting function $u_i$. The weighting functions are defined as $u_i(\mathbf{r}) = g_i(\mathbf{r}) / (\sum_i g_i(\mathbf{r}))$ where $g$ depends only on the distance to atom $i$. In this work, the default choice of $g_i$ implemented in FHI-aims was used, which is described in reference \cite{Stratmann1996AchievingQuadratures}. 

To assign atomic multipoles, the projected atomic multipoles were extracted from the last SC step. We have not seen different behaviour when instead using a Hirshfeld scheme for the partitioning functions $g_i$.

\clearpage
\appendix

\section{Appendix: Coarse-Graining the HK Energy Functional}
\label{appendix:cg_hkdft}

The energy-functional model used in the main text assigns an energy to a
coarse-grained charge density and minimizes that energy with respect to its coefficients. In this appendix, we use notation where the nuclear positions and applied fields are added as arguments, to be more explicit. The model has the form
\begin{align}
    \mathcal E_{\mathrm{model}}(\mathbf R;v_{\mathrm{app}})
    &=\min_{\mathbf p}\big[
    E_{\mathrm{local}}(\mathbf R)
    +G_{\mathrm{ML}}(\mathbf R,\mathbf p)\nonumber \\
    &+E_{\mathrm{Coulomb}}^{\mathrm{CG}}
    (\mathbf p;\mathbf R,v_{\mathrm{app}})\big].
    \label{eq:app_model_functional_intro}
\end{align}
The minimization is carried out subject to a total charge constraint. Here $E_{\mathrm{local}}$ is an extensive function of the local geometry,
$G_{\mathrm{ML}}$ is a learned functional of the geometry and charge
coefficients, and $E_{\mathrm{Coulomb}}^{\mathrm{CG}}$ describes long-range
electrostatics.
The purpose of this appendix is to derive this separation from the
Hohenberg--Kohn energy.

The derivation proceeds in four stages.  First, we define a
population-constrained DFT energy as a function of the coarse coefficients
$\mathbf p$.  Second, we expand that energy and its minimizing density about a
reference state.  Third, we state conditions under which the coarse density
reproduces the long-range electrostatic response of the DFT density.  Finally,
we use those conditions to collect all long-range terms into the explicit
coarse-grained Coulomb energy in Eq.~\eqref{eq:app_model_functional_intro}.
The electrostatic terms left over are local functions of the reference state
and its response, and are therefore absorbed into $E_{\mathrm{local}}$ and
$G_{\mathrm{ML}}$.

The Taylor expansions below are formal local expansions within a smooth sector
of the population-constrained energy landscape. We also assume that any
fixed-total-charge constraint has been imposed, so that variations
$\delta\mathbf p$ lie in the allowed charge-conserving subspace.

\subsection{Fine-grained and coarse-grained descriptions}

We begin by defining the fine-grained DFT energy at fixed nuclear geometry.
Let $\mathbf R=\{\mathbf R_i\}_i$ and $Z_i$ denote the nuclear positions and
charges.  We define the external potential for electrons, as well as the nuclear and applied fields parts as in the main text:
\begin{align*}
    - v_{\mathrm{ext}}^e = v_{\mathrm{ext}} = v_{\mathrm{app}} + v_{\mathrm{nuc}}
\end{align*}
The nuclear charge density, nuclear potential, and internuclear
energy are
\begin{align}
    \nu_{\mathbf R}(r)
    &=\sum_i Z_i\delta(r-\mathbf R_i),
    \label{eq:app_nuclear_density}\\
    v_{\mathrm{nuc}}(r;\mathbf R)
    &=\int\frac{\nu_{\mathbf R}(r')}
    {|r-r'|}\,dr',
    \label{eq:app_nuclear_potential}\\
    E_{\mathrm{nn}}(\mathbf R)
    &=\frac12\sum_{i\ne j}
    \frac{Z_iZ_j}{|\mathbf R_i-\mathbf R_j|}.
    \label{eq:app_enn}
\end{align}
The electron density $n(r)$ is positive by convention.  It is useful
to introduce the corresponding total charge density
\begin{equation}
    \widetilde n(r)=\nu_{\mathbf R}(r)-n(r),
    \label{eq:app_total_charge}
\end{equation}
which is positive for positive physical charge.  We write
\begin{align}
    E_{\mathrm H}[a]
    &=\frac12\iint\frac{a(r)a(r')}
    {|r-r'|}\,dr\,dr',
    \label{eq:app_hartree_definition}\\
    v_{\mathrm H}[a](r)
    &=\int\frac{a(r')}{|r-r'|}\,dr'
    \label{eq:app_hartree_potential}
\end{align}
for the Hartree functional and its potential.

Let $v_{\mathrm{app}}$ be an applied electrostatic potential defined for a
positive test charge.  Throughout this appendix, ``Coulomb energy'' denotes
the complete electrostatic contribution, including coupling to
$v_{\mathrm{app}}$.  The Born--Oppenheimer energy is separated as
\begin{equation}
    E_{\mathrm{tot}}^{\mathrm{DFT}}[n;\mathbf R,v_{\mathrm{app}}]
    =G[n]+E_{\mathrm{Coulomb}}^{\mathrm{DFT}}
    [n;\mathbf R,v_{\mathrm{app}}],
    \label{eq:app_bo_split}
\end{equation}
where $G[n]$ contains the kinetic and exchange--correlation contributions and
\begin{align}
    E_{\mathrm{Coulomb}}^{\mathrm{DFT}}
    [n;\mathbf R,v_{\mathrm{app}}]
    ={}&E_{\mathrm H}[n] + \int v_{\mathrm{ext}}^e(r)n(r)\,dr \nonumber \\
    &+ E_{\mathrm{nn}}(\mathbf R) + \int \nu_{\mathbf R}(r) v_{\mathrm{app}}(r)\,dr \nonumber\\
    =E_{\mathrm H}[n]&
    -\int v_{\mathrm{nuc}}(r;\mathbf R)n(r)\,dr
    +E_{\mathrm{nn}}(\mathbf R)
    \nonumber\\
    &+\int v_{\mathrm{app}}(r)
    [\nu_{\mathbf R}(r)-n(r)]\,dr.
    \label{eq:app_dft_coulomb}
\end{align}

In order to coarse grain the DFT energy, we introduce a mapping from the DFT charge density to a set of coarse-grained coefficients:
\begin{equation}
    \mathcal P_{\mathbf R}:\widetilde n\longmapsto\mathbf p,
    \qquad \mathbf p=\{p_k\}_k.
    \label{eq:app_projection}
\end{equation}
This map is generally many-to-one: different fine-grained densities may have
the same coarse coefficients.  Conversely, the model reconstructs a charge
density from $\mathbf p$ according to
\begin{equation}
    \rho_{\mathbf p}(r;\mathbf R)
    =\sum_k p_k\phi_k(r;\mathbf R).
    \label{eq:app_reconstruction}
\end{equation}
Our choice of basis functions $\phi_k$ does not need to have any connection to the charge assignment $\mathcal{P}$. At fixed geometry, we suppress the repeated $\mathbf R$ argument on
$\rho_{\mathbf p}$ and $\phi_k$ below. In general
$\rho_{\mathbf p}\ne\widetilde n$; the role of $\rho_{\mathbf p}$ is to retain only the long-range electrostatic information needed by the model.

We bring forward one part of the conditions discussed below because it is
important for the definitions that follow. For every admissible fine-grained
charge density with $\mathcal P_{\mathbf R}[\widetilde n]=\mathbf p$, we
require that the coarse-grained density reproduces its coupling to the class
of applied fields considered:
\begin{equation}
    \int \widetilde n(r)v_{\mathrm{app}}(r)\,dr
    =\int \rho_{\mathbf p}(r)v_{\mathrm{app}}(r)\,dr.
    \label{eq:app_field_coupling_condition}
\end{equation}
For a constant potential and a uniform electric field, this condition is met
when the total charge and dipole of the coarse-grained density match those of
the full charge density.

We can now define the DFT population-constrained energy by minimizing over all electron
densities which map to $\mathbf p$:
\begin{align}
    E_{\mathrm{pop}}^{\mathrm{DFT}}
    (\mathbf p;\mathbf R,v_{\mathrm{app}})
    &=\min_{n\mapsto\mathbf p}
    E_{\mathrm{tot}}^{\mathrm{DFT}}[n;\mathbf R,v_{\mathrm{app}}],
    \label{eq:app_constrained_energy}\\
    n[\mathbf p;\mathbf R,v_{\mathrm{app}}]
    &=\operatorname*{arg min}_{n\mapsto\mathbf p}
    E_{\mathrm{tot}}^{\mathrm{DFT}}[n;\mathbf R,v_{\mathrm{app}}].
    \label{eq:app_constrained_density}
\end{align}
Here $n\mapsto\mathbf p$ means
$\mathcal P_{\mathbf R}[\widetilde n]=\mathbf p$, as in the main text.  Thus
$E_{\mathrm{pop}}^{\mathrm{DFT}}(\mathbf p;\mathbf R,v_{\mathrm{app}})$ is an
ordinary function of the
coarse variables, even though it is obtained from a minimization over the full
electron density. The population-constrained DFT energy is the central object which we want to approximate with the model's coarse-grained energy functional. The density
$n[\mathbf p;\mathbf R,v_{\mathrm{app}}]$ describes how the fine-grained
electronic state changes when one of the coarse constraints is varied.

Equation~\eqref{eq:app_field_coupling_condition} means that the applied-field
term is constant during the constrained minimization. On the constraint
surface $n\mapsto\mathbf p$,
\begin{align}
    E_{\mathrm{tot}}^{\mathrm{DFT}}[n;\mathbf R,v_{\mathrm{app}}]
    =E_{\mathrm{tot}}^{\mathrm{DFT}}[n;\mathbf R,0]
    +\int \rho_{\mathbf p}(r)v_{\mathrm{app}}(r)\,dr.
    \label{eq:app_field_energy_split}
\end{align}
It follows that
\begin{align}
    E_{\mathrm{pop}}^{\mathrm{DFT}}
    (\mathbf p;\mathbf R,v_{\mathrm{app}})
    &=E_{\mathrm{pop}}^{\mathrm{DFT}}(\mathbf p;\mathbf R,0)
    +\int \rho_{\mathbf p}(r)v_{\mathrm{app}}(r)\,dr,
    \label{eq:app_field_factorization}\\
    n[\mathbf p;\mathbf R,v_{\mathrm{app}}]
    &=n[\mathbf p;\mathbf R,0]\equiv n[\mathbf p;\mathbf R].
    \label{eq:app_field_independent_density}
\end{align}
Thus the constrained density, and hence $G[n[\mathbf p;\mathbf R]]$, do not
have a separate dependence on the applied field.

Finally, define the coarse-grained Coulomb energy that will be retained explicitly in the model:
\begin{align}
    E_{\mathrm{Coulomb}}^{\mathrm{CG}}
    (\mathbf p;\mathbf R,v_{\mathrm{app}})
    =E_{\mathrm H}[\rho_{\mathbf p}]
    +\int v_{\mathrm{app}}(r)
    \rho_{\mathbf p}(r)\,dr.
    \label{eq:app_cg_coulomb}
\end{align}
At fixed $\mathbf R$ and $v_{\mathrm{app}}$, this is a function of the model's charge density coefficients $\mathbf p$. 
Combining Eqs.~\eqref{eq:app_field_factorization} and
\eqref{eq:app_cg_coulomb} also gives
\begin{align}
    &E_{\mathrm{pop}}^{\mathrm{DFT}}
    (\mathbf p;\mathbf R,v_{\mathrm{app}})
    -E_{\mathrm{Coulomb}}^{\mathrm{CG}}
    (\mathbf p;\mathbf R,v_{\mathrm{app}}) \nonumber\\
    &\qquad=E_{\mathrm{pop}}^{\mathrm{DFT}}(\mathbf p;\mathbf R,0)
    -E_{\mathrm H}[\rho_{\mathbf p}],
    \label{eq:app_field_independent_remainder}
\end{align}
so all applied-field dependence is contained in the explicit coarse-grained
Coulomb energy.

\subsection{Taylor expansion about a reference state}

Choose a set of reference charge-density coefficients $\mathbf p^0$, and consider perturbations about this point:
\begin{equation}
    \mathbf p=\mathbf p^0+\delta\mathbf p,
    \qquad n^0=n[\mathbf p^0;\mathbf R],
    \qquad \rho^0=\rho_{\mathbf p^0}.
    \label{eq:app_reference}
\end{equation}

Now consider the DFT energy in \eqref{eq:app_bo_split}. We can evaluate each part of the DFT energy at the population-constrained density, making $G$ and $E_{\mathrm{Coulomb}}^{\mathrm{DFT}}$ individually functions of the charge coefficients $\mathbf p$:
\begin{align}
    G(\mathbf p;\mathbf R)
    &= G[n[\mathbf p;\mathbf R]],
    \label{eq:app_G_of_p}\\
    E_{\mathrm{Coulomb}}^{\mathrm{DFT}}
    (\mathbf p;\mathbf R,v_{\mathrm{app}})
    &= E_{\mathrm{Coulomb}}^{\mathrm{DFT}}
    [n[\mathbf p;\mathbf R];\mathbf R,v_{\mathrm{app}}].
    \label{eq:app_Ec_of_p}
\end{align}
The energy change given a perturbation $\delta \mathbf p$ can then be written,
\begin{align}
    E_{\mathrm{pop}}^{\mathrm{DFT}}
    (\mathbf p^0+\delta\mathbf p;\mathbf R,v_{\mathrm{app}})
    &=E_{\mathrm{pop}}^{\mathrm{DFT}}
    (\mathbf p^0;\mathbf R,v_{\mathrm{app}})\nonumber \\
    &+\Delta E_{\mathrm{Coulomb}}^{\mathrm{DFT}}\nonumber \\
    &+\Delta G(\mathbf R,\mathbf p^0,\delta\mathbf p),
    \label{eq:app_energy_taylor}
\end{align}
after which we can Taylor expand each piece:
\begin{align}
    \Delta E_{\mathrm{Coulomb}}^{\mathrm{DFT}}
    ={}&\sum_k\delta p_k v_k^0
    +\frac12\sum_{kl}\delta p_k\delta p_l(C_{kl}+v_{kl}^0)
    \nonumber\\
    &+\frac{1}{3!}\sum_{klm}\delta p_k\delta p_l\delta p_m
    (C_{klm}+v_{klm}^0)+\cdots,
    \label{eq:app_delta_coulomb_series}\\
    \Delta G(\mathbf R,\mathbf p^0,\delta\mathbf p)
    ={}&\sum_k\delta p_k\overline\eta_k
    +\frac12\sum_{kl}\delta p_k\delta p_l\overline\eta_{kl}
    \nonumber\\
    &+\frac{1}{3!}\sum_{klm}\delta p_k\delta p_l\delta p_m
    \overline\eta_{klm}+\cdots.
    \label{eq:app_delta_G}
\end{align}
The coefficients in these two series will be derived below. Our objective is to rewrite the Taylor series for the DFT Coulomb energy in terms of the coarse-grained, model Coulomb energy, plus short-range corrections. To do so, we must also expand the population-constrained density, $n[\mathbf p]$, that enters both functionals.
A change in the population-constrained density is:
\begin{align}
    n[\mathbf p^0+\delta\mathbf p;\mathbf R](r)
    ={}&n^0(r)+\sum_k\delta p_k f_k(r)
    \nonumber\\
    &+\frac12\sum_{kl}\delta p_k\delta p_l f_{kl}(r)
    +\cdots,
    \label{eq:app_density_taylor}
\end{align}
where
\begin{align}
    f_k(r)
    &=\left.\frac{\partial n[\mathbf p;\mathbf R](r)}
    {\partial p_k}
    \right|_{\mathbf p^0},
    \label{eq:app_fukui_first}\\
    f_{k_1\ldots k_m}(r)
    &=\left.\frac{\partial^m
    n[\mathbf p;\mathbf R](r)}
    {\partial p_{k_1}\cdots\partial p_{k_m}}
    \right|_{\mathbf p^0}
    \label{eq:app_fukui_higher}
\end{align}
are the AIM Fukui function and its higher-order generalizations.  These
functions describe the full DFT density response to a change in the coarse-grained charge constraint. Since the nuclei are fixed during this response,
$\delta\widetilde n=-\delta n$.  We therefore also define
\begin{equation}
    \widetilde f_{k_1\ldots k_m}=-f_{k_1\ldots k_m},
    \label{eq:app_charge_fukui}
\end{equation}
so that $\widetilde f$ has the sign convention of a physical charge density.

\subsection{Derivatives of the constrained energy}

For $G$, the first derivative follows directly from the functional chain
rule:
\begin{equation}
    \frac{\partial G(\mathbf p;\mathbf R)}{\partial p_k}
    =\int\frac{\delta G}{\delta n(r)}
    f_k(r)\,dr.
    \label{eq:app_g_first}
\end{equation}
Taking a second derivative acts both on the functional derivative of $G$ and
on $f_k$.  This gives
\begin{align}
    \frac{\partial^2G(\mathbf p;\mathbf R)}{\partial p_k\partial p_l}
    ={}&\iint\frac{\delta^2G}
    {\delta n(r)\delta n(r')}
    f_k(r)f_l(r')\,dr\,dr'
    \nonumber\\
    &+\int\frac{\delta G}{\delta n(r)}
    f_{kl}(r)\,dr.
    \label{eq:app_g_second}
\end{align}
The same pattern continues at higher orders: functional derivatives of $G$
are contracted with products of ordinary and higher-order AIM Fukui
functions.  The general result is the functional Fa\`a di Bruno formula
\begin{align}
    \frac{\partial^mG(\mathbf p;\mathbf R)}
    {\partial p_{k_1}\cdots\partial p_{k_m}}
    ={}&\sum_{\pi\in\Pi_m}\int\!\cdots\!\int
    \frac{\delta^{|\pi|}G}
    {\delta n(r_1)\cdots\delta n(r_{|\pi|})}
    \nonumber\\
    &\times\prod_{B\in\pi}f_{k_B}(r_B)
    \,dr_1\cdots dr_{|\pi|}.
    \label{eq:app_faa_di_bruno}
\end{align}
Here $\Pi_m$ is the set of partitions of $\{1,\ldots,m\}$, $B$ is a block
in the partition, and $k_B$ denotes the indices collected in that block.  For
example, the two partitions of $\{1,2\}$ generate the two terms in
Eq.~\eqref{eq:app_g_second}.  We use the shorthand
\begin{equation}
    \overline\eta_{k_1\ldots k_m}
    \equiv\left.\frac{\partial^mG(\mathbf p;\mathbf R)}
    {\partial p_{k_1}\cdots\partial p_{k_m}}
    \right|_{\mathbf p^0}.
    \label{eq:app_eta_bar}
\end{equation}

The derivatives of the DFT Coulomb energy simplify because its dependence on
$n$ is at most quadratic.  At the reference state,
\begin{align}
    \frac{\delta E_{\mathrm{Coulomb}}^{\mathrm{DFT}}}
    {\delta n(r)}
    &=v_{\mathrm H}[n^0](r)
    -v_{\mathrm{nuc}}(r;\mathbf R)-v_{\mathrm{app}}(r),
    \label{eq:app_dft_coulomb_first_fd}\\
    \frac{\delta^2E_{\mathrm{Coulomb}}^{\mathrm{DFT}}}
    {\delta n(r)\delta n(r')}
    &=\frac{1}{|r-r'|},
    \label{eq:app_dft_coulomb_second_fd}
\end{align}
and all higher functional derivatives vanish.  Applying the same chain rule
as above, the first derivative is
\begin{align}
    \left.\frac{\partial E_{\mathrm{Coulomb}}^{\mathrm{DFT}}}
    {\partial p_k}\right|_{\mathbf p^0}
    ={}&\int[v_{\mathrm H}[n^0](r)
    -v_{\mathrm{nuc}}(r;\mathbf R)-v_{\mathrm{app}}(r)]
    f_k(r)\,dr
    \nonumber\\
    \equiv{}&v_k^0.
    \label{eq:app_vk_definition}
\end{align}
At second order, there are two contributions:
\begin{align}
    \left.\frac{\partial^2E_{\mathrm{Coulomb}}^{\mathrm{DFT}}}
    {\partial p_k\partial p_l}\right|_{\mathbf p^0}
    ={}&\iint\frac{f_k(r)f_l(r')}
    {|r-r'|}\,dr\,dr'
    \nonumber\\
    +\int[v_{\mathrm H}[n^0](r)&
    -v_{\mathrm{nuc}}(r;\mathbf R)-v_{\mathrm{app}}(r)]
    f_{kl}(r)\,dr
    \nonumber\\
    \equiv{}&C_{kl}+v_{kl}^0.
    \label{eq:app_second_coulomb_response}
\end{align}
Thus $C_{kl}$ denotes the Coulomb interaction between two first-order DFT
responses, whereas $v_{kl}^0$ couples the second-order response to the
reference potential.

For completeness, the same reasoning gives, for $m>2$,
\begin{align}
    \left.\frac{\partial^mE_{\mathrm{Coulomb}}^{\mathrm{DFT}}}
    {\partial p_{k_1}\cdots\partial p_{k_m}}
    \right|_{\mathbf p^0}
    =v_{k_1\ldots k_m}^0+C_{k_1\ldots k_m},
    \label{eq:app_higher_coulomb_response}
\end{align}
where
\begin{align}
    v_{k_1\ldots k_m}^0
    ={}&\int[v_{\mathrm H}[n^0](r)
    -v_{\mathrm{nuc}}(r;\mathbf R)-v_{\mathrm{app}}(r)]
    f_{k_1\ldots k_m}(r)\,dr,
    \label{eq:app_higher_v}\\
    C_{k_1\ldots k_m}
    ={}&\iint\frac{1}{|r-r'|}
    \sum_{(K,K^c)\in\Pi_{m,2}}
    f_{k_K}(r)f_{k_{K^c}}(r')
    \,dr\,dr'.
    \label{eq:app_higher_C}
\end{align}
Here $\Pi_{m,2}$ is the set of partitions of $\{1,\ldots,m\}$ into two
nonempty blocks.  Every term in Eq.~\eqref{eq:app_higher_C} is therefore a
Coulomb interaction between two density responses whose combined indices are
$k_1,\ldots,k_m$.

Equations~\eqref{eq:app_vk_definition}--\eqref{eq:app_higher_C} supply the
coefficients used in the DFT Coulomb series
Eq.~\eqref{eq:app_delta_coulomb_series}.  We will now determine which
parts of that series are reproduced by the explicit coarse-grained Coulomb
energy and which parts are short-ranged.

\subsection{Conditions for electrostatic coarse-graining}

For the following sections, we will introduce two definitions for `local' functions and functionals. Consider a point $r$ and a length scale $R$. Firstly, a scalar function $g$ is localised around $r_1$ if $g(r)=0$ whenever $|r-r_1| > R$. Secondly, take a functional $h[n](r)$ which depends on a function $n(r)$. We will say that $h$ is local with respect to $n$ if $h[n](r)$ only depends on $n(r')$ when $|r'-r| < R$. In other words, $h[n](r)$ can be determined from just the shape of $n$ near to $r$.

We require three conditions.  The first concerns the reference density, the second concerns its first-order response, and the third concerns higher-order
responses.

\subsubsection{Condition 1: Fidelity of the Coarse-Grained Reference Density}

Firstly, we need the coarse-grained density $\rho$ to properly represent the long-range part of the electric potential. This means that at a point $r$, the electric potential due to the full DFT charge density, $\widetilde n(r) = \nu_{\mathbf R}(r) - n(r)$, can be expressed using $\rho$ to account for all distant atoms, plus a local correction which accounts for only nearby atoms and density. We will write this mathematically as:
\begin{align}
    v_{\mathrm H}[\widetilde n^0](r) = v_{\mathrm H}[\rho^0](r) + \Delta[\widetilde n^0](r) \label{eq:appendix_density_condition}
\end{align}
Thus $\Delta[\widetilde n^0](r)$ is the difference between the full potential and that from the coarse-grained density. We require that $\Delta[\widetilde n^0](r)$ is a local functional of $\widetilde n^0$. In other words, we can reconstruct the full potential at $r$ with knowledge of $\rho^0$ everywhere, but the electron density $n^0(r)$ and the nuclear positions only nearby to $r$. This statement is equivalent to the one in the main text where we explicitly split the charge density into `near' and `far' components.

Secondly, Condition 1 also includes the applied-field coupling property stated
earlier in Eq.~\eqref{eq:app_field_coupling_condition}. In particular, for the
reference density it gives:
\begin{align}
    \int \widetilde n^0(r) v_{\mathrm{app}}(r)\,dr = \int \rho^0(r) v_{\mathrm{app}}(r)\,dr \label{eq:condition_1_field_part}
\end{align}
For the most common case of uniform electric fields, this is met as long as the total dipole and charge of the coarse-grained density match those of the full DFT charge density. For more complicated applied fields, additional conditions on higher total multipole moments would be required.

\subsubsection{Condition 2: fidelity of the first-order response}

\begin{figure}[tb]
    \centering
    \includegraphics[width=0.92\linewidth]{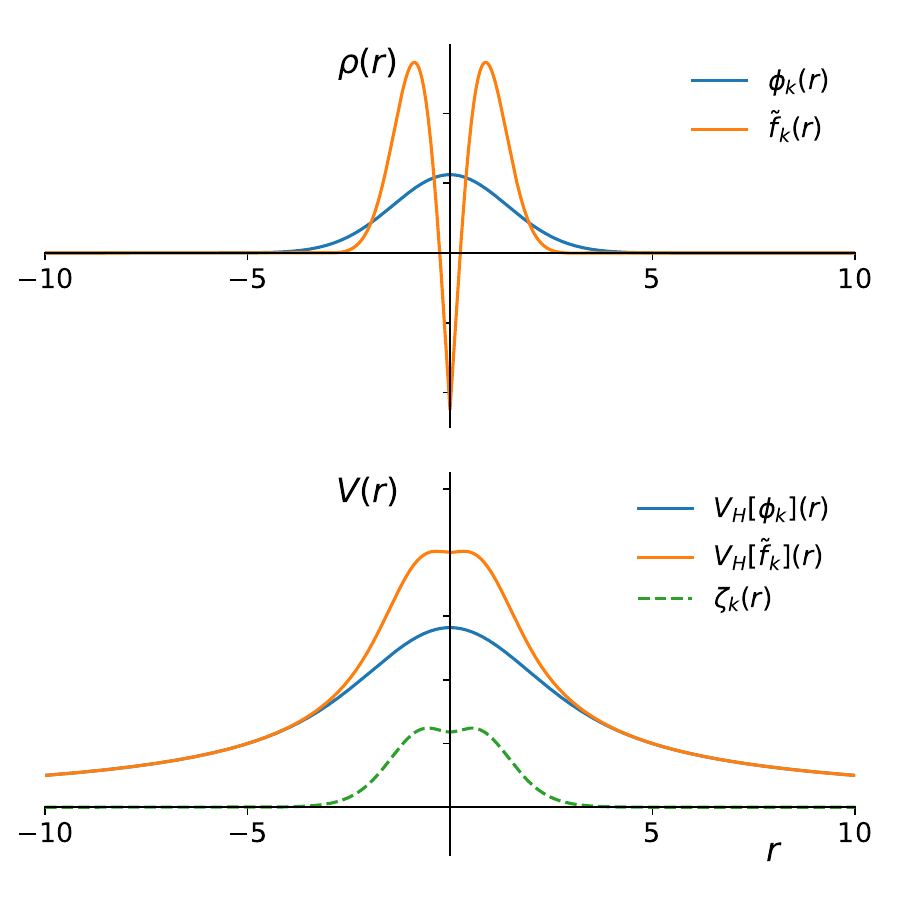}
    \caption{Two localised, spherically symmetric charge distributions
    $\widetilde f$ and $\phi$, together with their electrostatic potentials.
    If the two distributions have the same total charge, their far-field
    potentials agree and the difference $\zeta$ is confined to the region in
    which their detailed shapes differ.  For nonspherical functions, the same
    statement applies when the multipole moments agree at all orders.}
    \label{fig:Fukui_condition_si}
\end{figure}

We also need to match the response of the coarse-grained and fine-grained systems. Consider changing a single coefficient by a small amount $\delta p_k$. In
the model, the charge density changes by
\begin{equation}
    \delta\rho(r)=\delta p_k\phi_k(r),
\end{equation}
and hence the potential changes by
\begin{equation}
    \delta v_{\mathrm{CG}}(r)
    =\delta p_kv_{\mathrm H}[\phi_k](r).
    \label{eq:app_delta_v_cg}
\end{equation}
On the other hand, if we move the constraint $p_k$ by the same small amount, the constrained DFT electron density $n[\mathbf p]$ changes by $\delta n(r) = f_k(r) \delta p_k$, due to the definition of the AIM Fukui function. The corresponding potential of a positive test charge is then:
\begin{align*}
    \delta v_{\mathrm{DFT}}(r) = -\delta p_k \int\frac{f_k(r')}{|r-r'|}dr' = -\delta p_kv_{\mathrm H}[f_k](r)
\end{align*}
We require that these two potentials match at all points which are far from $r_k$, the location of the atom which $\phi_k$ is associated with. In practice this means points $r$ for which $|r-r_k| > R$. In other words, \textit{from a distance} a test charge cannot tell the difference between a charge density with the shape $-f_k$ and a charge density of the shape $\phi_k$, even though the actual shape of the two functions around the atom might be very different. We will state this by defining $\widetilde f_k = -f_k$ and writing
\begin{align}
    v_{\mathrm H}[\widetilde f_k](r) = v_{\mathrm H}[\phi_k](r) + \zeta_k(r),
    \label{eq:appendix_condition_Fukui_1}
\end{align}
in which $\zeta_k(r)$ is a function localised around $r_k$. This is illustrated in figure \ref{fig:Fukui_condition_si}. For $r$ near to $r_k$ we can see from the diagram that $\zeta_k(r)$ depends on the specific shape of $f_k$ and $\phi_k$.
We additionally require that both $\widetilde f_k$ and $\phi_k$ are localised
around $r_k$, in the sense defined above.

Another result follows from this condition. For any density $n(r)$ that vanishes for $|r-r_k|<R$, we have
\begin{align}
    \int \widetilde f_k(r)\, v_{\mathrm H}[n](r)\,dr
    =
    \int \phi_k(r)\, v_{\mathrm H}[n](r)\,dr .
    \label{eq:appendix_condition_Fukui_1a}
\end{align}
This follows because
\begin{align*}
    \int \widetilde f_k(r)\, v_{\mathrm H}[n](r)\,dr
    &=
    \iint \frac{ \widetilde f_k(r)n(r')}{|r-r'|}\,dr\,dr' \\
    &= \int n(r')\, v_{\mathrm H}[\widetilde f_k](r')\,dr',
\end{align*}
provided we can apply Fubini's theorem, which is true if $\iint |\widetilde f_k(r)\,n(r')|/|r-r'| \,dr\,dr' < \infty$.
An analogous relation holds with $\widetilde f_k$ replaced by $\phi_k$. Therefore, since $n(r)$ is non-zero only in the region $|r-r_k|>R$, and $v_{\mathrm H}[\widetilde f_k](r)=v_{\mathrm H}[\phi_k](r)$ holds there by locality of $\zeta_k(r)$, \eqref{eq:appendix_condition_Fukui_1a} follows immediately.

We can extend \eqref{eq:appendix_condition_Fukui_1a} to arbitrary densities $n$ (not necessarily zero around $r_k$) with another correction term $\tau$:
\begin{align}
    \int \widetilde f_k(r)\, v_{\mathrm H}[n](r)\,dr
    =
    \int \phi_k(r)\, v_{\mathrm H}[n](r)\,dr
    + \tau_k[n], \label{eq:appendix_condition_Fukui_with_c}
\end{align}
where the correction may be written as
\begin{align*}
    \tau_k[n]
    &=
    \int n(r)\,\bigl[v_{\mathrm H}[\widetilde f_k](r)-v_{\mathrm H}[\phi_k](r)\bigr]\,dr
    \\
    &=
    \int n(r)\,\zeta_k(r)\,dr .
\end{align*}
One can see that $\tau_k$ depends only on the overlap of $n(r)$ with $\zeta_k(r)$, which is localised around $r_k$. Hence, $\tau_k$ depends on the shape of $n(r)$ near to $r_k$.

At first order, differentiating Eq.~\eqref{eq:app_field_coupling_condition}
with respect to $p_k$ gives
\begin{equation}
    \int v_{\mathrm{app}}(r)\widetilde f_k(r)\,dr
    =\int v_{\mathrm{app}}(r)\phi_k(r)\,dr
    \label{eq:app_applied_fidelity}
\end{equation}
for the class of applied fields considered. Since $\rho_{\mathbf p}$ is linear
in $\mathbf p$, its higher derivatives vanish, and the same differentiation
at higher orders gives
\begin{equation}
    \int v_{\mathrm{app}}(r)
    \widetilde f_{k_1\ldots k_m}(r)\,dr=0,
    \qquad m\geq2.
    \label{eq:app_applied_fidelity_higher}
\end{equation}

\subsubsection{Condition 3: absence of higher-order response}

The first two conditions identify the long-range terms produced by the
reference charge and its linear response. We make the simplifying assumption
that all higher-order responses vanish:
\begin{equation}
    f_{k_1\ldots k_m}(r)=0, \qquad m\geq2.
    \label{eq:app_no_higher_response}
\end{equation}
This condition is overly restrictive, since nonzero higher-order responses
could also produce only local corrections, but it provides a simple and
transparent sufficient condition for the derivation below.

\subsection{Approximating the DFT Coulomb energy}

Recall from Eq.~\eqref{eq:app_energy_taylor} that the total perturbed energy is
decomposed into a Coulomb energy term and a kinetic, exchange-correlation term. We now show that $\Delta E_{\mathrm{Coulomb}}^{\mathrm{DFT}}$ can be written as
$\Delta E_{\mathrm{Coulomb}}^{\mathrm{CG}}$ plus local corrections, where the change in the model Coulomb energy is simply:
\begin{equation}
    \Delta E_{\mathrm{Coulomb}}^{\mathrm{CG}}
    = E_{\mathrm{Coulomb}}^{\mathrm{CG}}
    (\mathbf p;\mathbf R,v_{\mathrm{app}})
    -E_{\mathrm{Coulomb}}^{\mathrm{CG}}
    (\mathbf p^0;\mathbf R,v_{\mathrm{app}}).
    \label{eq:app_delta_cg_definition}
\end{equation}

This proceeds by taking the Taylor series for the DFT Coulomb energy, \eqref{eq:app_delta_coulomb_series}, and using the conditions above to evaluate the first two terms involving $v_k^0$ and $C_{kl}$. From Eqs.~\eqref{eq:app_total_charge} and
\eqref{eq:app_nuclear_potential},
\begin{equation}
    v_{\mathrm H}[n^0](r)-v_{\mathrm{nuc}}(r;\mathbf R)
    =-v_{\mathrm H}[\widetilde n^0](r).
    \label{eq:app_potential_rewrite}
\end{equation}
Using $\widetilde f_k=-f_k$ in Eq.~\eqref{eq:app_vk_definition} therefore
gives
\begin{align}
    v_k^0
    ={}&\int[v_{\mathrm H}[\widetilde n^0](r)
    +v_{\mathrm{app}}(r)]
    \widetilde f_k(r)\,dr.
    \label{eq:app_vk_total_charge}
\end{align}
We can use Condition 1 to separate the fine-grained reference potential into a
coarse-grained part and a local remainder:
\begin{align}
    v_k^0
    ={}&\int v_{\mathrm H}[\rho^0](r)
    \widetilde f_k(r)\,dr
    \nonumber\\
    &+\int\Delta[\widetilde n^0](r)
    \widetilde f_k(r)\,dr
    +\int v_{\mathrm{app}}(r)
    \widetilde f_k(r)\,dr.
    \label{eq:app_vk_condition_one}
\end{align}
We then apply Eq.~\eqref{eq:appendix_condition_Fukui_with_c} to the first term and
Eq.~\eqref{eq:app_applied_fidelity} to the last term.  The result is
\begin{align}
    v_k^0
    ={}&\int[v_{\mathrm H}[\rho^0](r)
    +v_{\mathrm{app}}(r)]\phi_k(r)\,dr
    +\ell_k^0,
    \label{eq:app_vk_decomposition}\\
    \ell_k^0
    \equiv{}&\tau_k[\rho^0]
    +\int\Delta[\widetilde n^0](r)
    \widetilde f_k(r)\,dr.
    \label{eq:app_lk}
\end{align}
The first term in Eq.~\eqref{eq:app_vk_decomposition} is the derivative of the
coarse-grained Coulomb energy at the reference state.  The correction
$\tau_k[\rho^0]$ depends only on $\rho^0$ near $r_k$, while the second
part of $\ell_k^0$ is an overlap between a local potential correction and the
response associated with $k$.  Hence $\ell_k^0$ is a short-range coefficient.

We next consider
\begin{equation}
    C_{kl}=\iint\frac{f_k(r)f_l(r')}
    {|r-r'|}\,dr\,dr'.
    \label{eq:app_Ckl_explicit}
\end{equation}
The two minus signs cancel, so this can be written as the interaction
between $\widetilde f_k$ and $\widetilde f_l$.  Applying
Eq.~\eqref{eq:appendix_condition_Fukui_with_c} first to $k$ and then to $l$ gives
\begin{align}
    C_{kl}
    ={}&\iint\frac{\phi_k(r)\phi_l(r')}
    {|r-r'|}\,dr\,dr'
    +\ell_{kl}^0,
    \label{eq:app_Ckl_decomposition}\\
    \ell_{kl}^0
    \equiv{}&\tau_k[\widetilde f_l]+\tau_l[\phi_k].
    \label{eq:app_lkl}
\end{align}
Because of the structure of the $\tau$, if $k$ and $l$ belong to distant centres the correction $\ell_{kl}^0$ vanishes. The only long-range part of $C_{kl}$ is therefore the Coulomb interaction between the two coarse basis functions.

Substituting Eqs.~\eqref{eq:app_vk_decomposition} and
\eqref{eq:app_Ckl_decomposition} into the linear and quadratic first-response
terms gives
\begin{align}
    &\sum_k\delta p_kv_k^0
    +\frac12\sum_{kl}\delta p_k\delta p_lC_{kl}
    \nonumber\\
    ={}&\sum_k\delta p_k
    \int[v_{\mathrm H}[\rho^0]+v_{\mathrm{app}}]
    \phi_k\,dr
    \nonumber\\
    &+\frac12\sum_{kl}\delta p_k\delta p_l
    \iint\frac{\phi_k(r)\phi_l(r')}
    {|r-r'|}\,dr\,dr'
    \nonumber\\
    &+\sum_k\delta p_k\ell_k^0
    +\frac12\sum_{kl}\delta p_k\delta p_l\ell_{kl}^0.
    \label{eq:app_first_two_reconstructed}
\end{align}
The first two terms on the right have a simple interpretation.  Using
$\delta\rho=\sum_k\delta p_k\phi_k$, they are
\begin{align}
    &\int[v_{\mathrm H}[\rho^0](r)
    +v_{\mathrm{app}}(r)]\delta\rho(r)\,dr
    +E_{\mathrm H}[\delta\rho].
    \label{eq:app_delta_rho_form}
\end{align}
If we expand the coarse-grained Hartree energy at
$\rho_{\mathbf p}=\rho^0+\delta\rho$, we find
\begin{align}
    E_{\mathrm H}[\rho^0+\delta\rho]
    ={}&E_{\mathrm H}[\rho^0]
    +\int v_{\mathrm H}[\rho^0](r)
    \delta\rho(r)\,dr
    \nonumber\\
    &+E_{\mathrm H}[\delta\rho],
    \label{eq:app_hartree_expansion}
\end{align}
while the applied-field term is linear:
\begin{equation}
    \int v_{\mathrm{app}}\rho_{\mathbf p}
    =\int v_{\mathrm{app}}\rho^0
    +\int v_{\mathrm{app}}\delta\rho.
    \label{eq:app_applied_expansion}
\end{equation}
Equations~\eqref{eq:app_delta_rho_form}--\eqref{eq:app_applied_expansion}
can therefore be used to show that
\begin{align}
    &\sum_k\delta p_kv_k^0
    +\frac12\sum_{kl}\delta p_k\delta p_lC_{kl}
    \nonumber\\
    ={}&\Delta E_{\mathrm{Coulomb}}^{\mathrm{CG}}
    +\sum_k\delta p_k\ell_k^0
    +\frac12\sum_{kl}\delta p_k\delta p_l\ell_{kl}^0.
    \label{eq:app_cg_difference_second_order}
\end{align}
This is the desired separation through the part of second order that depends
only on the first-order response. The DFT Coulomb series also contains $v_{kl}^0$ at second order and
$v_{k_1\ldots k_m}^0+C_{k_1\ldots k_m}$ at higher orders.  Every term beyond
the explicitly extracted interaction between two first-order responses
contains at least one higher-order Fukui function.  Under Condition 3, these
terms vanish. We collect the remaining local corrections into a combined
correction $L_{\mathrm{sr}}(\mathbf R,\mathbf p^0,\delta\mathbf p)$.

The full change in the fine-grained DFT Coulomb energy may then be written
\begin{equation}
    \Delta E_{\mathrm{Coulomb}}^{\mathrm{DFT}}
    =\Delta E_{\mathrm{Coulomb}}^{\mathrm{CG}}
    +L_{\mathrm{sr}}(\mathbf R,\mathbf p^0,\delta\mathbf p).
    \label{eq:app_central_difference}
\end{equation}
This equation states the main electrostatic result: under the three
conditions above, the nonlocal
part of the DFT Coulomb response is generated by the coarse-grained Coulomb
energy, while the difference is short-ranged.

\subsection{Locality of the Coulomb Energy Residual}

Before reassembling the full DFT energy, we need to show that the difference between the DFT and coarse-grained
Coulomb energies in the reference state is local. Since $\rho^0$
reproduces the long-range electrostatic effects of $\widetilde n^0$, we
would ideally compare $E_{\mathrm H}[\widetilde n^0]$ directly with
$E_{\mathrm H}[\rho^0]$.  This is difficult, however, because
$E_{\mathrm H}[\widetilde n^0]$ is ill defined for point nuclei due to the
Coulomb self-interactions. This problem does not occur in the actual DFT
Coulomb energy, where the electron-nuclear and internuclear contributions
are evaluated separately and nuclear self-interactions are omitted.

To make the comparison, consider instead a modified reference charge density
$\widetilde n_\epsilon^0$ in which each nuclear charge is represented by an
extremely narrow but finite-width function. The Hartree energy of this density is well
defined. We want to compute the DFT Coulomb energy of $\widetilde n_\epsilon^0$, which can now be written
\begin{align*}
    E_{\mathrm H}[\widetilde n_\epsilon^0] + \int \widetilde n^0_\epsilon(r) v_{\mathrm{app}}(r)\,dr.
\end{align*}
This differs from $E^{\mathrm{DFT}}_{\mathrm{Coulomb}}$ only in the self-energies of the finite-width nuclei. Using Condition 1, the Hartree part can be expressed as
\begin{align}
E_{\mathrm H}[\widetilde n_\epsilon^0]
&=
\frac12\int
\widetilde n_\epsilon^0(r)
v_{\mathrm H}[\widetilde n_\epsilon^0](r)\,dr
\nonumber \\
&=
\frac12\int
\widetilde n_\epsilon^0(r)
v_{\mathrm H}[\rho^0](r)\,dr
+
\frac12\int
\widetilde n_\epsilon^0(r)
\Delta[\widetilde n_\epsilon^0](r)\,dr.
\end{align}
where $\Delta[\widetilde n_\epsilon^0](r)$ is local. Now, due to the symmetry of the Coulomb kernel, we can exchange $\int a(r) v_{\mathrm H}[b](r)\,dr = \int b(r) v_{\mathrm H}[a](r)\,dr$, and repeat the process on the other copy of $\widetilde n_\epsilon^0$, giving
\begin{align}
E_{\mathrm H}[\widetilde n_\epsilon^0]
&=
E_{\mathrm H}[\rho^0]
+
\frac12\int
\left[
\widetilde n_\epsilon^0(r)+\rho^0(r)
\right]
\Delta[\widetilde n_\epsilon^0](r)\,dr.
\end{align}
Hence, for finite-width nuclei, the fine- and coarse-grained Hartree energies differ
only by local contributions.

Replacing the finite-width nuclear charge densities by point nuclei while
omitting their self-energies changes this expression only by local,
one-centre terms depending on the nuclear charges and the chosen finite-width
functions. It therefore follows that excluding applied fields, the DFT and model Coulomb energies differ only by local correction terms depending on properties of the reference state.

Finally, equality of the applied-field terms in the model and DFT Coulomb energies follows directly from the second part of Condition 1, as discussed around Eq.~\eqref{eq:condition_1_field_part}. Thus, $E_{\mathrm{Coulomb}}^{\mathrm{DFT}} - E_{\mathrm{Coulomb}}^{\mathrm{CG}}$ can be written as a sum of local functions of the reference state.

\subsection{Reconstruction of the model energy}

It remains to insert the electrostatic result into the total-energy
decomposition.  Combining Eqs.~\eqref{eq:app_energy_taylor} and
\eqref{eq:app_central_difference} gives
\begin{align}
    &E_{\mathrm{pop}}^{\mathrm{DFT}}
    (\mathbf p;\mathbf R,v_{\mathrm{app}})
    =\nonumber \\
    &\quad \big[E_{\mathrm{pop}}^{\mathrm{DFT}}
    (\mathbf p^0;\mathbf R,v_{\mathrm{app}})-E_{\mathrm{Coulomb}}^{\mathrm{CG}}
    (\mathbf p^0;\mathbf R,v_{\mathrm{app}})\big]\nonumber \\
    &\quad+\Delta G(\mathbf R,\mathbf p^0,\delta\mathbf p)
    +L_{\mathrm{sr}}(\mathbf R,\mathbf p^0,\delta\mathbf p)\nonumber\\
    &\quad+E_{\mathrm{Coulomb}}^{\mathrm{CG}}
    (\mathbf p;\mathbf R,v_{\mathrm{app}}).
    \label{eq:app_total_reconstruction}
\end{align}
Equation~\eqref{eq:app_total_reconstruction} has the structure of the model in
Eq.~\eqref{eq:app_model_functional_intro}. The term in square brackets is associated with $E_{\mathrm{local}}$, and it can be written:
\begin{align*}
    G(\mathbf p^0;\mathbf R) + E_{\mathrm{Coulomb}}^{\mathrm{DFT}}
    (\mathbf p^0;\mathbf R,v_{\mathrm{app}})
    -E_{\mathrm{Coulomb}}^{\mathrm{CG}}
    (\mathbf p^0;\mathbf R,v_{\mathrm{app}})
\end{align*}
Since we have previously shown that the difference in Coulomb energy terms is local, any long-range aspects of the local term arise only from $G$.

The constrained response
of $G$ and the local electrostatic corrections are represented together by
$G_{\mathrm{ML}}$. The choice of reference density $\mathbf p^0$ could simply be neutral atoms, but a geometry-dependent choice of reference $\mathbf p^0$ can be absorbed into the definitions of the two learned terms.

The result above is an order-by-order statement within the formal Taylor
expansion about $\mathbf p^0$. It applies within a smooth sector of the
population-constrained energy landscape and does not imply that the expansion
converges globally.

\section{Appendix: Equivalence of Methods Beyond Point Charges}
\label{appendix:exp_basis}

Start by restating the basis expansion of the charge density:
\begin{align*}
    \rho = \sum_i p_{i\alpha} \phi_{\alpha} (r-r_i)
\end{align*}
Then we suppose a general local form for the learned functional
\begin{align}
    G_\text{ML}(\mathbf{p}) = \sum_i g(\{p_{i\alpha}\}_\alpha) \ ,
\end{align}
In which $g$ is a smooth convex function of $\{p_{i\alpha}\}_\alpha$. Let $g_i := g(\{p_{i\alpha}\}_\alpha)$, then we write the Euler-Lagrange equation (in this case, without the total charge constraint):
\begin{align}
    0=\frac{\partial}{\partial p_{i\alpha}} & \left( \sum_i g(\{p_{i\alpha}\}_{\alpha}) + E_{\text{H}}(\rho) + \int v_\text{ext}(r) \rho(r) \right) \\
    &= \frac{\partial g_i}{\partial p_{i\alpha}} + \int \frac{\delta E_\text{H}}{\delta \rho(r)} \frac{\partial \rho(r)}{\partial p_{i\alpha}} dr + \int v_\text{ext} \frac{\partial \rho(r)}{\partial p_{i\alpha}} dr \\
    &= \frac{\partial g_i}{\partial p_{i\alpha}} + \int \left( v_\text{H}[\rho](r) + v_\text{ext} \right) \phi_{\alpha} (r-r_i) dr 
\end{align}

Finally, one can define $\hat{v}_{i\alpha} = \int \left( v_\text{H}[\rho](r) + v_\text{ext} \right) \phi_{\alpha} (r-r_i) dr$ as the relevant potential for the charge variable $p_{i\alpha}$. One can then rewrite the above as:
\begin{align}
    0= \frac{\partial g_i}{\partial p_{i\alpha}} + \hat{v}_{i\alpha} 
\end{align}
Thus, turning this into a fixed-point problem is possible if we can invert the following equations to solve for $p_{i\alpha}^{(t+1)}$.
\begin{align}
    \frac{\partial g_i}{\partial p_{i\alpha}}(\{p_{i\alpha}^{(t+1)}\}_\alpha) = - \hat{v}_{i\alpha}^{(t)}
\end{align}
Which is a set of potentially nonlinear equations coupled in $\alpha$ but not $i$. This is possible if $g_i$ is strictly convex, and given some restrictions on the domain of $\hat{v}_{i\alpha}$. We note that although strict convexity of $g_i$ is enough to invert the gradient map (and that the minimizer of the energy solves the resulting fixed-point equation), this does not guarantee that the fixed-point iteration will be a convergent iteration.

\section{Appendix: Hirshfeld Multipoles Reproduce Far-Field Potential}
\label{appendix:hirshfeld}

The condition on the coarse-grained density $\rho(r)$ appears quite natural, and it is interesting to see what charge partitioning schemes do or do not meet this condition. Consider, for instance, atomic multipoles determined by Hirshfeld partitioning. Denote a collection of atoms in open boundary conditions by $\{(z_i, r_i)\}_i$. Then, assume we have some partitioning functions $\{u_i\}_i$, with one defined for each atom, which sum to unity. 
\begin{align}
    \sum_i u_i(r) = 1
\end{align}
In the Hirshfeld method, these would be constructed by taking the free atom electron density for each atom, $g_i(|r-r_i|)$, and defining:
\begin{align}
    u_i = \frac{g_i(|r-r_i|)}{\sum_j g_j(|r-r_j|)}
\end{align}
Assume that there is some strict (but large) cutoff $R$ beyond which the partitioning functions $u_i$ are strictly zero\footnote{Using a cutoff risks missing a small amount of electron density which is not close to any atom, but in practice for large cutoff distances the error is small.}. Specifically, for each $i$, we have $u_i(r)=0$ when $|r-r_i|>R$. Now, for an electron density $n(r)$ define
\begin{align}
    n_i(r) = u_i(r) n(r),
\end{align}
so that $\sum_i n_i = n$. At a point $r$ we can split the Hartree plus nuclear potential as:
\begin{align}
    - v_\mathrm{H}[n](r) + v_\text{nuc}(r)  = \sum_i \left( - v_\mathrm{H}[n_i](r) + \frac{Z_i}{|r-r_i|} \right)
\end{align}
The `distant' atoms are those for which $|r-r_i|>R$. For each such $i$, the density $n_i(r)$ is confined around $r_i$, and hence we can expand the Hartree potential of this density (and nucleus $i$) at $r$ using a multipole expansion. By using enough multipole moments, one can therefore represent the potential at $r$ due to all distant atoms by using the partitioned multipole moments of those atoms. The is closely related to how the Hartree potential is represented in some DFT codes, such as FHI-aims \cite{Blum2009AbOrbitals}.

This does not work for some other charge partitioning schemes. It only works in this case because the full density is split up into a set of atom-centred densities which are reasonably well localised around atoms, and which sum to the full density. Of course, if one uses too few moments, for instance only partial charges, the representation may be very poor. We suspect this is one reason why the addition of atomic dipoles is beneficial in our experiments. 

\section{Appendix: Electrostatics Implementation}
\label{appendix:electrostatics_implementation}

\subsection{Periodic Electrostatics Implementation}

\subsubsection{Densities from Gaussian Type Orbitals}

In all of our models, electrostatic energies and features are based a Gaussian type orbital basis. In the main text we construct a charge density as:
\begin{align*}
    \rho(\mathbf{r}) = \sum_{ilm} p_{i,lm} \phi_{nlm}(\mathbf{r}-\mathbf{r}_i)
\end{align*}
This equation is ambiguous as to whether the density is periodic or not. Therefore, we will now make the periodicity explicit by writing:
\begin{align}
    \rho(\mathbf{r}) &= \sum_{\mathbf{a}\in\Lambda}\sum_{i=1}^{N_\text{atoms}}\sum_{lm} p_{i,lm} \phi_{nlm}(\mathbf{r}-\mathbf{r}_i-\mathbf{a})
    \label{eq:SI_electro_implementation_rho_realspace} \\
    &= \sum_{i=1}^{N_\text{atoms}}\sum_{lm} p_{i,lm} \left(\sum_{\mathbf{a}\in\Lambda} \phi_{nlm}(\mathbf{r}-\mathbf{r}_i-\mathbf{a})\right) \nonumber
\end{align}
Where $\Lambda$ is the lattice on which the density is periodic. The sum over $i$ now only counts one copy of each atom (not periodic replicas) and the sum over $\mathbf{a}$ accounts for the infinite periodic images.

In \eqref{eq:SI_electro_implementation_rho_realspace}, $\phi_{nlm}$ is the following Gaussian type orbital:
\begin{align}
    \phi_{nlm}(\mathbf{r}) = C_{l\sigma_n} e^{-\frac{r^2}{2\sigma_n^2}} r^{l} Y_{lm}(\hat{\mathbf{r}})
    \label{eq:SI_electro_implementation_phi}
\end{align}
The real spherical harmonic $Y_{lm}$ has the following normalization,
\begin{equation}
    \int Y_{lm}(\hat{\mathbf{r}})\,Y_{LM}(\hat{\mathbf{r}})\,d\hat{\mathbf{r}}
= \delta_{lL}\delta_{mM},
\end{equation}
and $C_{l\sigma_n}$ is chosen such that the each $\phi_{nlm}$ has its $lm$ electric multipole moment equal to one. Specifically, given the solid harmonic function $R_{lm}(\mathbf{r})$ defined as:
\begin{align}
    R_{lm}(\mathbf{r}) := \sqrt{\frac{4\pi}{2l+1}}\, r^{l} Y_{lm}(\hat{\mathbf{r}})
\end{align}
The $lm$ spherical multipole moment of a function $f(\mathbf{r})$ is defined as:

\begin{align}
    Q_{lm} = \int_{\mathbb{R}^3} f(\mathbf{r}) R_{lm}(\mathbf{r})\,d\mathbf{r}.
\end{align}

For $l=m=0$, this is just the total charge of $f$, and for $l=1$, $Q_{lm}$ is a permutation of the dipole moment. $C_{l\sigma_n}$ is such that $\int \phi_{nLM}(\mathbf{r}) R_{lm}(\mathbf{r})\,d\mathbf{r} = \delta_{lL} \delta_{mM}$.

\subsubsection{Fourier Series of the Density}

Throughout our implementation, we use the following definition for the Fourier series coefficients of a function $f:\mathbb{R}^3\rightarrow \mathbb{R}$, which is periodic on a lattice $\Lambda$ which has a unit cell volume $\Omega$:
\begin{align}
    \tilde{f}(\mathbf{k}) := \frac{(2\pi)^3}{\Omega} \int_{\text{unit cell}} e^{-i\mathbf{k}\cdot \mathbf{x}} f(\mathbf{x})d\mathbf{x} \qquad \mathbf{k} \in \Lambda^\star
\end{align}
In which $\Lambda^\star$ is the reciprocal lattice. The Fourier transform of an integrable function $g$ is defined in the usual way as $\tilde{g}(\mathbf{k}) = \int_{\mathbb{R}^3} e^{-i\mathbf{k}\cdot \mathbf{x}} g(\mathbf{x})d\mathbf{x}$.

Denote the Fourier transform of $\phi_{nlm}(\mathbf{r})$ by $\tilde{\phi}_{nlm}(\mathbf{k})$. The Fourier series of the density $\rho$ in \eqref{eq:SI_electro_implementation_rho_realspace} is:
\begin{align}
    \tilde{\rho}(\mathbf{k}) = \frac{(2\pi)^3}{\Omega}
\sum_{i,lm} p_{i,lm}\,\tilde{\phi}_{nlm}(\mathbf{k})\,e^{-i\mathbf{k}\cdot\mathbf{r}_i}  \qquad \mathbf{k} \in \Lambda^\star
    \label{eq:SI_electro_implementation_rho_kspace}
\end{align}
Computing $\tilde \rho$ requires knowledge of the Fourier transform $\tilde{\phi}_{nlm}$, which is known analytically \cite{JiyunKuang_1997}:
\begin{align}
    \tilde{\phi}_{nlm}(\mathbf{k}) =
4\pi\,C_{l,\sigma}\,(-i)^l\,Y_{lm}(\hat{\mathbf{k}})
\int_0^\infty r^{l+2} j_l(kr) e^{-\frac{r^2}{2\sigma^2}}\,dr.
    \label{eq:SI_electro_implementation_phi_tilde}
\end{align}
Define the radial integral as:
\begin{align}
    f_{l,\sigma}(k)
= 4\pi \int_0^\infty r^{l+2} j_l(kr) e^{-\frac{r^2}{2\sigma^2}}\,dr,
\end{align}
An expression was given for this in terms of the confluent hypergeometric function ${}_1F_1$ \cite{grisafi_thesis}, which in this case simplifies to just an exponential using the relation ${}_1F_1(a,a,z) = e^z$:
\begin{align}
    f_{l,\sigma}(k)
&= 4\pi \sqrt{\frac{\pi}{2}}\,k^l \sigma^{3+2l}\,
{}_1F_1\!\left(\frac{3}{2}+l,\frac{3}{2}+l,-\frac{(k\sigma)^2}{2}\right) \nonumber \\
&= 4\pi \sqrt{\frac{\pi}{2}}\,k^l \sigma^{3+2l}\,
e^{-\frac{(k\sigma)^2}{2}}
\label{eq:SI_electro_implementation_hypgeom}
\end{align}

Equations \eqref{eq:SI_electro_implementation_rho_kspace}--\eqref{eq:SI_electro_implementation_hypgeom} are the working equations in the implementation for generating the Fourier series from a set of atomic multipoles. 

\subsubsection{Coulomb Convolution}

Convolution with the Coulomb kernel in Fourier space is performed by
\begin{align}
    \tilde{v}(\mathbf{k}) = 4\pi \frac{\tilde{\rho}(\mathbf{k})}{k^2}, \quad \mathbf{k}\neq\mathbf{0},
    \label{eq:SI_electro_implementation_coulomb_kernel}
\end{align}
in which $\tilde{v}(\mathbf{k})$ is the Fourier series of the electric potential, and where $\tilde{v}(\mathbf{k}=0)$ is defined as zero. For the charged periodic supercells in the silicon dioxide example, no additional finite size corrections were applied, which was also the case for the DFT reference calculations. The results in both cases can be interpreted as having a uniform background charge cancelling any additional real charge.

\subsubsection{electrostatic energy}

The electrostatic energy per unit cell can be computed from the potential and density as $E=\frac{1}{2}\int_\Omega \rho(\mathbf{r}) v(\mathbf{r}) d\mathbf{r}$. Given our definition of the Fourier series coefficients, using Parseval's theorem gives:
\begin{align}
     E = \frac{1}{2}\frac{\Omega}{(2\pi)^6} \sum_{\mathbf{k} \in \Lambda^\star \backslash \mathbf{0}} \tilde{v}^\star(\mathbf{k}) \tilde{\rho}(\mathbf{k})
    \label{eq:SI_electro_implementation_coulomb_energy}
\end{align}

The notation $\mathbf{k} \in \Lambda^\star \backslash \mathbf{0}$ means to sum over all vectors in the reciprocal lattice, but omit the $\mathbf{k}=\mathbf{0}$ term. Overall, the procedure to evaluate the total energy from a set of multipoles is as follows: Firstly, compute a grid of $k$-vectors on the reciprocal lattice. This is done by choosing all points on the reciprocal lattice for which $|\mathbf{k}| < k_\text{cut}$ for some predetermined cutoff wavenumber $k_\text{cut}$. For each $k$-vector compute the Fourier series coefficient as in equation \eqref{eq:SI_electro_implementation_rho_kspace}, using the analytic form of $\tilde{\phi}$ in \eqref{eq:SI_electro_implementation_phi_tilde}. Finally, compute the energy by combining \eqref{eq:SI_electro_implementation_coulomb_kernel} and \eqref{eq:SI_electro_implementation_coulomb_energy}.

\subsubsection{Electrostatic Features}

For the features we need to compute:
\begin{align}
    v_{i,NLM} = \int v_\text{eff}(\mathbf{r}) \phi_{NLM}(\mathbf{r}-\mathbf{r}_i)\,d\mathbf{r}
    \label{eq:SI_electro_implementation_feature_integral_realspace}
\end{align}
Note the capital indices $NLM$. This is because $\{\phi_{NLM}\}_{NLM}$ is the set of functions used to produce the electrostatic features, which can be different from those used to construct the density ($\phi_{nlm}$). For instance, we use only one width $\sigma_n$ in the density construction, but several different widths to make electrostatic features. The potential is decomposed as $v_\text{eff}(\mathbf{r}) = v_\text{H}[\rho](\mathbf{r}) + v_\text{app}(\mathbf{r}) + \mu$ as described in the main text. Here we explain our implementation for just the first (Hartree) contribution to the features, and the effect of the applied fields and Fermi level is explained further below.

The electric potential features are computed using the Fourier series of $v(\mathbf{r})$, (equation \ref{eq:SI_electro_implementation_coulomb_kernel}), and the Fourier transform of $\phi_{nlm}$.

Since $\phi_{NLM}$ is non-periodic, we must first replace $\phi_{NLM}$ in \eqref{eq:SI_electro_implementation_feature_integral_realspace} with a periodic version:
\begin{align}
    \phi^\text{periodic}_{NLM}(\mathbf{r}) = \sum_{\mathbf{a}\in\Lambda} \phi_{NLM}(\mathbf{r}-\mathbf{a})
\end{align}
Note that this is the same as what appears in the periodic density \eqref{eq:SI_electro_implementation_rho_realspace}. One can then compute the Fourier series coefficients of $\phi^\text{periodic}_{NLM}(\mathbf{r})$, and use Parseval's theorem to compute the integral in \eqref{eq:SI_electro_implementation_feature_integral_realspace}. The result is that:
\begin{align}
    v_{i,NLM} = \frac{1}{(2\pi)^3}\sum_{\mathbf{k}\in\Lambda^\star \backslash \mathbf{0}} \tilde{v}(\mathbf{k})\,\Big(\tilde{\phi}_{NLM}(\mathbf{k})\,e^{-i\mathbf{k}\cdot\mathbf{r}_i}\Big)^{\star}
    \label{eq:SI_electro_implementation_feature_comp}
\end{align}
Where $\tilde{\phi}_{NLM}(\mathbf{k})$ is same object as in \eqref{eq:SI_electro_implementation_phi_tilde}, just with different index symbols to differentiate it from the density basis functions.

In summary, the overall process for computing the features is: Compute $\tilde\rho(\mathbf{k})$ in the same way as for the total energy and then use \eqref{eq:SI_electro_implementation_feature_comp} and \eqref{eq:SI_electro_implementation_phi_tilde} to get the features. 
\subsection{Real Space Electrostatics Implementation}

The open boundary conditions implementation is significantly simpler. Firstly, note that both the electrostatic energy and features can be expressed in terms of the following integrals:
\begin{align}
    \mathcal{T}_{inlm,jn'l'm'} &:=\iint \frac{\phi_{nlm}(\mathbf{r}-\mathbf{r}_i)\phi_{n'l'm'}(\mathbf{r}'-\mathbf{r}_j)}{|\mathbf{r}-\mathbf{r}'|} d\mathbf{r} d\mathbf{r}'.
\end{align}
In the density, we always just have one width and hence $n=1$. Given these integrals, the electrostatic energy is just:
\begin{align}
    E = \frac{1}{2}\sum_{ilm,jl'm'} p_{i,lm}p_{j,l'm'} \mathcal{T}_{inlm,jnl'm'}
    \label{eq:SI_electro_implementation_energy_realspace}
\end{align}
and the electrostatic features are:
\begin{align}
    v_{i,NLM}=\sum_{jl'm'} p_{j,l'm'} \mathcal{T}_{iNLM,jnl'm'} 
    \label{eq:SI_electro_implementation_features_realspace}
\end{align}

If all the $l$ indices are zero, the integral $\mathcal{T}_{inlm,jn'l'm'}$ to a simple one dimensional function:
\begin{equation}
\mathcal{T}_{in00,jn'00} = \frac{\mathrm{erf}\!\left(r_{ij}/(2\sigma_{\text{tot}})\right)}{r_{ij}},
\qquad
\sigma_{\mathrm{tot}}:=\sqrt{\frac{\sigma_{n}^2+\sigma_{n'}^2}{2}}.
\label{eq:SI_electro_implementation_erf}
\end{equation}

One could use electronic structure libraries for computing the $l>0$ variants, but given that we are mostly interested in $l=0,1$, we found an alternative method which parallelises well on a GPU. It can be shown that one can approximate an $l=1$ Gaussian orbital as a difference between two scaled $l=0$ functions, which are slightly displaced relative to one another \cite{mace_polar}. This allows us to compute all interactions between $l=1$ functions using a finite difference of a $l=0$ integrals. This parallelises quite well, and is the method used in our code to prepare $\mathcal{T}_{inlm,jn'l'm'}$ for both the energy and field computation.

\subsection{Applied Field and Fermi Level Terms}

The electric field features have contributions from the applied potential and the Fermi level, which are not captured in the above discussion. In all of our examples, the external field is just a homogeneous field. Denote this field by $\mathbf{F}$ such that $v_\text{app}(\mathbf{r})=-\mathbf{F}\cdot\mathbf{r}$. Then, the contribution from the applied field and the Fermi level $\mu$ can be written:
\begin{align*}
    v_{i,NLM}^{\text{app, }\mu}=\int_{\mathbb{R}^3} (-\mathbf{F}\cdot\mathbf{r}+\mu) \phi_{NLM}(\mathbf{r}-\mathbf{r}_i)\,d\mathbf{r}
\end{align*}
This is simply rearranged into an expression involving the total charge and dipole moment of the function $\phi_{NLM}$:
\begin{align*}
    v_{i,NLM}^{\text{app, }\mu}&=(\mu -\mathbf{F}\cdot\mathbf{r}_i)\int_{\mathbb{R}^3} \phi_{NLM}(\mathbf{r})\,d\mathbf{r} \\
    & \quad - \mathbf{F} \cdot \int_{\mathbb{R}^3} \mathbf{r} \,\phi_{NLM}(\mathbf{r})\,d\mathbf{r}
\end{align*}
The multipole moments of $\phi_{NLM}$ are known and hence one can easily compute these contributions to the features and then add them to the contributions from the Hartree potential in either periodic or open boundary conditions. 

\subsection{Self Interaction Terms}
\label{sec:methods:self_interaction}

In both the real space and periodic implementations, one must choose how to treat `self-interaction' terms in both the energy and the electrostatic features. For instance, consider computing the energy in open boundary conditions using \eqref{eq:SI_electro_implementation_energy_realspace}. Note that the terms with $i=j$ are included in the sum, these are the self interaction terms. One can pre-compute the coefficients $\mathcal{T}_{inlm,in'l'm'} = \delta_{ll'}\delta_{mm'} T_{l,nn'}$ for an appropriate range of $l$ and $n$ using numerical methods, instead of using \eqref{eq:SI_electro_implementation_erf}.

To be consistent we also need to include the self interaction energy in the periodic implementation. This is easy because the expression for the energy \eqref{eq:SI_electro_implementation_coulomb_energy} is the energy of the full charge density interacting with itself, and therefore already contains the self interaction terms. We have observed generally better behaviour from models when including self interaction energies, which is consistent with the experience of some other developers of electrostatic MLIPs \cite{eMLP}.

For the electrostatic features, one can also consider self interaction terms. This time, rather than the energy of a distributed charge interacting with itself, we are interested in the component of an electrostatic feature $v_{i,NLM}$ which comes from the multipoles on atom $i$. Again, for a given choice of basis one can compute these terms in advance using numerical methods. In order to remove these terms, in the real space implementation it is sufficient to just restrict the sum \eqref{eq:SI_electro_implementation_features_realspace} to not include $i=j$. In the periodic implementation, one must instead compute the features with \eqref{eq:SI_electro_implementation_feature_comp} and then subtract
\begin{align}
    v^\text{self-int.}_{i,NLM} = \sum_{lm} p_{i,lm} \mathcal{T}_{iNLM,inlm},
\end{align}
to get the features without the self interaction. Note that, again, there is no sum over $n$ because we only use one Gaussian width for expanding the charge density.

\subsection{Dipole Corrections for the Energy and Features}

Whenever working with slabs, we apply dipole corrections to both the electrostatic energy and the electrostatic features \cite{Bengtsson1999}. For the energy, this simply involves adding
\begin{align*}
    E^\text{dipole correction} = \frac{2\pi|P_z|^2}{\Omega}
\end{align*}
to the electrostatic energy computed via \eqref{eq:SI_electro_implementation_coulomb_energy}, in which $P_z$ is the total dipole in the $z$ (non-periodic) direction and $\Omega$ is the total volume of the simulation cell. Similarly, to correct the electrostatic features one just needs to introduce a dipole correction electric field ($F_z^\text{dipole correction}$) at each step of the SC cycle and in addition to existing applied fields. The strength of this field is \cite{Bengtsson1999}:
\begin{align}
    F_z^\text{dipole correction} = -\frac{4\pi
    P_z}{\Omega}
\end{align}
The field (defined for positive test charges) is opposite in direction to the total dipole. This field enters into the electric field features in exactly the same way as other applied fields as described above. 

\end{document}


\title{Supplementary Information for: Design Space of Self--Consistent Electrostatic Machine Learning Interatomic Potentials}

\date{\today}

\maketitle

\tableofcontents

\section{Aluminium and Water Slabs}

\subsection{Model Training Schedules and Hyperparameters}

\begin{table}[ht]
\centering
\caption{Summary of all trained models. The `Architecture' Column describes the complexity of either the fixed point update or the energy functional layer.}
\renewcommand{\arraystretch}{1.25}
\setlength{\tabcolsep}{6pt}

\begin{tabular}{
l
@{\hspace{20pt}} l
@{\hspace{20pt}} l
@{\hspace{20pt}} c
@{\hspace{20pt}} l
}
\toprule
\makecell{\textbf{Fit}\\\textbf{Name}} &
\makecell{\textbf{Training}\\\textbf{Schedule}} &
\makecell{\textbf{Architecture}} &
\makecell{\textbf{Atomic}\\\textbf{multipole}\\\textbf{order}} &
\makecell{\textbf{Remove}\\\textbf{from}\\\textbf{loss}} \\
\midrule
\multicolumn{5}{l}{\textbf{Energy Models}} \\
\midrule

\textbf{energy\_A} & implicit & quadratic & 1 & -- \\
\textbf{energy\_B} & implicit & nonlinear & 1 & -- \\
\textbf{energy\_C} & implicit & manybody  & 1 & -- \\
\textbf{energy\_D} & implicit & quadratic & 1 & all but $E$, $F$ \\
\textbf{energy\_E} & implicit & quadratic & 0 & $q_{lm}$ weight $=1$ \\

\midrule
\multicolumn{5}{l}{\textbf{Fixed-Point Models}} \\
\midrule

\textbf{fixedpoint\_A} & direct + implicit & linear    & 1 & -- \\
\textbf{fixedpoint\_B} & direct + implicit & nonlinear & 1 & -- \\
\textbf{fixedpoint\_C} & direct + implicit & manybody  & 1 & -- \\

\textbf{fixedpoint\_D} & shortcut-scf & linear    & 1 & -- \\
\textbf{fixedpoint\_E} & shortcut-scf & nonlinear & 1 & -- \\
\textbf{fixedpoint\_F} & shortcut-scf & manybody  & 1 & -- \\

\textbf{fixedpoint\_G} & direct + implicit & linear & 1 & all but $E$, $F$ \\
\textbf{fixedpoint\_H} & shortcut-scf      & linear & 1 & all but $E$, $F$ \\

\textbf{fixedpoint\_I} & direct + implicit & linear & 0 & $q_{lm}$ \\
\textbf{fixedpoint\_J} & shortcut-scf      & linear & 0 & $q_{lm}$ \\

\midrule
\multicolumn{5}{l}{\textbf{MACE}} \\
\midrule

\textbf{mace\_A} & LocalCharges &  & -- & all but $E$, $F$ \\

\midrule
\multicolumn{5}{l}{\textbf{MACE-LocalCharges}} \\
\midrule

\textbf{localcharges\_A} & LocalCharges &  & 1 & -- \\
\textbf{localcharges\_B} & LocalCharges &  & 1 & all but $E$, $F$ \\
\textbf{localcharges\_C} & LocalCharges &  & 0 & $q_{lm}$ weight $=1$ \\

\midrule
\multicolumn{5}{l}{\textbf{MACE-QEq}} \\
\midrule

\textbf{qeq\_A} & LocalCharges &  & 0 & $q_{lm}$ weight $=1$ \\
\textbf{qeq\_B} & LocalCharges &  & 0 & all but $E$, $F$ \\

\bottomrule
\end{tabular}

\label{table:all_models}

\end{table}

Table \ref{table:all_models} contains a summary of all the models used in the results of the main text and the supplementary information, with the exception of the training dynamics studies in main the text figures 7 and 8. The column called training schedule refers to a set of loss weights and other training parameters, which can be found in tables \ref{table:energy_schedule_slabs}--\ref{table:local_charges_schedule_slabs}. In table \ref{table:all_models}, the `implicit' training schedule for energy functional models refers to table \ref{table:energy_schedule_slabs}, `direct + implicit' refers to table \ref{table:fixedpoint_schedule_slabs_implicit}, `shortcut-SCF' refers to table \ref{table:fixedpoint_schedule_slabs_backprop} and `LocalCharges' refers to table \ref{table:local_charges_schedule_slabs}. It is standard practice when training MACE models to train in two stages. Firstly, using one set of weights and a comparatively large learning rate, and then switching to use a smaller learning rate, and an increased weight on the total energy per atom. This can be seen in many of the training schedules below.  

One can see that the different models sometimes require different losses due to their design, but wherever possible we have kept loss and training parameters identical. Some of the fits are not used in figures in the main text, but we have provided them for comparison. All of the fixed-point models were trained once using direct training followed by implicit differentiation, and the again using the shortcut-SCF method. For instance, one can see in the table that fixed-point models D, E and F are the same as A, B and C, but trained with the other method. In the results of the main text, we used the direct followed by implicit differentiation method where possible. The exception is figure 10 in the main text (the importance of body order), where we used the models trained with the shortcut-SCF method because the many-body fixed-point model trained with implicit differentiation showed unstable training dynamics. This is discussed in section \ref{sec:SI_training_method_comp}. 

\begin{table}[ht]
\centering
\caption{Energy Functional Models Implicit Training Schedule for Al/\ce{H_2O}}
\renewcommand{\arraystretch}{1.25}
\setlength{\tabcolsep}{6pt}

\begin{tabular}{
l
@{\hspace{20pt}} l
@{\hspace{20pt}} c
@{\hspace{20pt}} c|c|c|c|c
@{\hspace{20pt}} c
@{\hspace{20pt}} c
}
\toprule
 & \textbf{Method} & \textbf{Epochs} & \multicolumn{5}{c}{\textbf{Loss weights}} & \textbf{Batch} & \textbf{Learning} \\
 \cmidrule(lr){4-8}
 &                 &                 & $E$ & $F$ & $q_{ilm}$ & $Q_\text{tot}$ & $P$ & \textbf{Size} & \textbf{Rate} \\
\midrule

\textbf{Stage 1}
& \makecell[l]{Implicit\\Differentiation}
& 0--200
& 100
& 100
& 100
& 0
& 1{,}000{,}000
& 1
& 0.01 \\
\bottomrule
\end{tabular}
\label{table:energy_schedule_slabs}
\end{table}

\begin{table}[ht]
\centering
\caption{Fixed Point Models Direct + Implicit Training Schedule for Al/\ce{H_2O}}
\renewcommand{\arraystretch}{1.25}
\setlength{\tabcolsep}{6pt}

\begin{tabular}{
l
@{\hspace{20pt}} l
@{\hspace{20pt}} c
@{\hspace{20pt}} c|c|c|c|c
@{\hspace{20pt}} c
@{\hspace{20pt}} c
}
\toprule
 & \textbf{Method} & \textbf{Epochs} & \multicolumn{5}{c}{\textbf{Loss weights}} & \textbf{Batch} & \textbf{Learning} \\
 \cmidrule(lr){4-8}
 &                 &                 & $E$ & $F$ & $q_{ilm}$ & $Q_\text{tot}$ & $P$ & \textbf{Size} & \textbf{Rate} \\
\midrule

\textbf{Stage 1}
& Direct
& 0--200
& 100
& \multirow{2}{*}{100}
& \multirow{2}{*}{100}
& \multirow{2}{*}{10{,}000}
& \multirow{2}{*}{1{,}000{,}000}
& 2
& 0.01 \\

\textbf{Stage 2}
& \makecell[l]{Implicit\\Differentiation}
& 200--230
& 1000
&  &  &  &
& 2
& 0.001 \\
\bottomrule
\end{tabular}
\label{table:fixedpoint_schedule_slabs_implicit}
\end{table}

\begin{table}[ht]
\centering
\caption{Fixed Point Models Shortcut SCF Training Schedule for Al/\ce{H_2O}}
\renewcommand{\arraystretch}{1.25}
\setlength{\tabcolsep}{6pt}

\begin{tabular}{
l
@{\hspace{20pt}} l
@{\hspace{20pt}} c
@{\hspace{20pt}} c|c|c|c|c
@{\hspace{20pt}} c
@{\hspace{20pt}} c
}
\toprule
 & \textbf{Method} & \textbf{Epochs} & \multicolumn{5}{c}{\textbf{Loss weights}} & \textbf{Batch} & \textbf{Learning} \\
 \cmidrule(lr){4-8}
 &                 &                 & $E$ & $F$ & $q_{ilm}$ & $Q_\text{tot}$ & $P$ & \textbf{Size} & \textbf{Rate} \\
\midrule

\makecell[l]{\textbf{Warm-up}\\\textbf{(Stage 0)}}
& Direct
& 0--10
& 100
& \multirow{3}{*}{100}
& \multirow{3}{*}{100}
& \multirow{3}{*}{10{,}000}
& \multirow{3}{*}{1{,}000{,}000}
& 2
& 0.01 \\

\textbf{Stage 1}
& Shortcut SCF
& 10--200
& 100
&  &  &  & 
& 2
& 0.01 \\

\textbf{Stage 2}
& Unrolled SCF
& 200--230
& 1000
&  &  &  &
& 2
& 0.001 \\
\bottomrule
\end{tabular}
\label{table:fixedpoint_schedule_slabs_backprop}
\end{table}

\begin{table}[ht]
\centering
\caption{LocalCharges Training Schedule for Al/\ce{H_2O}}
\renewcommand{\arraystretch}{1.25}
\setlength{\tabcolsep}{6pt}

\begin{tabular}{
l
@{\hspace{20pt}} l
@{\hspace{20pt}} c
@{\hspace{20pt}} c|c|c|c|c
@{\hspace{20pt}} c
@{\hspace{20pt}} c
}
\toprule
 & \textbf{Method} & \textbf{Epochs} & \multicolumn{5}{c}{\textbf{Loss weights}} & \textbf{Batch} & \textbf{Learning} \\
 \cmidrule(lr){4-8}
 &                 &                 & $E$ & $F$ & $q_{ilm}$ & $Q_\text{tot}$ & $P$ & \textbf{Size} & \textbf{Rate} \\
\midrule

\textbf{Stage 1}
& --
& 0--200
& 100
& \multirow{2}{*}{100}
& \multirow{2}{*}{100}
& \multirow{2}{*}{10{,}000}
& \multirow{2}{*}{1{,}000{,}000}
& 2
& 0.01 \\

\textbf{Stage 2}
& --
& 200--230
& 1000
&  &  &  & 
& 2
& 0.001 \\
\bottomrule
\end{tabular}
\label{table:local_charges_schedule_slabs}
\end{table}

For the analysis of training dynamics, the energy functional and fixed-point models used the schedules in tables \ref{table:training_dynamics_energy} and \ref{table:training_dynamics_field} respectively.

\begin{table}[ht]
\centering
\caption{Training Schedule for Energy Functional models when testing training dynamics on the metal water system (main text figure 7). The loss on the gradient of the functional ($\nabla_q E$) is omitted when training with implicit differentiation since it is zero by assumption.}
\renewcommand{\arraystretch}{1.25}
\setlength{\tabcolsep}{6pt}

\begin{tabular}{
l
@{\hspace{20pt}} l
@{\hspace{20pt}} c
@{\hspace{20pt}} c|c|c|c|c
@{\hspace{20pt}} c
@{\hspace{20pt}} c
}
\toprule
 & \textbf{Method} & \textbf{Epochs} & \multicolumn{5}{c}{\textbf{Loss weights}} & \textbf{Batch} & \textbf{Learning} \\
 \cmidrule(lr){4-8}
 &                 &                 & $E$ & $F$ & $q_{ilm}$ & $\nabla_{q}E$ & $P$ & \textbf{Size} & \textbf{Rate} \\
\midrule

\textbf{Stage 1}
& --
& 0--200
& 100
& 100
& 100
& 1,000
& 1{,}000{,}000
& 1
& 0.01 \\
\bottomrule
\end{tabular}
\label{table:training_dynamics_energy}
\end{table}

\begin{table}[ht]
\centering
\caption{Training Schedule for Fixed Point models when testing training dynamics (main text figure 7).}
\renewcommand{\arraystretch}{1.25}
\setlength{\tabcolsep}{6pt}

\begin{tabular}{
l
@{\hspace{20pt}} l
@{\hspace{20pt}} c
@{\hspace{20pt}} c|c|c|c|c
@{\hspace{20pt}} c
@{\hspace{20pt}} c
}
\toprule
 & \textbf{Method} & \textbf{Epochs} & \multicolumn{5}{c}{\textbf{Loss weights}} & \textbf{Batch} & \textbf{Learning} \\
 \cmidrule(lr){4-8}
 &                 &                 & $E$ & $F$ & $q_{ilm}$ & $Q_\text{tot}$ & $P$ & \textbf{Size} & \textbf{Rate} \\
\midrule

\textbf{Warn-up}
& direct
& 0--10
& \multirow{2}{*}{100}
& \multirow{2}{*}{100}
& \multirow{2}{*}{100}
& \multirow{2}{*}{10{,}000}
& \multirow{2}{*}{1{,}000{,}000}
& \multirow{2}{*}{4}
& \multirow{2}{*}{0.01} \\

\textbf{Stage 1}
& --
& 10--200
& 
& 
& 
& 
& 
& 
&  \\
\bottomrule
\end{tabular}
\label{table:training_dynamics_field}
\end{table}

\subsubsection{Figure Contents}

A summary of which figures use which fits (in order from left to right for bar charts) is:
\begin{enumerate}
    \item figure 9: mace\_A, localcharges\_B, qeq\_B, fixedpoint\_G, energy\_D, localcharges\_A, qeq\_A, fixedpoint\_A, energy\_A. 
    \item figure 10: mace\_A, localcharges\_A, qeq\_A, fixedpoint\_D, fixedpoint\_E, fixedpoint\_F, energy\_A, energy\_B, energy\_C.
    \item figure 11: qeq\_A, fixedpoint\_B, energy\_B.
    \item figure 12: energy\_A, fixedpoint\_A
    \item figure 13: localcharges\_A, localcharges\_C, qeq\_A, fixedpoint\_I, fixedpoint\_A, energy\_E, energy\_A.
\end{enumerate}

\subsection{Performance Differences due to Training Method}
\label{sec:SI_training_method_comp}

\begin{figure}
    \centering
    \includegraphics[width=0.9\linewidth]{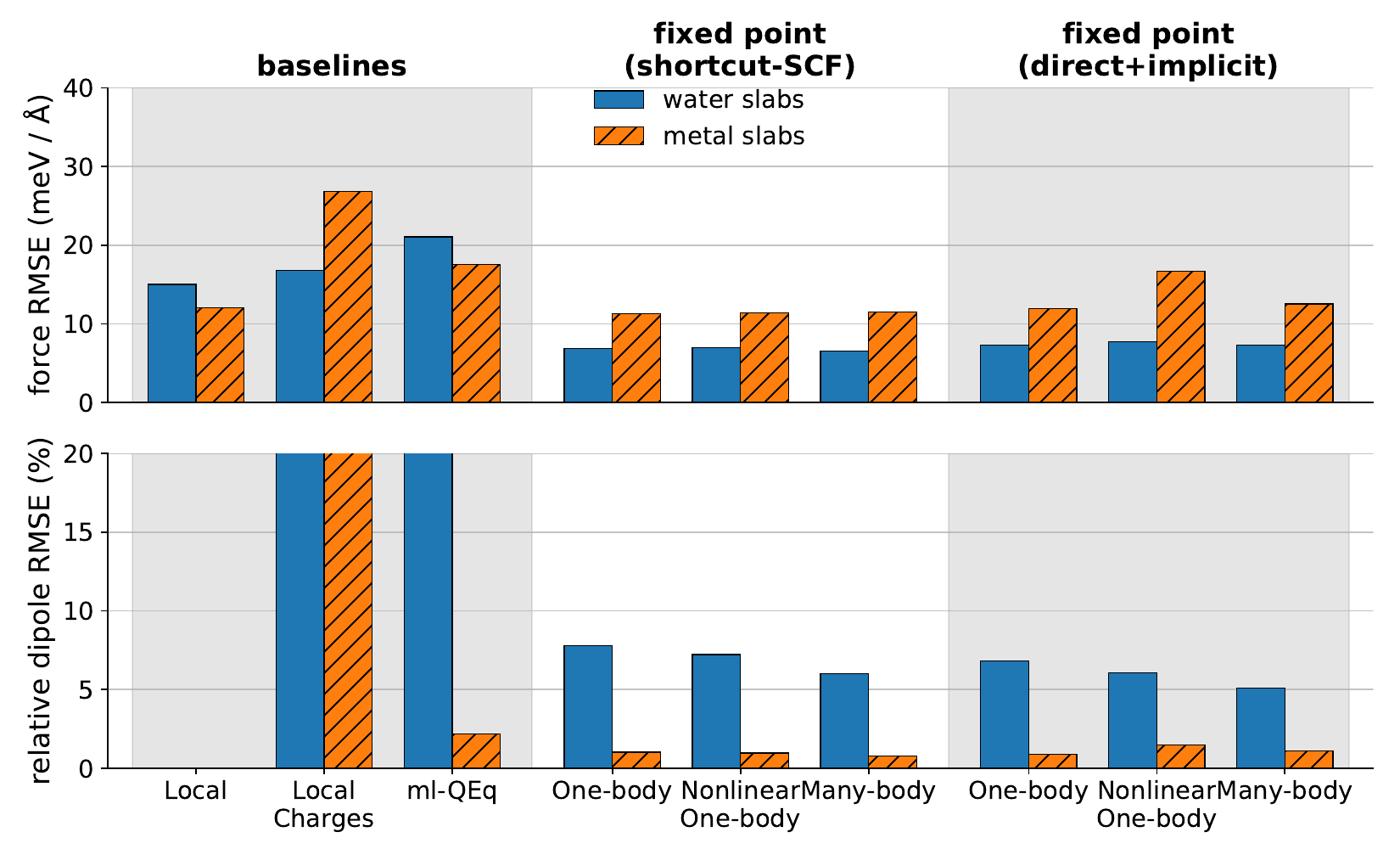}
    \caption{Difference in force and dipole error for fixed-point models when trained with either shortcut-SCF or direct+implicit training.}
    \label{fig:S_backprop_vs_implicit_errors}
\end{figure}

\begin{figure}
    \centering
    \includegraphics[width=0.4\linewidth]{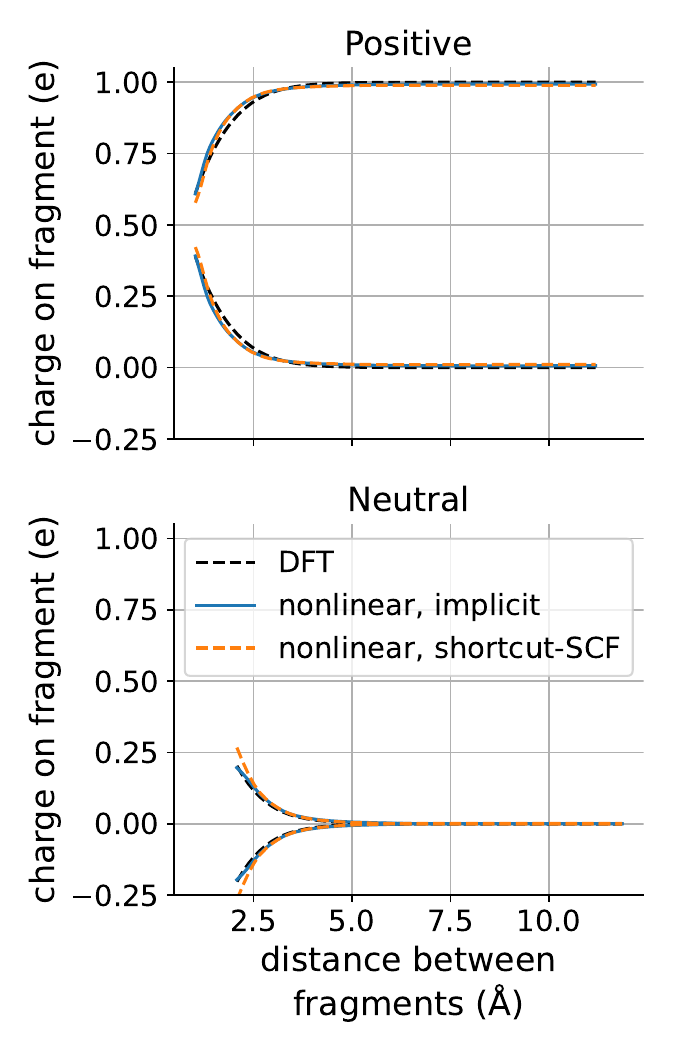}
    \caption{Difference in charge localization for fixed-point models when trained with either shortcut-SCF or direct+implicit training. One can see that shortcut-SCF training is marginally worse in both cases.}
    \label{fig:S_backprop_vs_implicit_frag}
\end{figure}

\begin{figure}
    \centering
    \includegraphics[width=0.9\linewidth]{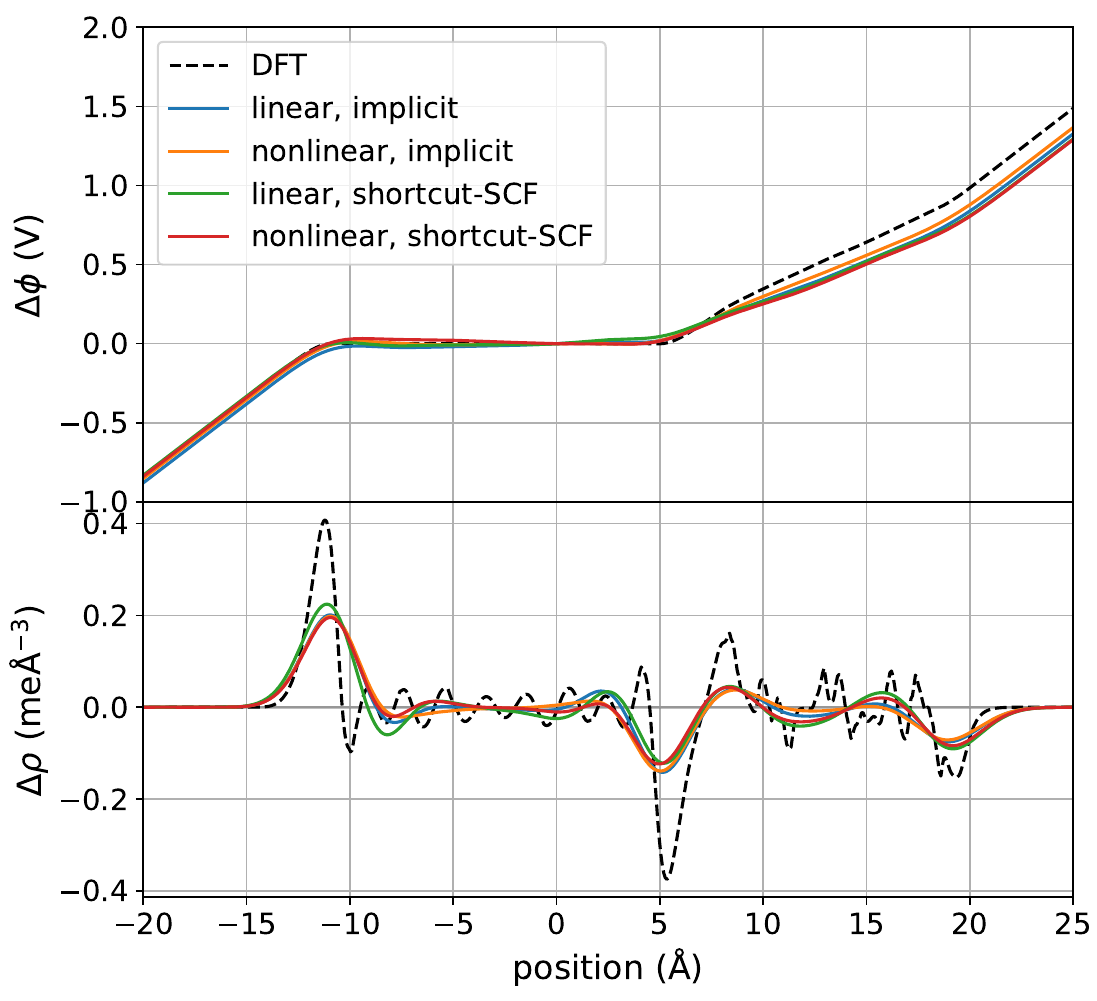}
    \caption{Difference in finite field response for fixed-point models when trained with either shortcut-SCF or direct+implicit training. The legend indicates the type of update function and the training schedule. One can see that the shortcut-SCF trained models seem to have more exaggerated differences in charge density, and slightly worse screening.}
    \label{fig:S_backprop_vs_implicit_response}
\end{figure}

We trained all fixed-point models using both shortcut-SCF schedule (table \ref{table:fixedpoint_schedule_slabs_backprop}) and the direct followed by implicit differentiation schedule (table \ref{table:fixedpoint_schedule_slabs_implicit}). This leads to some small differences. In our experience, on the the metal and water dataset, using direct training gave better physicality and extrapolation but slightly worse errors, particularly in atomic forces. This is shown in figures \ref{fig:S_backprop_vs_implicit_errors}--\ref{fig:S_backprop_vs_implicit_response}. Note that for the silicon dioxide test, we observed that direct training gave dramatically worse errors, and hence only used shortcut-SCF training. The figures come with several additional notes. Firstly, in figure \ref{fig:S_backprop_vs_implicit_errors}, the many-body fixed-point models were sometimes ill behaved, showing erratic training dynamics. This does not seem to be case with simpler, one-body fixed-point updates. When doing the final 30 epochs of implicit differentiation training, the loss of the many-body model began to rapidly increase. We found that training with a constant Fermi level, instead of at constant charge, solved the problem. Secondly, when unrolling the SCF loop the many-body model also suffered from some numerical instabilities, causing the training loss to suddenly diverge. On this occasion, the problem was fixed by simply restarting the fit with a different seed, but this illustrates the difficulties in training many-body fixed-point models. We expect that introducing some architectural bias towards stable fixed-point functions may improve the behaviour, which will be needed for more complicated models. 

\subsection{Training Using a Truncated Self Consistency Loop}
\label{appendix:shortscf}

\begin{figure*}
    \centering
    \includegraphics[width=0.9\linewidth]{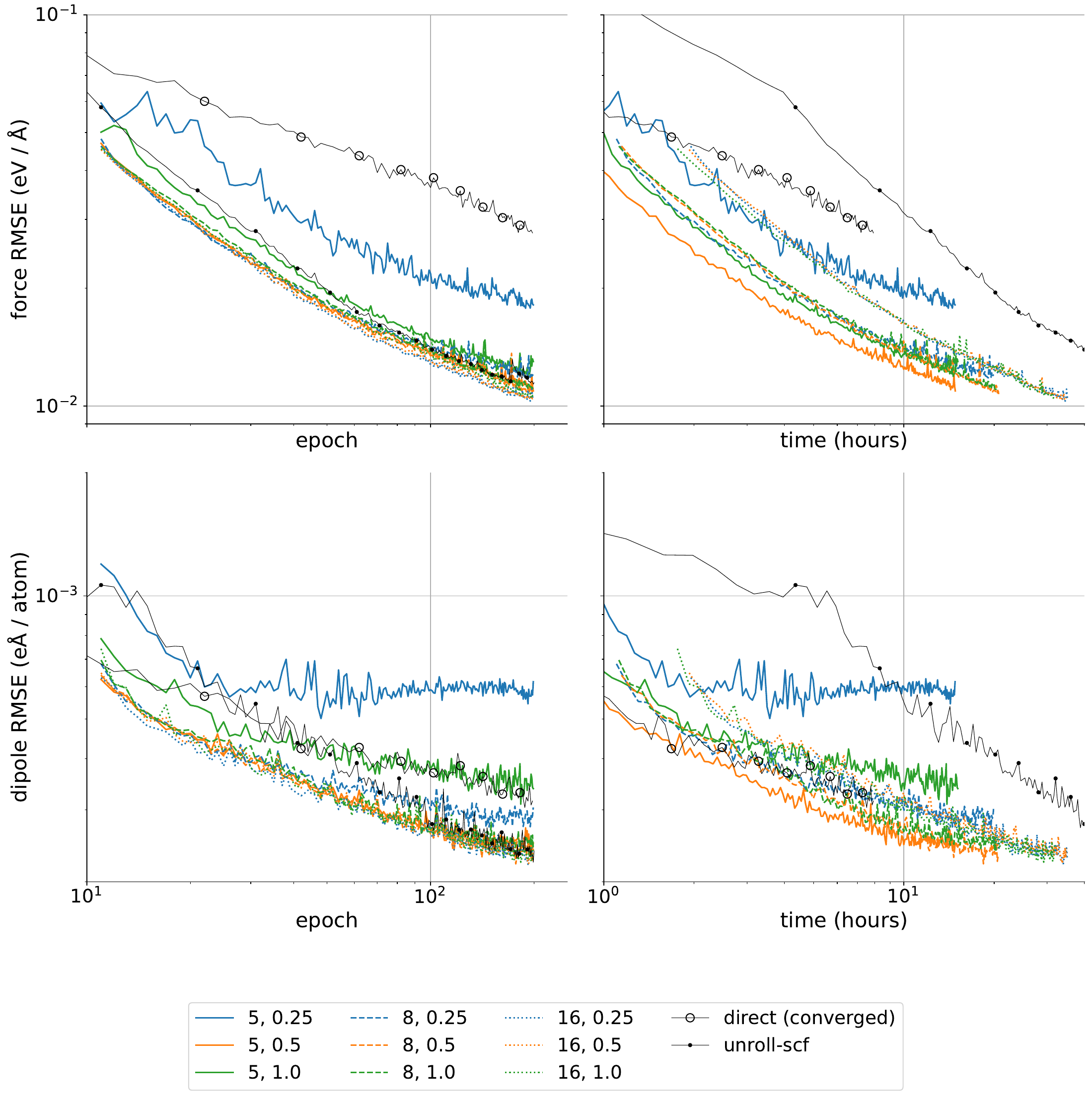}
    \caption{Training a fixed-point model with a truncated SC-cycle. The legend indicates the number of SC-steps, and the mixing parameter. For example `8, 0.5' means do exactly 8 SC steps with $\lambda=0.5$.}
    \label{fig:scf:shortcut_scf_noema}
\end{figure*}

Figure \ref{fig:scf:shortcut_scf_noema} shows the training dynamics of the fixed-point model when trained using a truncated SC-cycle. 

\subsection{The Effect of Field Features in a fixed-point Model}

\begin{figure*}
    \centering
    \includegraphics[width=0.85\linewidth]{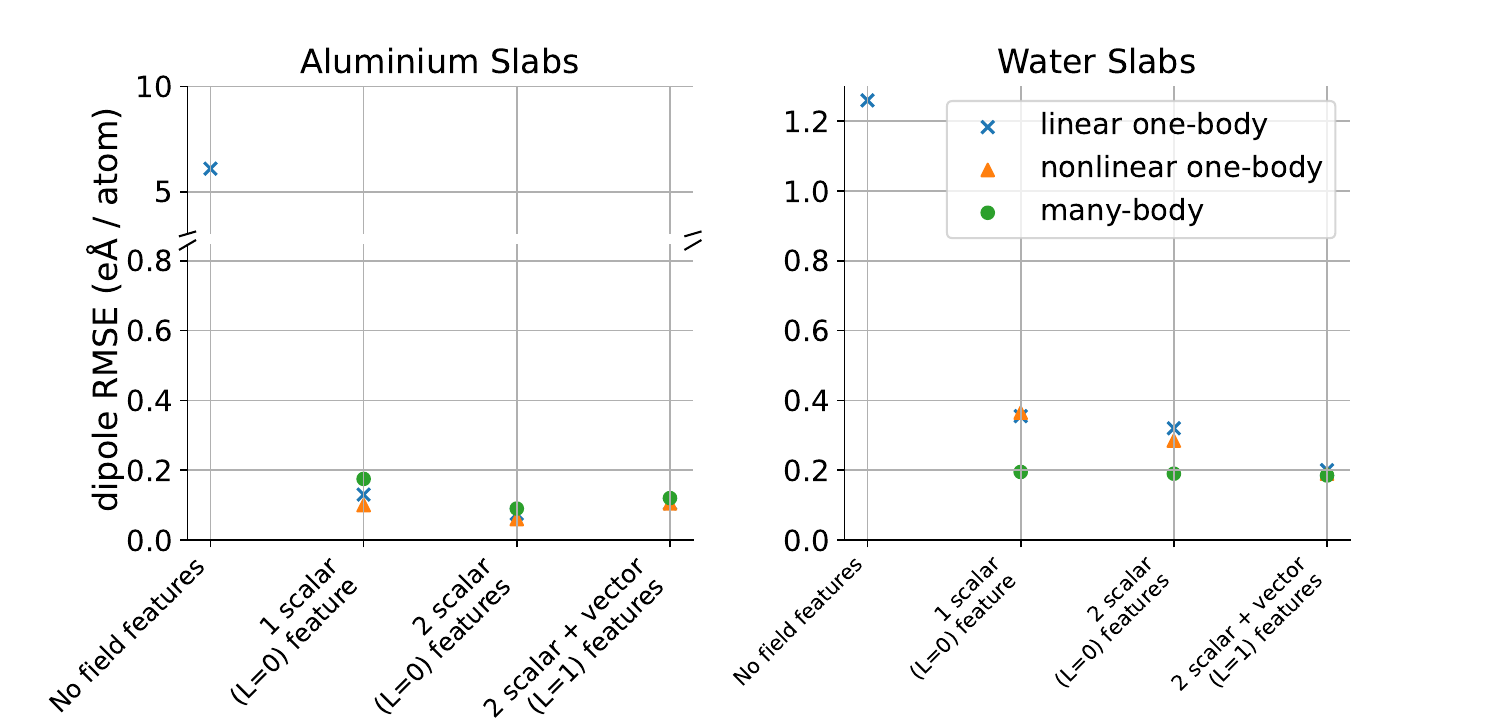}
    \caption{fixed-point model accuracy as a function of update function complexity and the number of electrostatic features. The column `No field features' is the prediction of the MACE-LocalCharges model since this is analogous to the purely local part of the fixed-point model.}
    \label{fig:scf:fixed_point_chars}
\end{figure*}

We now investigate how some of the architectural choices in the fixed-point model effect the performance. This model has some additional flexibility compared to the energy functional, since one can choose to have a rich description of the effective potential $v(r)$ by extracting many equivariant potential features. The effective potential features are computed as:
\begin{align*}
    v_{i,nlm} = \frac{1}{\mathcal{N}_{lm}}\int d\mathbf{r} \ \phi_{nlm}(\mathbf{r} - \mathbf{r}_i) v(\mathbf{r})
\end{align*}
In which $\phi_{nlm}$ is a Gaussian type orbital. One can make a richer description of $v$ by increasing the range of $n$ and $l$. Figure \ref{fig:scf:fixed_point_chars} shows how the accuracy of the total dipole prediction changes when using a different number of these features. The exact choices for $\sigma$ and $l$ which correspond to the different points in the figure are given in Table \ref{table:scf:field_feat_options}. The experiment was performed for the one-body linear, one-body nonlinear and many-body update functions. All the fits were done using direct training for 200 epochs followed by implicit differentiation for 30 epochs.

One can see that as the number of electric potential features is increased, the accuracy improves for both the one-body linear and one-body nonlinear models, with the best accuracy being achieved with multiple scalar and vector features. On the other hand, the many-body model is able to reach the same accuracy with only a single scalar feature. It is therefore possible to trade-off the complexity of the update function with the richness of the charge density description. In practice, we find that using more field features and a simpler update function is always less computationally expensive. 

One can also see clear differences between the metal and water subsets. Good accuracy for the metal slabs is achieved in all models with only one scalar feature. Using more electric potential features is relatively cheap, and therefore in all other fixed-point model fits in this study, we use 2 scalar and 2 vector features corresponding to the last line in Table \ref{table:scf:field_feat_options}.

\begin{table}[h]
    \centering
    \begin{tabular}{l  c c}
        \hline
        \textbf{Description} & $\sigma$ & $l$ \\
        \hline
        one scalar     & 2.0 & 0 \\
        two scalar         & 1.5, 2.5 & 0  \\
        two scalar + vector & 1.5, 2.5 & 0, 1  \\
        \hline
    \end{tabular}
    \caption{Choices of $\sigma$ of $l$ used in Figure \ref{fig:scf:fixed_point_chars}}
    \label{table:scf:field_feat_options}
\end{table}

\newpage
\pagebreak

\section{Oxygen Vacancies in Silicon Dioxide}
\subsection{Dataset Contents}

Configurations were generated by MD simulations with Hookian constraint on Si-Si bond length. We ran simulations with MACE-MATPES where the Si-Si bond distance across the oxygen vacancy was constrained at values between 2.4 \AA \space and 4.75 \AA. The Hookian spring stiffness was set to 5 eV$/\text{\AA}^2$. In total, we ran 12 simulations, each of length 100 ps for the smaller 143 atom supercell, and another 12 simulations for the larger 242 atom supercell, at 400 K.

\begin{figure}
    \centering
    \includegraphics[width=0.65\linewidth]{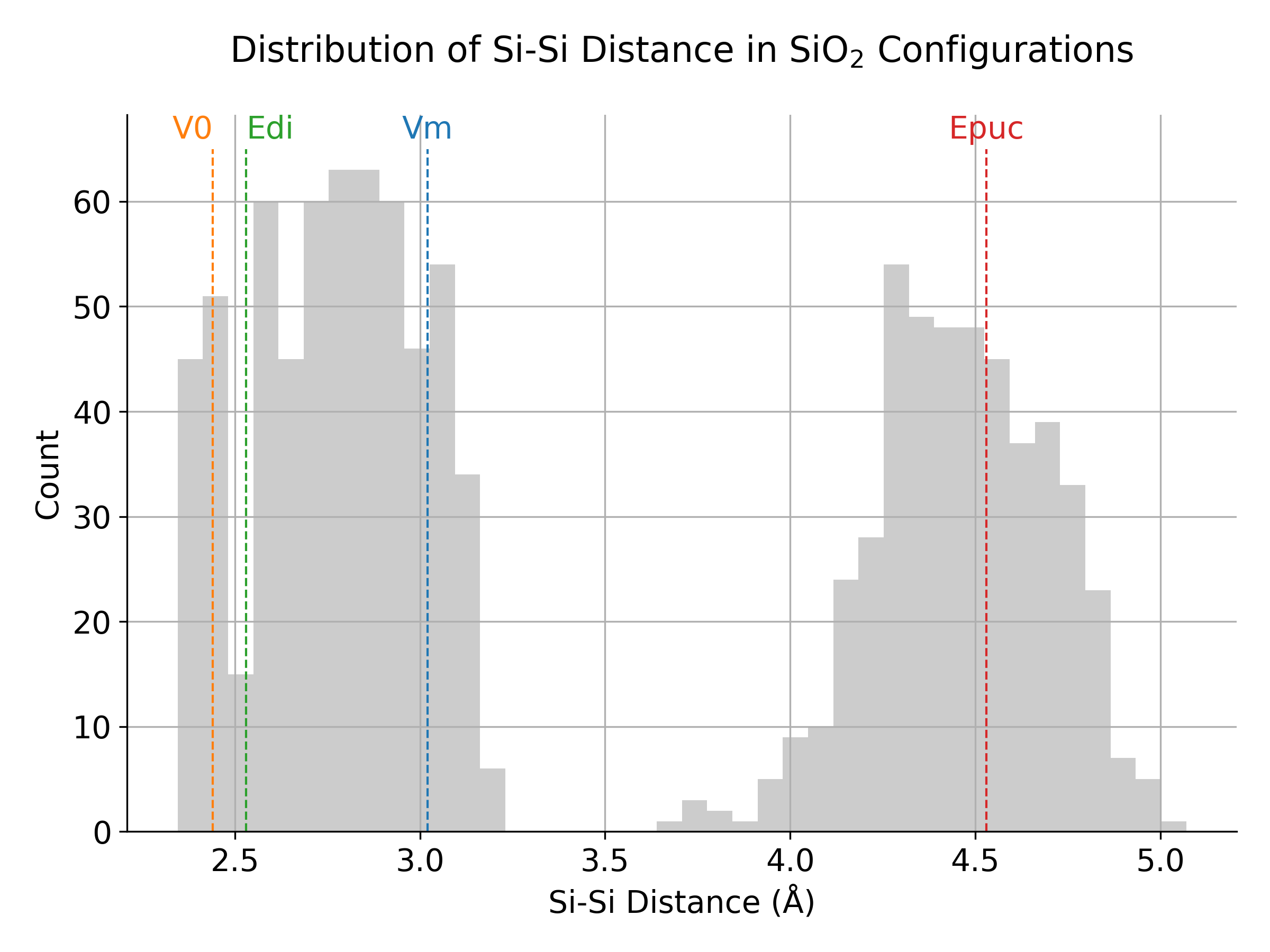}
    \caption{Histogram of Si--Si distances in the training set for the silicon dioxide example.}
    \label{fig:distance_distribution}
\end{figure}

We found that for bond constraints less than 3.5 \AA, the structures generally resembled a Si-Si dimer geometry, whereas for longer bond lengths the structures always looked like the puckered vacancy in which one oxygen atom becomes 3-fold coordinated. We sampled configurations randomly from all these trajectories, choosing (i) 300 frames with bond lengths less than 3.5 \AA \space (200 samples of 143 atom supercells and 100 samples from 242 atom supercells) and (ii) 374 with Si-Si distances greater than 3.5 \AA \space (250 from 143 atom supercells and 124 from 242 atom supercells). The distribution of Si-Si distances from this procedure is shown in figure \ref{fig:distance_distribution}. One can compare this distribution to the Si--Si bond length in the DFT relaxed defect structures in table \ref{table:dft_defect_distances}.

Finally, in addition to the configurations sampled from molecular dynamics, we added two copies of each of the DFT relaxed defect geometries, in both the smaller and larger supercells. 

For performing the DFT calculations, the long bond length configurations were all evaluated with +1 total charge, since they all resemble the positive puckered oxygen vacancy. The remaining configurations were all labelled in all three charge states (-1, 0, +1). The DFT relaxed defect geometries were added in only the charge state associated with that geometry.

\begin{table}[]
    \centering
    \begin{tabular}{c c}
        \toprule
        \quad Defect \quad &  Si-Si Distance (\AA) \\
        \midrule
        $V_\text{O}$            & 2.44 \\
        $V_\text{m}$               & 2.53 \\
        $E_\text{di}$         & 3.02 \\
        $E_\text{puck}$        & 4.53 \\
        \bottomrule
    \end{tabular}
    \caption{Si-Si distances for various defects after DFT relaxations with our DFT settings in a 242 atom supercell.}
    \label{table:dft_defect_distances}
\end{table}

\subsection{Loss weights and Training schedule}

Four kinds of models were trained for the silicon dioxide example. The training schedules for the global embedding MACE, MACE-QEq, fixed-point models and energy models are given in tables \ref{table:schedule_sio2_embedding}, \ref{table:schedule_sio2_qeq}, \ref{table:schedule_sio2_fp}, \ref{table:schedule_sio2_energy} respectively. One can note that the Fermi level is included in the loss for the fixed-point models.

\begin{table}[ht]
\centering
\caption{Global Embedding MACE Training Schedule for \ce{SiO2}}
\label{table:schedule_sio2_embedding}

\renewcommand{\arraystretch}{1.25}
\setlength{\tabcolsep}{6pt}

\begin{tabular}{
l
@{\hspace{20pt}} l
@{\hspace{20pt}} c
@{\hspace{20pt}} c|c|c|c|c
@{\hspace{20pt}} c
@{\hspace{20pt}} c
}
\toprule
 & \textbf{Method} & \textbf{Epochs}
 & \multicolumn{5}{c}{\textbf{Loss weights}}
 & \textbf{Batch} & \textbf{Learning} \\
\cmidrule(lr){4-8}
 &                 & 
 & $E$ & $F$ & $q_{ilm}$ & $Q_\text{tot}$ & $\mu$
 & \textbf{Size} & \textbf{Rate} \\
\midrule

\textbf{Stage 1}
& --
& 0--250
& 100
& 100
& 0
& 0
& 0
& 4
& 0.005 \\

\textbf{Stage 2}
& --
& 250--300
& 100
& 100
& 0
& 0
& 0
& 4
& 0.002 \\
\bottomrule
\end{tabular}
\end{table}

\begin{table}[ht]
\centering
\caption{MACE-QEq Training Schedule for \ce{SiO2}}
\label{table:schedule_sio2_qeq}

\renewcommand{\arraystretch}{1.25}
\setlength{\tabcolsep}{6pt}

\begin{tabular}{
l
@{\hspace{20pt}} l
@{\hspace{20pt}} c
@{\hspace{20pt}} c|c|c|c|c
@{\hspace{20pt}} c
@{\hspace{20pt}} c
}
\toprule
 & \textbf{Method} & \textbf{Epochs}
 & \multicolumn{5}{c}{\textbf{Loss weights}}
 & \textbf{Batch} & \textbf{Learning} \\
\cmidrule(lr){4-8}
 &                 & 
 & $E$ & $F$ & $q_{ilm}$ & $Q_\text{tot}$ & $\mu$
 & \textbf{Size} & \textbf{Rate} \\
\midrule

\textbf{Stage 1}
& --
& 0--200
& 10
& 1000
& 1000
& \multirow{2}{*}{0}
& \multirow{2}{*}{0}
& 4
& 0.005 \\

\textbf{Stage 2}
& --
& 200--250
& 1000
& 100
& 1000
& 
& 
& 4
& 0.002 \\
\bottomrule
\end{tabular}
\end{table}

\begin{table}[ht]
\centering
\caption{Fixed Point Models Training Schedule for \ce{SiO2}}
\label{table:schedule_sio2_fp}

\renewcommand{\arraystretch}{1.25}
\setlength{\tabcolsep}{6pt}

\begin{tabular}{
l
@{\hspace{20pt}} l
@{\hspace{20pt}} c
@{\hspace{20pt}} c|c|c|c|c
@{\hspace{20pt}} c
@{\hspace{20pt}} c
}
\toprule
 & \textbf{Method} & \textbf{Epochs}
 & \multicolumn{5}{c}{\textbf{Loss weights}}
 & \textbf{Batch} & \textbf{Learning} \\
\cmidrule(lr){4-8}
 &                 & 
 & $E$ & $F$ & $q_{ilm}$ & $Q_\text{tot}$ & $\mu$
 & \textbf{Size} & \textbf{Rate} \\
\midrule

\makecell[l]{\textbf{Warm-up}\\\textbf{(Stage 0)}}
& Direct
& 0--10
& 100
& \multirow{3}{*}{100}
& \multirow{3}{*}{1000}
& \multirow{3}{*}{10{,}000}
& \multirow{3}{*}{1000}
& 4
& 0.01 \\

\textbf{Stage 1}
& Shortcut SCF
& 10--250
& 100
&  &  &  &
& 4
& 0.005 \\

\textbf{Stage 2}
& Unrolled SCF
& 250--300
& 1000
&  &  &  &
& 4
& 0.002 \\
\bottomrule
\end{tabular}
\end{table}

\begin{table}[ht]
\centering
\caption{Energy Functional Models Training Schedule for \ce{SiO2}}
\label{table:schedule_sio2_energy}

\renewcommand{\arraystretch}{1.25}
\setlength{\tabcolsep}{6pt}

\begin{tabular}{
l
@{\hspace{20pt}} l
@{\hspace{20pt}} c
@{\hspace{20pt}} c|c|c|c|c
@{\hspace{20pt}} c
@{\hspace{20pt}} c
}
\toprule
 & \textbf{Method} & \textbf{Epochs}
 & \multicolumn{5}{c}{\textbf{Loss weights}}
 & \textbf{Batch} & \textbf{Learning} \\
\cmidrule(lr){4-8}
 &                 & 
 & $E$ & $F$ & $q_{ilm}$ & $Q_\text{tot}$ & $\mu$
 & \textbf{Size} & \textbf{Rate} \\
\midrule

\textbf{Stage 1}
& Unrolled SCF
& 0--400
& 100
& \multirow{2}{*}{100}
& \multirow{2}{*}{1000}
& \multirow{2}{*}{0}
& \multirow{2}{*}{0}
& 4
& 0.01 \\

\textbf{Stage 2}
& Unrolled SCF
& 400--500
& 1000
&  &  &  &
& 4
& 0.01 \\
\bottomrule
\end{tabular}
\end{table}